\documentclass[fleqn,usenatbib]{mnras}

\usepackage{newtxtext,newtxmath}

\usepackage[T1]{fontenc}

\DeclareRobustCommand{\VAN}[3]{#2}
\let\VANthebibliography\thebibliography
\def\thebibliography{\DeclareRobustCommand{\VAN}[3]{##3}\VANthebibliography}

\usepackage{graphicx}	
\usepackage{amsmath}	
\usepackage[dvipsnames,svgnames]{xcolor}
\usepackage{siunitx}
\usepackage{color}
\usepackage{makecell}
\usepackage{xspace} 

\definecolor{darkgreen}{rgb}{0.0,0.5,0.0}
\definecolor{darkredGary}{rgb}{0.65,0.1,0.35}
\definecolor{magenta}{rgb}{0.8,0,0.8}
\definecolor{purple}{rgb}{0.5,0,0.5}
\definecolor{gray}{rgb}{0.5,0.6,0.7}
\definecolor{orange}{rgb}{1.0,0.3,0.0}
\definecolor{teal}{rgb}{0,0.45,0.35}
\definecolor{Black}{rgb}{0,0,0}

\newcommand{\Compact}{Compact}
\newcommand{\Compacts}{Compacts}
\newcommand{\CompactMB}{Compact$_\mathrm{MB}$}
\newcommand{\CompactsMB}{Compacts$_\mathrm{MB}$}
\newcommand{\CompactMBs}{Compacts$_\mathrm{MB}$}
\newcommand{\CompactSB}{Compact$_\mathrm{SB}$}
\newcommand{\CompactsSB}{Compacts$_\mathrm{SB}$}
\newcommand{\CompactSBs}{Compacts$_\mathrm{SB}$}

\newcommand{\Normals}{Normals}
\newcommand{\subfind}{\textsc{subfind}\xspace}
\newcommand{\sublink}{\textsc{SubLink}\xspace}
\newcommand{\rockstar}{\textsc{rockstar}\xspace}

\defcitealias{DeAlmeida2024AA}{Paper~I}

\newcommand{\msun}{\,\mathrm{M}_\odot}
\newcommand{\lmsun}{\log(M_\star/{\rm M}_\odot)}

\usepackage{upgreek}
\renewcommand{\pi}{\uppi}
\title[Formation of compact satellite dwarf galaxies]{Satellite compaction pathways: environmental drivers shaping dwarf galaxy corpulence in the TNG50 simulation}

\author[A. P. De Almeida et al.]{
Abhner P. de Almeida,$^{1,2}$\thanks{Email: abhner.almeida@usp.br}
Gary A. Mamon,$^{2}$
and Gastão B. Lima Neto$^{1}$
\\
$^{1}$Instituto de Astronomia, Geof\'isica e Ci\^encias Atmosf\'ericas (Universidade de S\~ao Paulo), R. do Mat\~ao, 1226, S\~ao Paulo - SP, 05508-090, Brasil \\
$^{2}$Institut d'Astrophysique de Paris (UMR 7095: CNRS \& Sorbonne Universit\'e), 98 bis Bd Arago, F-75014 Paris, France 
}

\date{Accepted 2026 June 08. Received 2026 June 02; in original form 2026 March 20}

\pubyear{2026}

\begin{document}
\label{firstpage}
\pagerange{\pageref{firstpage}--\pageref{lastpage}}
\maketitle

\begin{abstract}
We explore the physical mechanisms driving dwarf galaxy corpulence, focusing on those that end up as compact satellites.
We select dwarf galaxies at $z=0$ with $\lmsun$ between 8.4 and 9.2 from the TNG50 hydrodynamical simulation after excluding systems flagged as potentially spurious. \Compact{} dwarfs are defined according to the $z$=0 size--mass relation as those on the lower envelope of its main branch or on its lower-size secondary branch, while `Normals' lie on the main branch spine.
We identify two robust compaction pathways and a third, more tentative, channel: 1) Compact satellites that remain rich in dark matter (DM) inhabit poorer environments having fewer mergers, favouring the accretion of lower-angular-momentum gas. This allows gas inflows that drive concentrated inner star formation and compaction, as previously found for centrals. 
2) Most DM-poor satellites (which typically end up red and metal-rich for their stellar mass) undergo compaction mainly caused by tidal stripping of outer stars.  Their compaction is faster when gas is present, by at least 15 per cent after correcting for the stronger tidal field. 3) For most of our few very metal-rich DM-poor Compact satellites, the major compaction phase begins with a starburst driven by ram pressure compression near first pericentre, even if much of the compaction often occurs during subsequent tidal stripping. As a result, compact dwarf satellites in TNG50 arise through distinct pathways. We discuss how numerical effects can affect this conclusion.
\end{abstract}




\section{Introduction}

Group and cluster environments can have different effects on galaxy evolution, mainly driven by gravitational processes involving collisional (gas) and non-collisional (dark matter and stars) components. For the former, ram-pressure stripping \citep{Gunn1972ApJ} is the primary process associated with the removal of interstellar gas, caused by the pressure from the intra-cluster (or intra-group) hot plasma. For the latter, tides from the host are the main mechanism disrupting galaxies and removing their external material in clusters \citep{Merritt1983ApJ} and groups \citep{Mamon87}. Other physical processes can also be important, such as collisional tides \citep{Richstone1976ApJ}, harassment \citep{Moore1996Natur}, starvation \citep*{Larson+80}, and feedback from a nearby galaxy with an active galactic nucleus \citep{Dashyan+19}.

These different processes are crucial to explain, for example, the observed segregations of galaxies within groups and clusters according to their morphology  \citep[e.g.,][]{Melnick+77, Dressler+80, Whitmore&Gilmore91} and star formation \citep{Balogh+97,Dressler+99,Balogh1999ApJ}. The latter segregation is related to quenching of star formation due to the loss of star-forming gas. Environmental interactions are also key to the assembly of the brightest cluster galaxy (e.g., \citealt{deLucia2007MNRAS}), the appearance of the diffuse intra-cluster light \citep{Contini+14}, as well as the chemical enrichment of the intra--cluster gas \citep*{DeLucia+04}.

The environment is also expected to affect galaxy \emph{corpulence}, which we use hereafter to refer to how compact or diffuse a galaxy is at fixed stellar mass. Environment is a likely culprit to explain the bimodality of the sizes of low-mass stellar systems \citep{Misgeld2011MNRAS}. Observations of compact dwarf galaxies, both ultra-compact dwarf galaxies (UCDs) and compact ellipticals (cEs), in galaxy clusters or in close proximity to massive galaxies, suggest formation scenarios associated with tidal stripping of larger elliptical galaxies (for cEs) and dwarf galaxies (for UCDs). In these scenarios, only the core of these objects remains \cite[e.g.,][]{Faber1973ApJ, Chilingarian&Mamon08, Brodie2011AJ, Pfeffer2013MNRAS, Kim2020ApJ, Wang2023Nature}. The tidal stripping scenario is also related to the formation of dwarf early-type galaxies (dEs), which can be the remnants of late-type galaxies \cite[e.g.,][]{Kormendy2012ApJS, Janz2016MNRAS}; and the formation of some compact galaxies as the remains of spheroidal dwarf galaxies (dSphs) \citep[e.g.,][]{Zinnecker1988, DOnghia2016, Ibata2019}. 

In general, tidal stripping acts by removing material from the outer layers of a galaxy, leaving behind only the core, with physical properties similar to those observed in galaxies that have retained their outer material. In this context, numerical  simulations (both $N$-body and hydrodynamical) are valuable tools to verify the tidal stripping scenario in the formation of dwarf galaxies, as we can track the galaxy's orbit and observe the removal of outer material. Tidal interactions have been probed with idealized simulations (e.g., \citealt*{Bassino1994ApJ}, \citealt*{Bekki2001}, \citealt{Klimentowski2009}, \citealt{Pfeffer2013MNRAS}), as well as with large-scale cosmological simulations \citep[e.g.,][]{Pfeffer2014, Fattahi2018MNRAS}, which add another layer of complexity due to the hierarchical assembly of structures.

In addition to tidal stripping,  gas dynamical processes have also been identified as a mechanism for forming compact stellar systems. In particular, \cite{Du+19} used an idealized simulation showing that tidal compression of the satellite gas leads to inner star formation, whose subsequent stellar feedback leads to metal-rich compact galaxies.  \cite{Bian2025ApJ}, from the same team, acquired statistics on galaxies that end up as very metal-rich compact satellites, using the TNG50 cosmological simulation, and deduced that the inner star formation was the consequence of ram pressure compression of its gas from the host. 

Our goal is to reach a global view of the different physical mechanisms leading to the present-day compact and diffuse dwarf galaxies in the present-day Universe.  In \citeauthor{DeAlmeida2024AA} (\citeyear{DeAlmeida2024AA}, hereafter \citetalias{DeAlmeida2024AA}{}), we showed that central dwarf galaxies become compact because they live in poor environments at $z \sim 1$, suffer fewer mergers thereafter, and are thus the only galaxies with pronounced gas infall, leading to concentrated star formation. In this paper, we explore the evolution of  galaxies that end up as satellite compact dwarfs, focusing on the different physical mechanisms causing their compaction and their link with the final properties of these galaxies.

This paper is organized as follows. In Sect.~\ref{sec:sim&sample}, we present the TNG50 simulation as well as  our  selection of compact galaxies and a control sample of normal galaxies in the  same range of stellar masses. In Sect.~\ref{sec:03} we discuss the different environmental impact across different populations in our sample. In Sects.~\ref{sec:evol} and ~\ref{sec:details}, we discuss the evolution of populations  that are respectively weakly affected by the environment and those that are significantly influenced by it, exploring the physical  mechanisms driving their size evolution. We study the colour and metallicity evolution of our compact galaxies in Sect.~\ref{sec:color}. We discuss our results in Sect.~\ref{sec:discuss} and summarise them in Sect.~\ref{Conclusion}.

\section{Simulation and sample}
\label{sec:sim&sample}

\subsection{IllustrisTNG-50}
\label{sec:TNG}

To understand the evolution of compact satellite dwarf galaxies, we use the Illustris TNG50-1 (hereafter, TNG50) simulation  \citep{Nelson2019MNRAS, Nelson2019ComAC, Pillepich2019MNRAS}, which is the best resolved  run in the suite of cosmological magneto-hydrodynamical simulations IllustrisTNG \citep[hereafter, TNG][]{Springel2018MNRAS, Pillepich2018bMNRAS, Marinacci2018MNRAS, Naiman2018MNRAS, Nelson2018MNRAS}. These simulations reproduce the evolution of the Universe in a variety of box sizes and resolutions, following the evolution and interaction of different components (gas, dark matter, stars, and black holes) with the {\sc AREPO} magneto-hydrodynamics moving mesh code \citep{Springel10}.  TNG employs sub-grid recipes for star formation, stellar feedback, metal enrichment, and black hole (BH) physics (seeding and feedback) for the baryonic component \citep[e.g.,][]{Vogelsberger2013MNRAS, Weinberger2017MNRAS, Pillepich2018MNRAS}. The initial conditions for the simulations are based on the cosmological parameters from \cite{Planck2016AA}: a flat $\Lambda$CDM cosmology with a Hubble constant of $H_0 = 67.74\, \mathrm{km \,s^{-1} \, Mpc^{-1}}$, matter density $\Omega_{\rm m} = 0.3089$, baryon density $\Omega_{\rm b} = 0.0486$, power spectrum normalisation $\sigma_8 = 0.8159$, and primordial spectral index $n_s = 0.9667$. 

TNG50 is limited to a volume of $51.7^3\, \mathrm{Mpc^3}$ with dark matter (DM) and target baryonic mass resolutions of $m_\mathrm{DM} = 4.5 \times 10^5 \msun$ and $m_\mathrm{bar} = 8.5 \times 10^4 \msun$, respectively, where `target' refers to the reference gas-cell mass maintained by {\sc AREPO} refinement and derefinement; and low-redshift softening lengths of 288 pc (for stars and DM) and  74 pc (for gas). For dwarf galaxies with stellar masses $M_\star > 10^{8.4} {\rm M_\odot}$, this provides good mass resolution with the least massive systems containing of $\sim$700 DM particles and at least $\sim$3500 stellar particles. This does not, however, imply uniform structural convergence, and the smallest systems must still be interpreted with caution when their  half-stellar--mass radii approach the low-redshift stellar softening scale.

For our study, galaxies are subhalos identified using the \subfind algorithm \citep{Springel2001MNRAS, Dolag2009MNRAS}, which can be either in a group with other subhalos or isolated. The main halo (which can be a group or an isolated galaxy) is identified by the Friends-of-Friends algorithm \citep{Davis1985ApJ}. When discussing satellites, we will use `group' for their halo. We select and study galaxies using the database provided by the TNG collaboration.\footnote{https://www.tng-project.org/data/} Additional details can be found in \citetalias{DeAlmeida2024AA}.

\subsection{Sample selection}\label{sec:sample}

\begin{figure}
  \centering
  \includegraphics[width=\hsize]{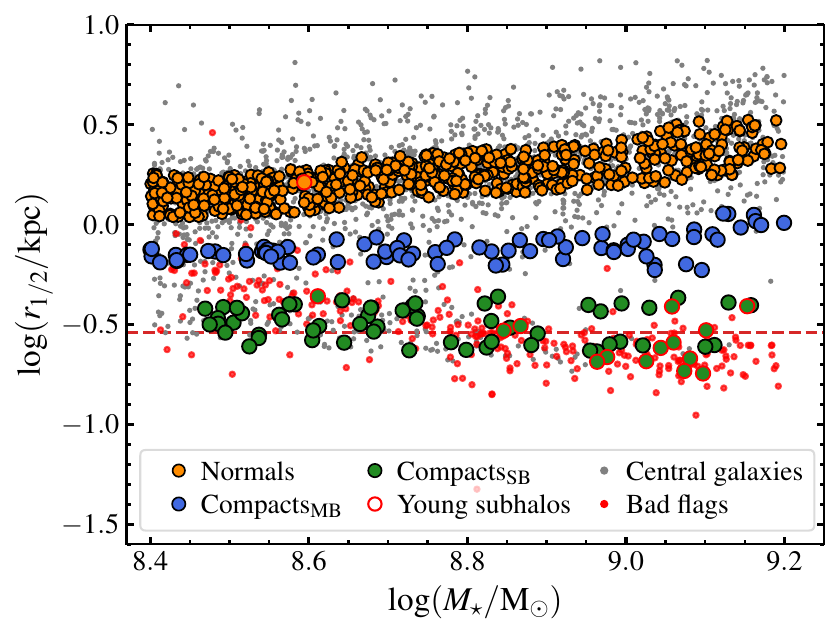}
  \caption{Present-day half-stellar-mass radius vs. stellar mass (within twice the half-stellar-mass radius), highlighting our adopted samples of satellite \Compacts{} and \Normals{}. The orange, blue, and green circles represent the different  samples, respectively: \Normals{}, \CompactsMB{} (main branch), and \CompactsSB{} (secondary branch).  The grey dots indicate central subhalos, while the red dots show all the bad-flag galaxies. The symbols with red circles mark the young galaxies ($z_\mathrm{birth} < 1$), which are shown for context but are excluded from the analysis. The  red dashed line indicates the $z$=0 softening length of the stellar and dark matter particles of TNG50.}
   \label{fig:sizevsmass_classes}
   \end{figure}

We select dwarf galaxies with stellar masses measured within two half-stellar-mass radii ($r_{1/2}$) between $10^{8.4} < M_\star/{\rm M_\odot} < 10^{9.2}$ and define three populations according to their location in the present-day stellar size–mass relation, as shown in Fig.~\ref{fig:sizevsmass_classes}. This relation presents a `main branch' with size increasing with mass, and a lower `secondary branch', where size decreases with mass.  `\CompactsSB{}' are those located on the secondary branch, with  $r_{1/2}$ below 447~pc. `\CompactsMB{}' are those taken from the lower envelope of the main branch, below its 5th  percentile.  The `Normals' represent a control sample located on the spine of the main branch (within its 25th to 75th percentiles). Throughout this work, we define `compactness' to describe how small a galaxy is, at given stellar mass, relative to the present-day stellar size–mass relation. By contrast, we use `compaction' to refer to the evolutionary decrease in  half-stellar--mass radius that drives a galaxy towards one of our compact classes.

We conservatively ensure that we analyse only bona fide galaxies, by restricting our sample to good-flag subhalos (with \texttt{SubhaloFlag}=1) as in \citetalias{DeAlmeida2024AA}. In TNG, a bad-flag subhalo typically corresponds to non-cosmological baryonic fragments or transient structures forming within one virial radius of a massive host and with very low dark-matter fractions \citep{Nelson2019ComAC}.  As in \citetalias{DeAlmeida2024AA}, we define the galaxy `birth' epoch as the earliest snapshot identified by the \sublink  main progenitor merger tree. We hereafter refer to the redshift of this snapshot as the galaxy \sublink  birth epoch, $z_\mathrm{birth}$, and to the corresponding lookback time as the \sublink age. Galaxies formed at $z_\mathrm{birth} < 1$ (hereafter young galaxies) are excluded from our analysis, in order to focus on systems with well-sampled evolutionary tracks over sufficiently long timescales. 
 
We define satellite galaxies as those that are in a group at $z=0$, by comparing the galaxy ID with the {\tt GroupFirstSub}. We include backsplash systems, i.e. galaxies that have recently passed through the inner regions of a cluster but are now beyond the virial radius (e.g., \citealt{Balogh2000, Sales2007}). For this, we check central galaxies at $z = 0$ that previously had one or several passages through a group.  Hereafter, we will refer to `satellites' any galaxy that ends up as a satellite or is a backsplash, as defined above, even during the epochs when it is a central galaxy, unless specified otherwise. All galaxies that are not satellites are centrals.

\begin{table}
\caption{Subsamples of satellite dwarfs}
\centering
    \begin{tabular}{l S[table-format=3.0] S[table-format=3.0] S[table-format=3.0]}
    \hline
    \hline
      & \multicolumn{1}{c}{Normals}
      & \multicolumn{1}{c}{\CompactsMB{}} 
      & \multicolumn{1}{c}{\CompactsSB{}} \\
\hline
All & 611 & 83 & 57 \\
DM-rich & 603 & 29 & 41 \\
DM-poor & 8 & 54 & 16 \\
\hline
    \end{tabular}
    \label{tab:sample}
\end{table}
To avoid biases driven by stellar mass, we ensure that the three size classes have statistically indistinguishable stellar mass distributions at $z = 0$, following the same procedure adopted in \citetalias{DeAlmeida2024AA}. Taking the \CompactsMB{} population as a reference, we iteratively resample the \CompactsSB{} and \Normals{} populations until their stellar-mass distributions match those of the \CompactsMB{} sample. This yields 611 Normals, 83 \CompactMBs\ and 57 \CompactSBs{} satellite galaxies (see Table~\ref{tab:sample}). These correspond to $47$, $63$, and $40$ per cent with respect to the total population of each class, respectively. Only $7$, $2$ and $3$ per cent of satellite Normals, \CompactsMB{} and \CompactsSB{} are backsplash galaxies, respectively. We also note that the \CompactsSB{} lie close to the low-redshift TNG50 softening scale, with 20 of the 57 lying below it.

Bad-flag systems represent only a very small fraction of the Main Branch population. In particular, among \CompactsMB{} satellites, only 9 objects (3 per cent of all bad-flag subhalos) are classified as bad-flag, while for Normal satellites this fraction is negligible, with only 2 objects (0.7 per cent). Therefore, the evolutionary trends discussed in this work are overwhelmingly defined by genuine cosmological galaxies. On the other hand, most \CompactsSB{} satellites are classified as bad-flag (230 objects, 82 per cent). Appendix~\ref{sec:samplebad} shows that the selected `old' (i.e. born at $z > 5$) \CompactsSB{} are born as centrals and have dark-matter fractions at birth $>80$ per cent, consistent with a standard cosmological origin. This clearly distinguishes them from young \CompactsSB{} and bad-flag subhalos, which share very similar properties, in particular extremely low dark-matter fractions at birth and formation within the host halo. Throughout this paper, the term \CompactsSB{} therefore refers exclusively to this old, good-flag, cosmological subset. We refer the reader to Sect.~\ref{sec:limits} for the discussion on the limits of the simulation.

\subsection{Proxies for tidal stripping}
\label{sec:tidalproxies}

Satellite galaxies can experience a wide range of environmental interactions, depending on when they were accreted, their orbital properties, and the density profile of their host halos. This could lead to systems that have undergone multiple pericentric passages and prolonged exposure to the host potential, while others may not yet have experienced significant environmental processing.  It is therefore important to distinguish between weakly and strongly processed satellites.

\begin{figure}
    \centering
\includegraphics[width=\columnwidth]{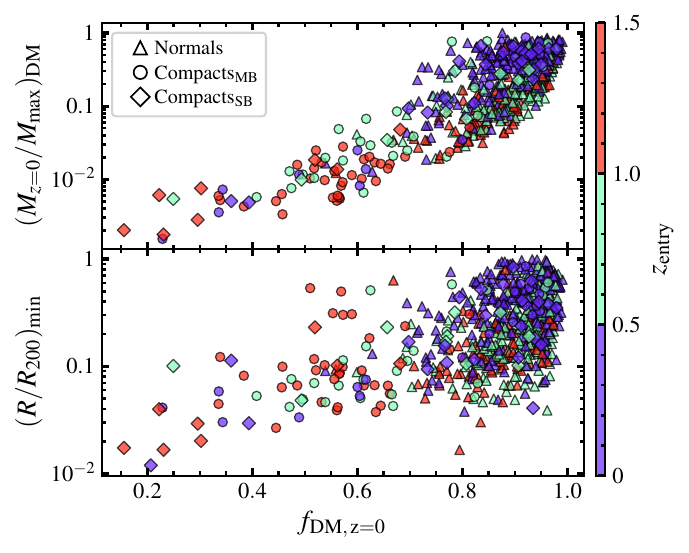}
    \caption{Dark matter mass at $z = 0$ normalised by the maximum dark matter (\emph{top}) and minimum pericentre distance (\emph{bottom}) versus dark matter fraction at $z = 0$, for satellites: Normals (triangles), \CompactsMB{}  (circles) and \CompactsSB{}  (diamonds). The colours indicate the redshift of entry into the first host. The mass fractions are for the full subhalo.
} 
  \label{fig:DMFracMaxTot}
\end{figure}

Since the extended dark-matter halo is the least bound and most extended component of a galaxy, it is the first to respond to the host tidal field. Its truncation provides a direct measure of the integrated strength of tidal stripping. Figure \ref{fig:DMFracMaxTot} illustrates several proxies for the tidal effects of the host on satellite galaxies. The fraction of remaining DM mass, $M_{\rm DM}(z=0)/M_{\rm DM,max}$, is directly related to tidal stripping, but it is not observable. One could also consider the easily observable relative location within the group, but this is not an ideal proxy as galaxies often have elongated orbits, making their $z$=0 relative radial coordinate a poor proxy for their tidal history. The minimum $(R/R_\mathrm{200})$ during the subhalo history (where $R_\mathrm{200}$ is the radius enclosing a mean mass density 200 times the critical density of the Universe at that epoch for the host halo) is a proxy for the strength of the  tidal field, but this quantity is not observable. The colour coding in Fig.~\ref{fig:DMFracMaxTot} further shows that lower present-day DM fractions are generally associated with earlier first penetration within the host virial radius, hereafter `entry'.

 We therefore adopt the (somewhat observable) present-day DM fraction, $f_{\rm DM,\,z=0}$, as a practical indicator of tidal truncation, and classify satellites as DM-rich or DM-poor according to whether $f_{\rm DM,\,z=0}$ is above or below 0.7. Although this threshold is not meant to represent a sharp physical boundary, Appendix~\ref{sec:fdmthreshold} shows that the main environmental contrasts between the two populations remain qualitatively stable for nearby threshold choices. Lower values (about 0.5) are too restrictive and quickly lead to statistically unstable DM-poor samples, whereas higher values (about 0.8) become more inclusive but start to dilute part of the environmental signal. For reference, the minimum DM fraction of central galaxies in our mass range is 0.82.  Only 8 (1.3 per cent) \Normals{} are DM-poor, whereas 54 (65 per cent) \CompactsMB{} and 16 (28 per cent) \CompactsSB{}  are DM-poor, highlighting the much stronger impact of environmental stripping on compact satellites, particularly those on the main branch. The relatively small fraction of DM-poor systems among \CompactsSB{} suggests that strong environmental stripping is unlikely to be the dominant pathway placing galaxies on the secondary branch.

\subsection{Environmental conditions} 
\label{subsec:environment}

Satellite galaxies can enter their final group either as central galaxies or as satellites. Those entering as satellites are likely to have been pre-processed in their previous hosts \citep[e.g.,][]{Benavides2020MNRAS,Donnari2021MNRAS}. Table~\ref{tab:SnapEntry} displays the environmental conditions of the three size classes of dwarf satellites. The fraction of pre-processed systems rises from $\sim5$–$15$ per cent among DM-rich satellites to $\sim30$ per cent among DM-poor Compact satellites. While the fraction of pre-processed systems is moderate for DM-rich systems (respectively 13, 5 and 15 per cent for Normals, \CompactsSB{} and \CompactsMB{}), pre-processed systems account for roughly one-third of DM-poor satellites. This higher fraction of pre-processed galaxies among DM-poor satellites is statistically significant when compared to their DM-rich counterparts ($P=0.04$ and 0.01 for \CompactsMB{} and \CompactsSB{}, respectively)\footnote{Here and throughout this article, we assess whether two distributions differ significantly by using 5000 random shuffles. We also report the medians of each population to help interpret the direction of the difference and check that the medians differ with the Wilcoxon and Mood tests. Except where noted, the Wilcoxon and Mood tests agree with the shuffling test}. This indicates that a non-negligible fraction of DM-poor systems have already experienced environmental processing before entering their final host. This is consistent with recent observational evidence suggesting that compact ellipticals in the Coma cluster may have been pre-processed in infalling groups prior to joining the central cluster \citep{Sharonova2025ApJ}.

To analyse the evolution of their environment, we first compare the entry redshift (i.e. the redshift of the first infall into the first host). DM-poor compact satellites enter their environments significantly earlier than their DM-rich counterparts. For \CompactsMB{}, DM-poor ones enter their hosts much earlier (median $z\simeq 1.2$) compared to their DM-rich counterparts ($z\simeq 0.5$, $P=0.003$). The same is true for \CompactsSB{}: $z\simeq 1.2$ vs. 0.2 for DM-poor and DM-rich ones, respectively. The small number of DM-poor \Normals{} prevents a statistically meaningful comparison in that case. The systematically earlier entry of DM-poor galaxies naturally explains their stronger tidal stripping.

\begin{table}
     \caption{Environment conditions of satellites}
    \label{tab:SnapEntry}
    \centering
    \tabcolsep=3pt
    \begin{tabular}{lccc}
    \hline
    \hline 
     & Normals & \multicolumn{2}{c}{Compacts} \\ 
    \cline{3-4}
    & & (MB) & (SB) \\
    \hline
Pre-Processed fraction \\ 
\quad DM-rich & 0.13 & \textit{0.15} & \textit{0.05} \\
\quad DM-poor & 0.25 & \textit{0.33} & \textit{0.31} \\
\quad All & 0.14 & 0.26 & 0.12 \\
\multicolumn{4}{l}{\rule{0pt}{2.5ex}Median redshift of entry} \\
\quad DM-rich & 0.58 & \textit{0.50} & \textit{\textbf{0.21}} \\
\quad DM-poor & 0.59 & \textit{1.20} & \textit{1.16} \\

\multicolumn{4}{l}{\rule{0pt}{2.5ex}Minimum pericentre: $(R/R_{200})_\mathrm{pericentre}^\mathrm{min}$} \\ 
\quad DM-rich & {0.25} & \textit{0.31} & \textit{0.32}\\
\quad DM-poor & {0.13} & \textit{0.07} & \textit{0.07}\\
\multicolumn{2}{l}{\rule{0pt}{3ex}Average orbital radius: $\overline{(R/R_\mathrm{200})}$} \\ 
\quad DM-rich &  0.83 & \textit{0.89} & \textit{0.72} \\
\quad DM-poor &  0.57 & \textit{0.43} & \textit{0.41} \\
\multicolumn{2}{l}{\rule{0pt}{3ex}Host group mass: $\overline{\log (M_{200} / \msun)}$ [16th--84th]} \\ 
\quad DM-rich 
& \makecell[c]{\textbf{12.68}\\{\scriptsize [11.77, 13.57]}}
\rule{0pt}{1ex}
& \makecell[c]{\textit{12.33}\\{\scriptsize [11.31, 13.06]}}
\rule{0pt}{1ex}
& \makecell[c]{11.87\\{\scriptsize [11.27, 13.46]}} \\
\rule{0pt}{4ex}
\!\quad DM-poor 
& \makecell[c]{12.00\\{\scriptsize [11.61, 12.94]}}
\rule{0pt}{1ex}
& \makecell[c]{\textit{13.00}\\{\scriptsize [12.01, 13.52]}}
\rule{0pt}{1ex}
& \makecell[c]{12.39\\{\scriptsize [11.74, 13.53]}} \\

      \hline
    \end{tabular}
    
    \vspace{0.5ex}
\begin{minipage}{\linewidth}
\footnotesize
\raggedright
    Notes:
         MB and SB refer to the main and secondary branches. The bold numbers highlight significantly different distributions ($P<0.05$) from the other two columns (class sizes), using 5000 random shuffles, while the italic ones highlight significantly different distributions between rows (DM-rich versus DM-poor). For $\overline{\log(M_{200}/\msun)}$, we report the median together with the 16th--84th percentile range.
    \end{minipage}
\end{table}

We also compare the global environment (host virial mass) and local environment (pericentre in units of host virial radius) for the different size classes. To do this, we compute the time-averaged orbital radius in running host virial units and the time-averaged running host mass.\footnote{Running host refers to the host at the given epoch.} The time averages are from entry into the first host to the present. The median, time-averaged, relative locations and host masses are also provided in Table~\ref{tab:SnapEntry}. While this approach simplifies the characterization of the environment, it provides a clear and consistent framework to assess the large-scale conditions experienced by satellites during their evolution after entry. Also, since TNG50 spans a relatively small cosmological volume, the most massive host halos are only sparsely sampled. To better contextualise the statistical representation of the high-mass end, Table~\ref{tab:SnapEntry} also reports the 16th--84th percentile ranges of $\overline{\log(M_{200}/\msun)}$.

DM-poor \CompactsMB{} satellites live in significantly higher-mass groups, with median values of $\overline{\log (M_{200} / \msun)} =$  $13.00$ vs. $12.33$  for their DM-rich counterparts, with corresponding 16th--84th percentile ranges of $[12.01,13.52]$ and $[11.31,13.06]$ ($P=0.002$, $P=0.01$ with the Wilcoxon test, but $P=0.06$ with the Mood test). In general, the host-mass distributions in Table~\ref{tab:SnapEntry} lie predominantly in the range expected for Milky Way to group environments. Both DM-poor \CompactsMB{} and \CompactsSB{} lie deeper in their hosts, with median $\overline{(R/R_\mathrm{200})} \simeq 0.42$ vs. $0.89$ and $0.72$ than their DM-rich counterparts ($P=10^{-5}$ and $0.001$, respectively). Although \CompactsMB{} have slightly higher median host masses than \CompactsSB{} for both DM-rich and DM-poor populations, these differences are not statistically significant ($P = 0.22$ and $0.18$, respectively).

For comparison, Table~\ref{tab:SnapEntry} also shows the median minimum pericentric distances (Fig.~\ref{fig:DMFracMaxTot}). As expected from the selection criteria, DM-poor satellites reach deeper into the potential well of their host group. Both DM-poor \CompactsMB{} and \CompactsSB{} have significantly smaller pericentric distances than their DM-rich counterparts, with median minimum $(R/R_\mathrm{200}) \simeq 0.07$ vs.\ 0.31 and 0.32, respectively ($P = 0.001$ for both samples).

To summarise, Fig.~\ref{fig:DMFracMaxTot} shows that DM-poor satellites are those that lose more DM mass (as selected). Table~\ref{tab:SnapEntry} reinforces this result, showing that, compared to DM-rich \Compact{} satellites, DM-poor ones (i) enter their first hosts earlier; (ii)  enter their final hosts earlier; (iii)  live in significantly more massive hosts (marginally so for the \CompactsSB{}); and (iv)  penetrate deeper into their hosts.

\section{DM-rich versus DM-poor satellites}
\label{sec:03}

\subsection{Environmental impact on the galaxy content}
\label{sec:environImpact}

We assess the impact of the environment on the properties of satellites by  examining the evolution across time of their components. For this, as in \citetalias{DeAlmeida2024AA}, we first compare the median evolution of each parameter for the main progenitor branch of dwarf galaxies of  different  size classes. At each snapshot, medians are only shown when 5 or more galaxies of a given size-class and final DM-fraction class are present.   We use bootstrap resampling to determine the uncertainties associated with these medians.  While median trends may obscure individual evolutionary paths, they offer a first insight into the general evolutionary trends of each population. We will explore the individual histories in Appendix~\ref{sec:AppFigs}.

Figure \ref{fig:sizeEvolMass} shows the normalised mass (in terms of the maximum mass) evolution for the three different galaxy components (DM, gas, and stars) and for baryonic content (gas plus stars), splitting between DM-rich and DM-poor satellites. The masses correspond to the masses of particles gravitationally bound to the subhalo, as identified by \subfind{}. Each mass component is normalised by the maximum value reached by that same galaxy along its main progenitor history before the median is computed. As a result, the median curves do not necessarily reach unity, because the epochs of maximum mass differ among galaxies. 

The top right panel of Fig.~\ref{fig:sizeEvolMass} indicates that DM-poor Compacts end up with only $\sim$1 per cent of their maximum DM mass. Thus, while these DM-poor satellites were selected to end up with less than 70 per cent of their mass in DM, the  tidal stripping of their DM was most violent. DM-poor satellite Normals end up with 3 per cent of their maximum gas and DM mass, which shows that these galaxies also experience severe ram pressure and tidal stripping, but not as strongly as Compacts. In contrast, DM-rich satellites only have a slow decrease in DM mass at later times (left panels of Fig.~\ref{fig:sizeEvolMass}).

\begin{figure}
    \centering
        \includegraphics[width=\hsize]{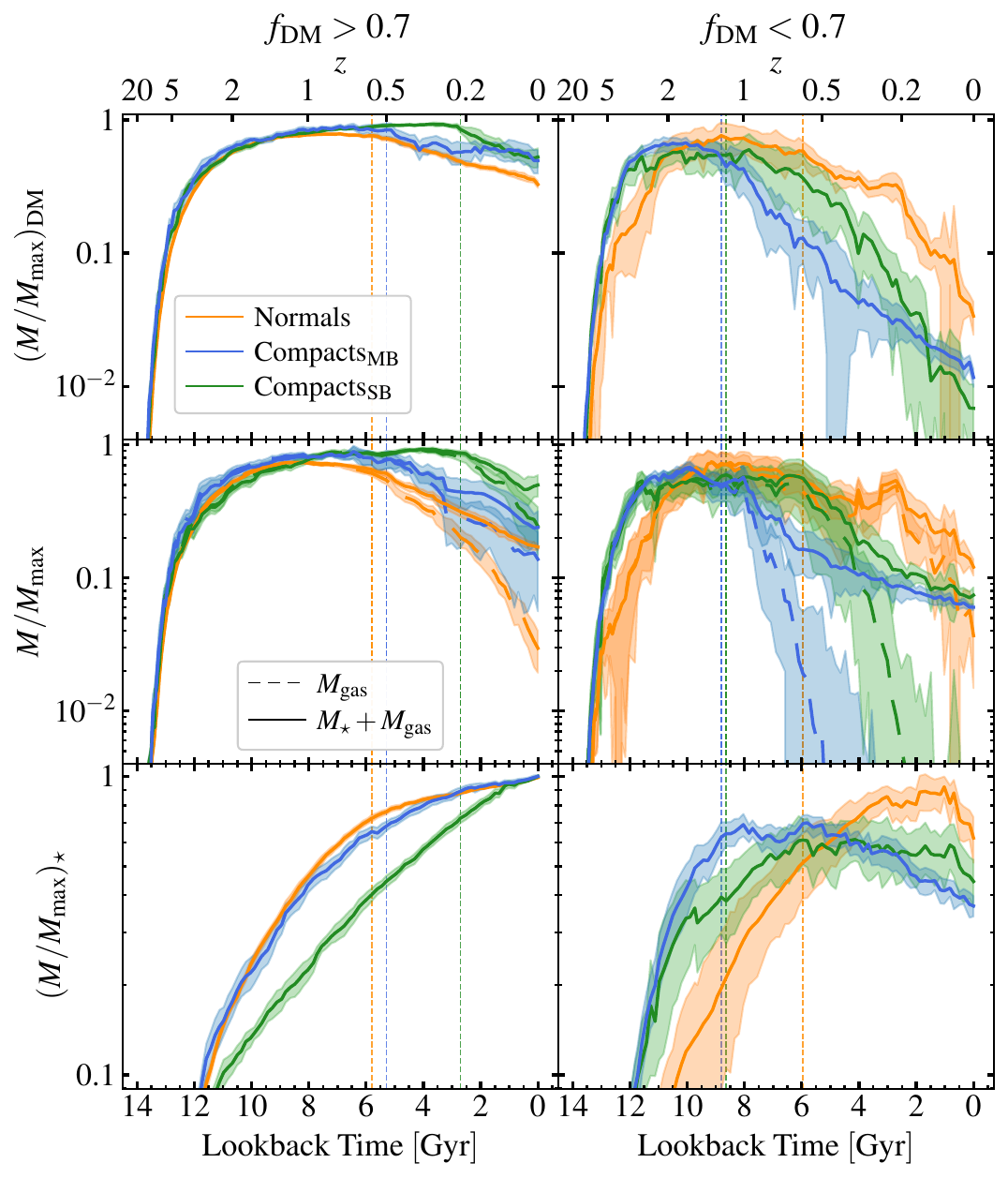}
    \caption{Normalised mass evolution (in terms of maximum mass) for the DM (\emph{top}), gas and total baryons (\emph{middle}) in dashed and solid lines, respectively, and stellar components (\emph{bottom}), for DM-rich (\emph{left}) and DM-poor (\emph{right}) satellites. The colours highlight the size classes: Normals (orange), \CompactsMB{} (blue), and \CompactsSB{} (green). The lines represent the medians, and are only shown when there are at least five galaxies at that epoch. Because each galaxy is normalised by its own maximum before the median is computed, the median curves do not necessarily reach unity. The shaded region indicates the uncertainty on the median, estimated using bootstraps. The vertical lines mark the median entry epoch of each subsample.
    }
  \label{fig:sizeEvolMass}
\end{figure}

The stellar component of DM-poor galaxies is also reduced (contrary to that of DM-rich satellites, which typically reach their maximum stellar mass at $z=0$), but considerably less so than the DM. By $z=0$, DM-poor \CompactsMB{} and \CompactsSB{} typically retain only $\sim30$--$40$ per cent  of their maximum stellar mass. The gas component, however, is the most strongly depleted, with roughly three quarters of DM-poor compact satellites losing all their gas before $z=0$, whereas only a small fraction (of roughly ten per cent) of their DM-rich counterparts end up gas-free. For the DM-poor populations, especially \CompactsMB{} and \CompactsSB{}, the gas mass declines more rapidly than baryonic mass, but baryonic mass also decreases substantially, indicating that the gas depletion is not due solely to star formation, but also reflects significant net baryonic mass loss, consistent with bound gas removal by ram pressure, possibly aided by tidal stripping.

These results highlight the different environmental effects on DM-rich and DM-poor satellites. DM-poor galaxies undergo intense tidal and ram pressure stripping, losing nearly all their gas, most of their DM mass, and a significant fraction of their stellar mass. The evolution of baryon mass further shows that the decline in gas is not solely due to star formation, but also reflects substantial net baryonic mass loss. In contrast, DM-rich satellites retain most of their baryons, experience only mild DM mass loss, and reach their maximum stellar mass by the end of the simulation. 

\subsection{When do satellites become \Compact{}?}

\begin{figure}
   \centering
\includegraphics[width=\columnwidth]{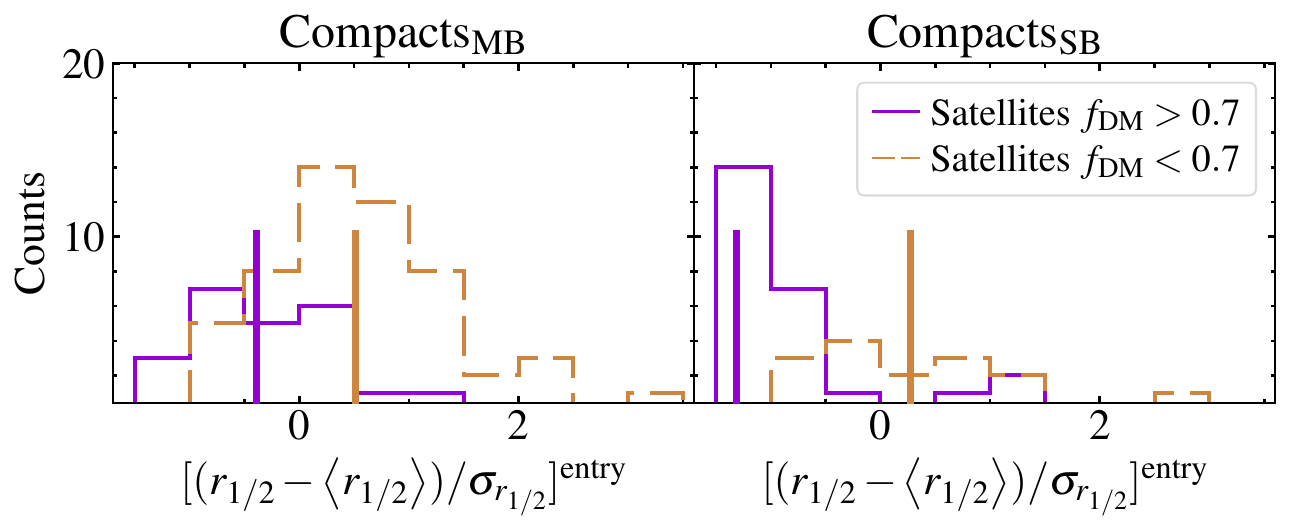}
    \caption{Distributions of the relative size at first entry into virial radius ($R_\mathrm{200}$) of the first host for \CompactsMB{} (\emph{left}) and \CompactsSB{} (\emph{right}) satellites. The relative size is defined as the difference between the size at entry and the median size at entry of all main-branch galaxies, normalised by the corresponding standard deviation. The vertical segments denote the median values. The purple solid and light brown dashed  lines are for DM-rich and  DM-poor satellites, respectively.}
  \label{fig:SizeDelta}
\end{figure}

Sections~\ref{sec:tidalproxies}, \ref{subsec:environment}, and \ref{sec:environImpact} show that DM-rich galaxies interact later, are more likely to reside in the outer regions of less massive groups, and experience less environmental impact on their content compared to DM-poor galaxies. Therefore, before delving into the details of the different evolutionary paths of DM-rich and DM-poor galaxies, we begin by asking whether satellites were already compact when they entered the virial radius of their first host, that is, when they develop a clear offset toward smaller sizes relative to the \Normals{} at fixed stellar mass. Figure~\ref{fig:SizeDelta} compares the distributions of  the relative size, defined as the difference between the size at entry and the median size at entry for all galaxies in the main branch (mostly Normals) normalised by the standard deviations of all main-branch sizes at entry, for both \Compact{} classes, splitting each between DM-rich and DM-poor satellite galaxies. Figure~\ref{fig:SizeDelta} is therefore primarily designed to distinguish whether present-day compact satellites were already structurally offset from the median of the main-branch population when they entered their first host, or instead underwent compaction mainly after environmental processing.

Figure~\ref{fig:SizeDelta}  shows that DM-rich  \Compact{} satellites typically have smaller sizes at entry than  DM-poor  \Compacts{}. Indeed, DM-rich and DM-poor \CompactsMB{} typically enter their first host with respective median relative sizes of --0.39 and 0.51 deviations from the median. These distributions are statistically different ($P \sim 10^{-5}$). A similar trend is observed for \CompactsSB{}, where DM-rich and DM-poor \Compacts{} satellites enter their first host with respective median relative sizes of --1.31 and 0.28 deviations ($P=0.0007$).

These results indicate that DM-rich \Compacts{} are already compact at entry, whereas DM-poor satellites enter their first hosts bloated relative to the median of main-branch galaxies. This shows that the environment  is important for the compaction of DM-poor satellites. For \CompactsMB{}, this indicates that the compactness of the DM-rich population is largely established before entry, whereas the compactness of the DM-poor population must be built up mainly after entry.

The comparison also remains relevant for \CompactsSB{}, since many DM-poor galaxies still have entry sizes closer to the main-branch population and therefore acquire a substantial part of their final displacement toward the secondary branch only after entry, whereas the DM-rich ones are already far from the median of the main branch at entry. DM-rich  \CompactSB{} satellites enter their first hosts significantly smaller than their \CompactMB{} counterparts ($P = 0.0005$). On the other hand, there is no significant difference between the entry-size distributions of the two classes of DM-poor \Compact{} satellites.

These results suggest that most DM-rich \CompactMB{} satellites  and nearly all \CompactSB{} ones were already typically compact at entry, whereas DM-poor \Compact{} satellites enter their first host with sizes close to the median of the main-branch, highlighting a stronger environmental impact on the compaction of the latter population.

\section{How do DM-rich satellites undergo compaction?}
\label{sec:evol}

Section~\ref{sec:03} shows that, contrary to DM-poor dwarf satellites, DM-rich ones see the stellar mass of their main progenitor increase throughout cosmic time (bottom left panel of Fig.~\ref{fig:sizeEvolMass}). This is likely due to continued stellar growth, which we examine below in terms of both star formation and ex-situ assembly. As shown in Table~\ref{tab:SnapEntry}, these galaxies typically enter later and into less massive hosts, are therefore less affected by cumulative environmental processing, and already have smaller sizes at entry relative to the median of the main-branch population (Fig.~\ref{fig:SizeDelta}). We now wonder whether the compaction of DM-rich Compact satellites is related to star formation similarly to what we found for central Compact dwarfs in \citetalias{DeAlmeida2024AA}.

\subsection{Evolution of size and specific star formation rate}
\label{sec:MassSize}

\begin{figure}
    \centering
        \includegraphics[width=\columnwidth]{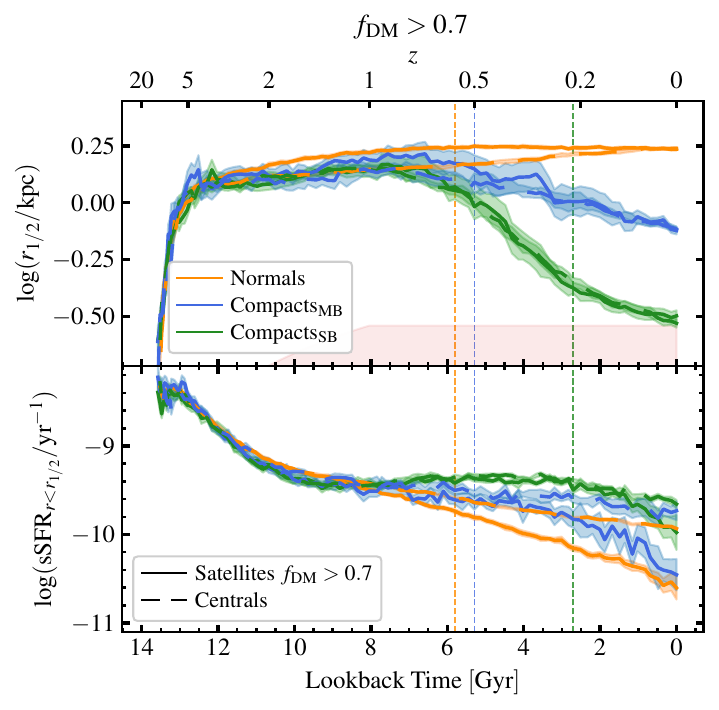}
    \caption{Median evolution for the half-mass radius of the stellar component (\emph{top}) and the specific star formation rate (sSFR) in the inner region (\emph{bottom}), for DM-rich satellite galaxies (solid lines). For comparison, the median trends for central Compact galaxies from Paper~I are overlaid as dashed lines. The shaded region around the medians indicates the uncertainty on the median, estimated using bootstraps. The red shaded region in the top panel indicates the redshift-dependent softening length of TNG50-1. The vertical lines mark the median entry epoch of each subsample.}
  \label{fig:sizeEvol}
\end{figure}

The top panel of Fig.~\ref{fig:sizeEvol} displays the evolution of the half-stellar-mass radius of galaxies that end up DM-rich and Compact. Both \CompactsMB{} and \CompactsSB{} show a compaction of their stellar component respectively after $z \sim 0.7$ and $z \sim$ $0.9$. Figure~\ref{fig:sizeEvol} and the bottom-left panel of Fig.~\ref{fig:sizeEvolMass} show together that the compaction of DM-rich galaxies occurs  while they simultaneously increase in stellar mass. This trend is similar to that of central \Compacts{} \citepalias{DeAlmeida2024AA}, which decrease in size by concentrated star formation during their evolution.

The bottom panel of Fig.~\ref{fig:sizeEvol}  shows the median evolution of the specific star formation rate (sSFR) in the inner region, defined by the running half-stellar-mass radius $r_{1/2}(t)$). To ensure numerical stability, we set a floor value of $10^{-14}$ yr$^{-1}$ for the sSFR of quenched galaxies throughout our analysis. Since DM-rich satellites end up compact while they increase in stellar mass, one expects that star formation should be important for these galaxies. DM-rich \CompactsSB{} can maintain their inner sSFR almost constant between $z \sim 1$ and $z \sim 0.2$ and higher than in DM-rich Normals, which are continuously quenching. In contrast, DM-rich \CompactsMB{} are slightly quenching, but slower than Normals.  These trends for inner star formation in DM-rich satellites are the same as those reported for centrals in \citetalias{DeAlmeida2024AA}. This trend remains if we use a fixed aperture instead of an evolving one to define the inner region, as shown in the top panel of Fig.~\ref{fig:fixed_aperture_ssfr_ratio}.

\subsection{DM-rich satellites versus centrals}
\label{subsec:dmrichcentral}

The qualitatively similar evolution in stellar mass, size, and inner sSFR of DM-rich satellites and centrals pushes us to make more direct comparisons. We use the centrals studied in \citetalias{DeAlmeida2024AA}, which were selected as in Sect.~\ref{sec:sample}.

Figure~\ref{fig:sizeEvol} also provides a  direct comparison of size and inner sSFR evolution between centrals and DM-rich satellites. Compared with central Normals, satellite Normals remain somewhat larger until $z \sim 0.2$, while their inner sSFR declines faster after $z=1.5$, consistent with a mild environmental effect. For \Compacts{}, however, there is a remarkably similar median size evolution between DM-rich satellites and centrals, strongly suggesting that the compaction mechanism is fundamentally the same. This is consistent with DM-rich Compacts entering their groups already compact (Fig.~\ref{fig:SizeDelta}). The main differences arise at late times, where the inner sSFR of \CompactSB{} satellites decreases faster than their central counterparts. Both classes of DM-rich \Compacts{} show sustained inner star formation at late epochs, contrary to their Normal counterparts. These results confirm that the evolution of DM-rich satellites is primarily a continuation of their pre-entry, central-like history, with environmental effects introducing only secondary modifications.

\begin{figure}
    \centering
\includegraphics[width=\columnwidth]{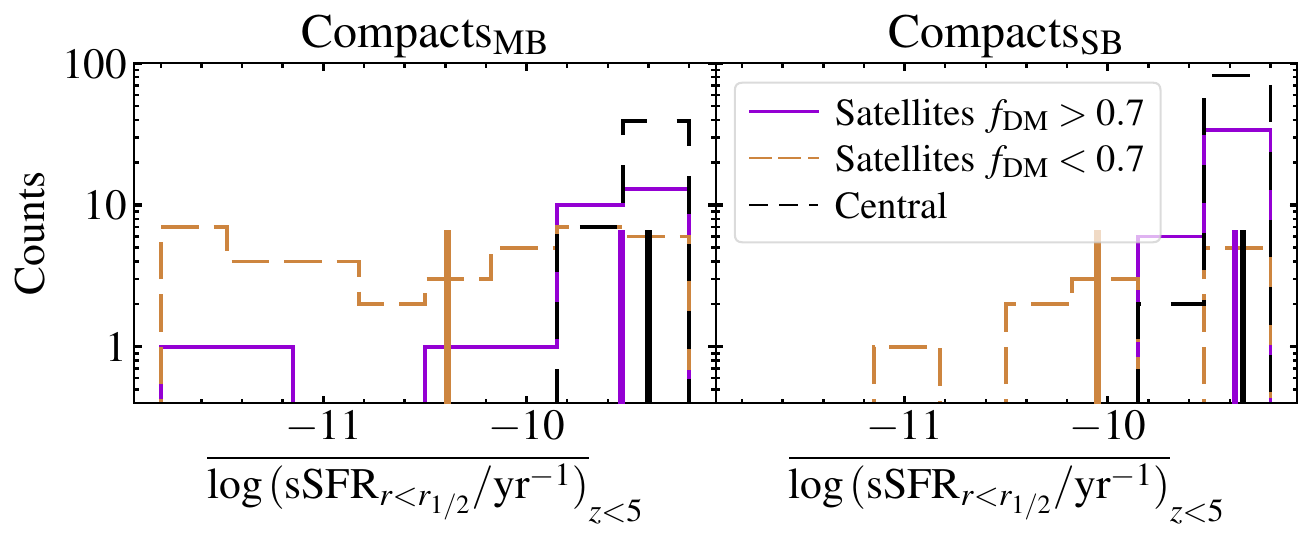}
    \caption{Distributions of the mean sSFR within the running half-mass radius after $z = 5$ for \CompactsMB{} (\emph{left}) and \CompactsSB{} (\emph{right}), both split between DM-rich (solid purple), DM-poor (dashed light brown) and centrals (dashed black). The vertical segments denote the median values. 
    }
  \label{fig:sSFRHist}
\end{figure}

Figure~\ref{fig:sSFRHist} displays the distributions of the mean inner log sSFR of each galaxy averaged over $z<5$, chosen to avoid the highly stochastic early phase of star formation.  DM-poor satellites are also shown to highlight the differences with the DM-rich. DM-rich satellites and centrals exhibit very similar sSFR distributions, with median values of $-9.54$ and $-9.40$ for \CompactsMB{}, and $-9.37$ and $-9.34$ for \CompactsSB{}, respectively. For DM-poor satellites, the median values are respectively 40 and 7 times lower than for their \CompactMB{} and \CompactSB{} DM-rich counterparts. These results indicate that DM-rich satellites typically have much higher inner sSFR than DM-poor satellites, with  median values similar to central Compacts. 

Although \CompactsSB{} exhibit higher mean inner sSFR than \CompactsMB{}, their present-day quenched fractions are broadly similar at fixed DM content. Among \CompactsSB{}, 20 and 75 per cent of the DM-rich and DM-poor populations, respectively, are quenched at $z=0$, compared to 31 and 80 per cent for \CompactsMB{}. Thus, the main difference is not that \CompactsSB{} are generally unquenched at $z=0$, but that they maintain higher mean inner sSFR over their evolution. 

\begin{figure}
    \centering
        \includegraphics[width=\columnwidth]{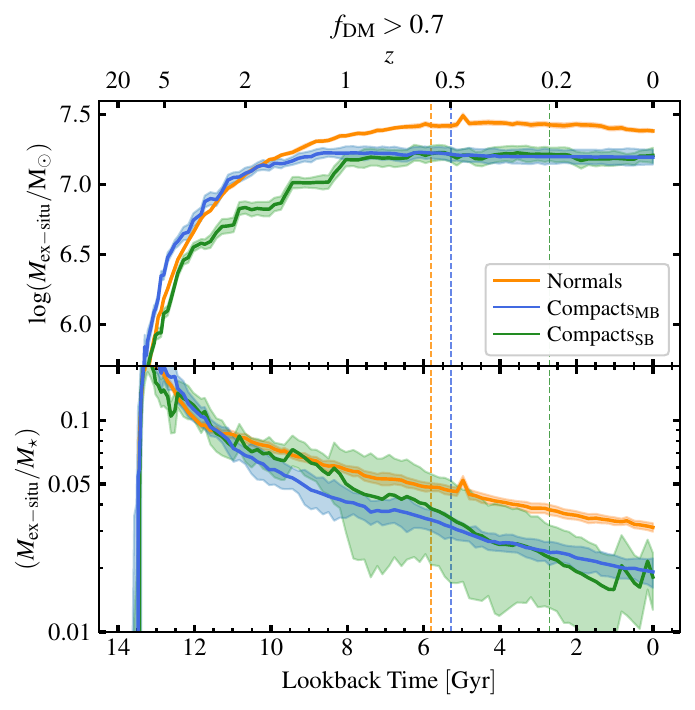}
    \caption{Same as Fig.~\ref{fig:sizeEvol}, but for the total ex-situ stellar mass (\emph{top}), and for the ex-situ stellar mass fraction (\emph{bottom}). The earlier times ($z \gtrsim 5$) should be interpreted with the usual caution, owing to the uncertainties in resolution and progenitor identification at high redshift.}
  \label{fig:ExSituMass}
\end{figure}

We also examine whether DM-rich satellites share with centrals the quiet merger histories previously identified as a prerequisite for compaction (\citetalias{DeAlmeida2024AA}). Figure~\ref{fig:ExSituMass} displays the evolution of ex-situ stellar mass and of the ex-situ stellar mass fraction, for DM-rich satellites. The ex-situ stellar mass is obtained from the {\sf Stellar Assembly TNG} supplementary data catalog \citep{RodriguezGomez2016MNRAS, Rodriguez2017MNRAS}. DM-rich \Compacts{} show no growth in ex-situ stellar mass after $z\approx 1.2$ (\CompactsMB{}) and $z\approx1.0$ (\CompactsSB{}), well before their entry into their host group. These early epochs of suppressed mergers are comparable to those that we found in \citetalias{DeAlmeida2024AA} for their central counterparts. In contrast, the growth in ex-situ stellar mass of Normals typically stops much later ($z\approx 0.6$), in fact at the typical epoch of their entry into their host. At earlier times, however, the three DM-rich populations display broadly similar ex-situ fractions, indicating comparably important merger-driven assembly during their initial growth.

At $z=0$, DM-rich \Compacts{} contain only $\sim2$ per cent of their stellar mass in ex-situ material, compared to $\sim4$ per cent for \Normals{}. Thus, the lower ex-situ fractions of \Compacts{} at late times do not appear to result from a strongly different early ex-situ contribution, but rather from the fact that their later ex-situ growth is weaker and stops earlier. This lower ex-situ mass fraction for \Compacts{} relative to \Normals{} after $z \sim 1$ has strong statistical significance (see the uncertainties in the median evolutions in Fig.~\ref{fig:ExSituMass}), as also observed for centrals in \citetalias{DeAlmeida2024AA}{}. These $z=0$ ex-situ fractions are also broadly consistent with recent cosmological hydrodynamic results. For example, \citet{Celiz2025cAA} find typical total accreted fractions of only $\sim2$--$3$ per cent in TNG50 dwarfs with $M_\star \sim 10^8\,\msun$, comparable to our DM-rich \Compacts{}. By contrast, the larger and more widely varying accreted fractions reported for low-stellar mass haloes \citep[e.g.][]{GonzalezJara2025,Tau2025a} refer to the outer regions of galaxies and are therefore not directly comparable to our total ex-situ fractions.

One may also wonder whether the lower ex-situ fractions of DM-rich \Compacts{} compared to those of Normals are simply a consequence of their present-day bound mass. However, the present-day DM masses of DM-rich \Normals{}, \CompactsMB{} and \CompactsSB{} have similar medians and are not significantly different ($\log (M_{\rm DM,z=0}/\msun) \simeq 10.21$, 10.33 and 10.38, respectively). By contrast, \Normals{} reached somewhat larger maximum DM masses in the past, with median $\log (M_{\rm DM,max}/\msun) = 10.74$, compared to 10.67 for \CompactsMB{} and 10.63 for \CompactsSB{}.  Since $M_{\rm DM,max}$ is much less affected by subsequent tidal stripping than the present-day bound DM mass, it provides a more informative proxy of the pre-stripping halo scale. This is consistent with DM-rich \Normals{} having experienced somewhat stronger early halo growth and, plausibly, somewhat stronger ex-situ assembly. The lower ex-situ fractions of DM-rich \Compacts{} are therefore more naturally interpreted as an assembly-history effect combined with weaker late-time ex-situ growth.

Altogether, the close similarity in inner sSFR, star-formation concentration, size evolution, and merger history demonstrates that DM-rich satellite \Compacts{} follow the same compaction channel as central compact dwarfs, and that their compactness is primarily set by internal processes rather than by environmental stripping.

\subsection{Radial profiles}
\label{sec:profiles}

\begin{figure}
    \center
    \includegraphics[width=0.96\columnwidth]{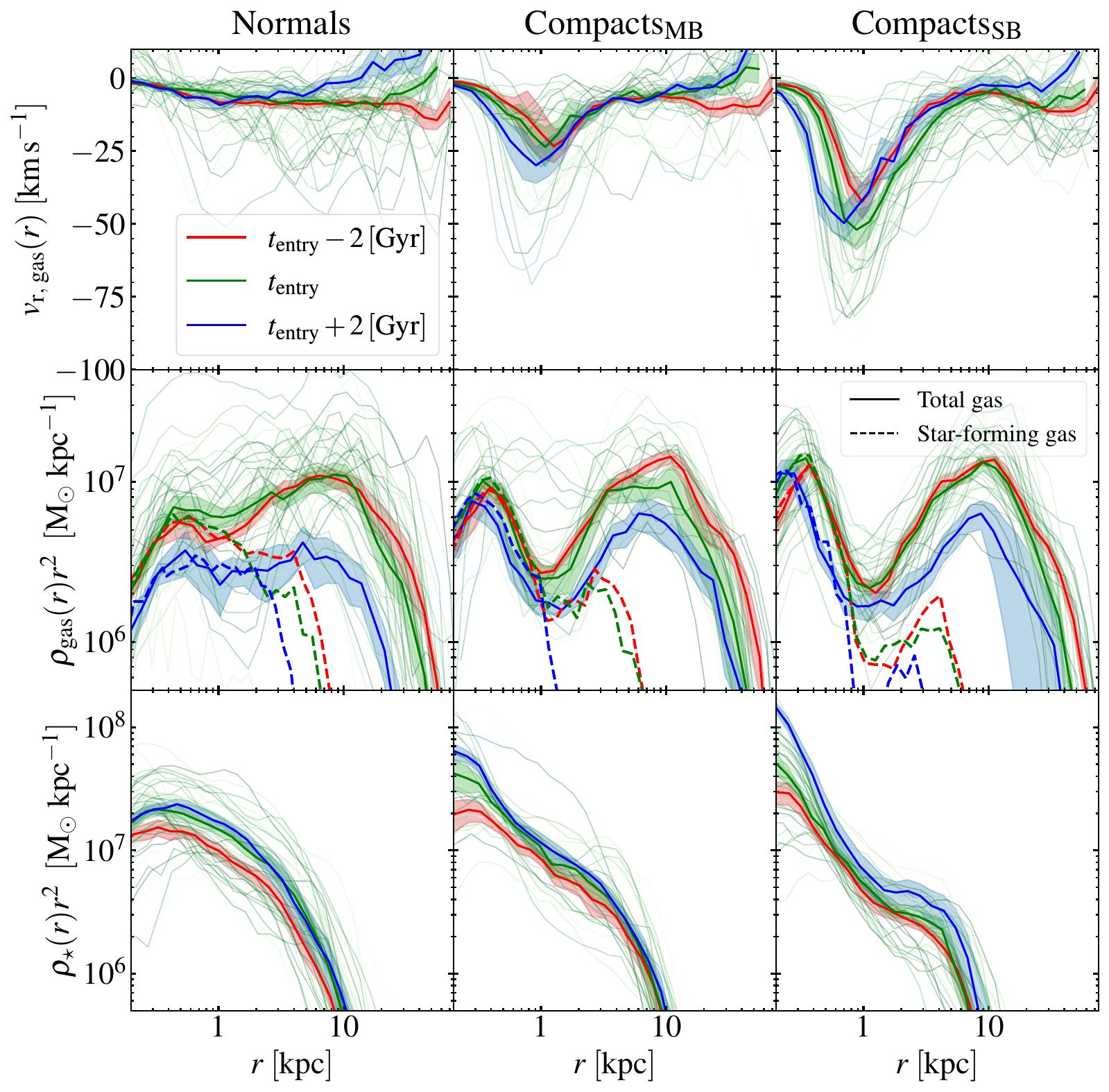}
    \caption{Radial profiles of gas radial velocity (\emph{top}), gas mass density (\emph{middle}) and stellar mass density (\emph{bottom}) for DM-rich satellites: \Normals{},  \CompactsMB{}, and \CompactsSB{} (respectively in the first, second, and third columns), at different epochs relative to the first entry to the sphere of radius $R_\mathrm{200}$ of their running host.  Medians are shown as {thick lines} and individual galaxies as {thin lines} (only for entry time, to avoid confusion). In the gas density panels, solid curves show the total gas profile, while dashed curves show the star-forming gas profile. The density profiles are multiplied by $r^2$ to better highlight the differences.
    }
  \label{fig:FirstProfileGas}
\end{figure}

Figure \ref{fig:FirstProfileGas} shows the  radial profiles of gas radial velocity measured with respect to the centre of the satellite, defined as the position of the particles with the minimum (negative) gravitational potential, as well as the total and star-forming gas and stellar mass density for DM-rich satellites of the three size classes at three different epochs: 2 Gyr before first entry, at entry, and 2 Gyr after entry. This allows us to examine the galaxy's condition prior to entry and compare it to its condition after entry. Radial profiles are computed in spherical radial bins centred on the galaxy. Stellar profiles are built from the stellar particles bound to the subhalo identified by \subfind at each epoch, extending to the outermost bound stellar particle included in the profile, while gas profiles are built analogously from the bound gas cells.

The top panel of Fig.~\ref{fig:FirstProfileGas} shows that DM-rich \Compacts{} have a clear trend of experiencing efficient gas infall toward the satellite centre before entry and maintaining it afterward without significant change, unlike DM-rich \Normals{}, which typically do not show gas infall (except for a few isolated cases that do not affect the overall population-level trend). We interpret this as evidence for more sustained centrally directed gas motions in the inner regions of DM-rich \Compacts{}. The absence of gas infall in Normals is consistent with their more merger-rich histories, which may have increased the angular momentum of their gas. At larger radii, however, the radial-velocity profiles should be interpreted more cautiously, since they may include recycled or redistributed gas within the subhalo, whose detailed behaviour may also be affected by the decoupled-wind implementation of stellar feedback in TNG50 \citep[e.g.][]{Hemler2021MNRAS}.  This  behaviour found only in DM-rich \Compact{} satellites was also found for the \Compact{} centrals (\citetalias{DeAlmeida2024AA}). The more centrally directed gas motions in  \Compacts{}, whether centrals or DM-rich satellites, is responsible for sustaining a higher gas density in the inner region and a lower gas density in the intermediate region compared to \Normals{} (second row of Fig.~\ref{fig:FirstProfileGas}). 

To clarify whether the central gas enhancement is associated with the star-forming phase, Fig.~\ref{fig:FirstProfileGas} also shows the radial profile of star-forming gas\footnote{Gas particles with attribute \texttt{StarFormationRate} $>0$.} on top of the total gas profile. The two profiles track each other most closely in the inner regions, particularly for \Compacts{}, indicating that a substantial fraction of the central gas excess in compact satellites lies in the star-forming phase. At larger radii, the profiles diverge more strongly, showing that the outer gas reservoir is predominantly non-star-forming. This supports the interpretation that the compaction of these systems is associated not only with a centrally enhanced gas reservoir, but specifically with a more centrally concentrated star-forming gas component. Consequently, the galaxy maintains its reservoir for star formation in the inner region, explaining the continuous increase in stellar density at the galaxy centre (bottom row of Fig.~\ref{fig:FirstProfileGas}). This result is related to the concentrated star formation, as was found for the central galaxies in \citetalias{DeAlmeida2024AA}. This suggests that DM-rich satellites have the same compaction mechanism as centrals.  
Similar to their central counterparts, \CompactsMB{} exhibit less efficient gas infall than \CompactsSB{}. This explains their greater final sizes compared to DM-rich \CompactsSB{}. 

In summary, we find that DM-rich Compact satellites ended up with small sizes for the same reasons as their central counterparts: they were the centrals of poorer hosts at $z \sim 1$, leading to fewer mergers after that epoch, and are thus the only galaxies that show gas infall, leading to inner star formation.

\section{How do DM-poor satellites undergo compaction?}
\label{sec:details}

Sections ~\ref{sec:03} and \ref{sec:evol} show that DM-rich satellites interact  little with the environment and exhibit   concentrated star formation, similar to what we found for the central galaxies in \citetalias{DeAlmeida2024AA}. In contrast, the compaction of DM-poor Compact  satellites in TNG50 seems to depend more on the environment. 

\subsection{Physical channels for DM-poor satellites compaction}
\label{sec:tides}

Tidal stripping efficiently removes the loosely bound outer layers of satellites, while the inner regions remain relatively protected until repeated pericentric passages \citep[e.g.,][]{Faber1973ApJ,Hayashi+03,Chilingarian&Mamon08}. Gas-related environmental processes can further help compaction by suppressing star formation at large radii and/or concentrating it toward the centre. In particular, (i) \emph{Ram pressure stripping} removes the outer gas reservoir \citep{Gunn1972ApJ}, (ii) \emph{Ram pressure thinning} lowers the star-forming gas fraction at intermediate radii, and (iii) in dense environments, tides \citep{Dekel+03}, ram pressure \citep{Steyrleithner2020MNRAS}, or both \citep{Du+19,Bian2025ApJ} may compress gas and temporarily enhance central star formation.

While tidal stripping has an immediate effect on the decrease in size, gas dynamical processes have a delayed effect on compaction. Ram pressure stripping and ram pressure thinning reduce  the future star formation rate in the outer and intermediate regions, while compression of gas by tides or ram pressure will enhance star formation in the inner regions. We first investigate whether the compaction of dwarfs in TNG50 is faster when gas is still present in the satellite.

\subsection{When do DM-poor satellites become a \Compact{}?}
\label{sec:whenCompact}

\begin{figure}
    \centering
    \includegraphics[width=\hsize]{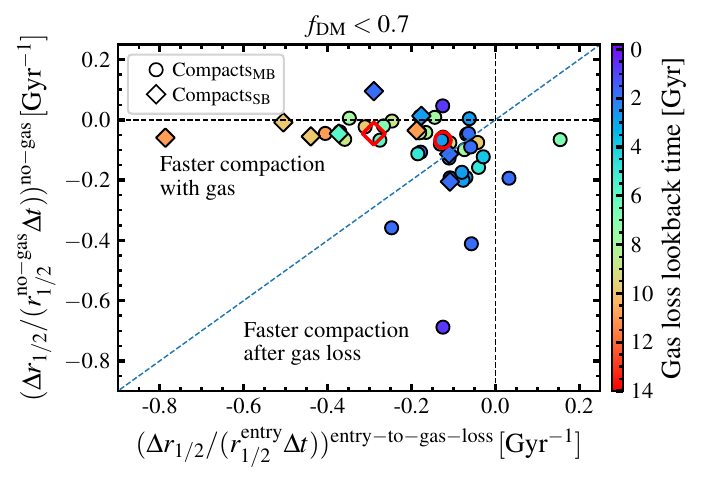}
    \caption{Mean rate of relative size change during the no-gas epoch versus during the entry-to-gas-loss epoch. Only shown are the three-quarters of the DM-poor satellites that lose their gas before $z=0$. Circles and diamonds represent \CompactsMB{} and \CompactsSB{}, respectively. Larger open symbols indicate the median positions of each population. The colours represent the gas loss lookback times. The dashed blue line represents equality; galaxies above it compact faster before complete gas loss. The dashed black lines represent $x = 0$ and $y = 0$.}
  \label{fig:HistDecreases}
\end{figure}

Figure \ref{fig:HistDecreases} compares the rate of relative compaction of the DM-poor \Compact{} satellites between entry and the epoch of total gas loss (hereafter `entry-to-gas-loss epoch') and during the subsequent gas-free phase (hereafter, `no-gas epoch'), but only for those \Compact{} satellites that lose their gas before the last snapshot of the simulation. Negative relative size changes correspond to compaction.  Only 16 (46 per cent) DM-poor \CompactsMB{}  and 2 (22 per cent) DM-poor \CompactsSB{}  that lose their gas before $z = 0$ show more compaction during the no-gas epoch (symbols below the line of equality). 

For most DM-poor \Compact{} satellites, the compaction rate is significantly more pronounced during entry-to-gas-loss than after gas loss. The median relative size change rates are $-0.125\,\mathrm{Gyr^{-1}}$ for \CompactsMB{} and $-0.184\,\mathrm{Gyr^{-1}}$ for \CompactsSB{} while gas is still present, compared to $-0.068\,\mathrm{Gyr^{-1}}$ and $-0.035\,\mathrm{Gyr^{-1}}$, respectively, after complete gas loss. The difference between the entry-to-gas-loss and no-gas phases is statistically significant for both Compact classes ($P=0.007$ for \CompactsMB{} and $P=0.001$ for \CompactsSB{}). Because \CompactsSB{} often approach the TNG50 softening scale at late times (Fig.~\ref{fig:sizevsmass_classes} and \ref{fig:sizeEvol}), we focus primarily on \CompactsMB{} when isolating the physical drivers.

These results indicate that, once the different phase durations are accounted for, compaction is typically faster before complete gas loss than during the subsequent no-gas phase. However, this timing alone does not isolate gas physics, since the strongest tidal stripping may often occur early as well. Moreover, the colours in Fig.~\ref{fig:HistDecreases} show that galaxies that shrink fastest during the entry-to-gas-loss phase also tend to lose their gas earlier, pointing to stronger environmental processing (ram pressure and tides acting together), or faster gas consumption in the inner regions. We will return to this distinction below by examining the stellar mass redistribution and the evolution of the inner sSFR.

\subsection{Inner vs outer stellar mass change of DM-poor satellites} 
\label{sec:DeltaMassInnerOuter}

\begin{figure}
    \center
    \includegraphics[width=1\hsize]{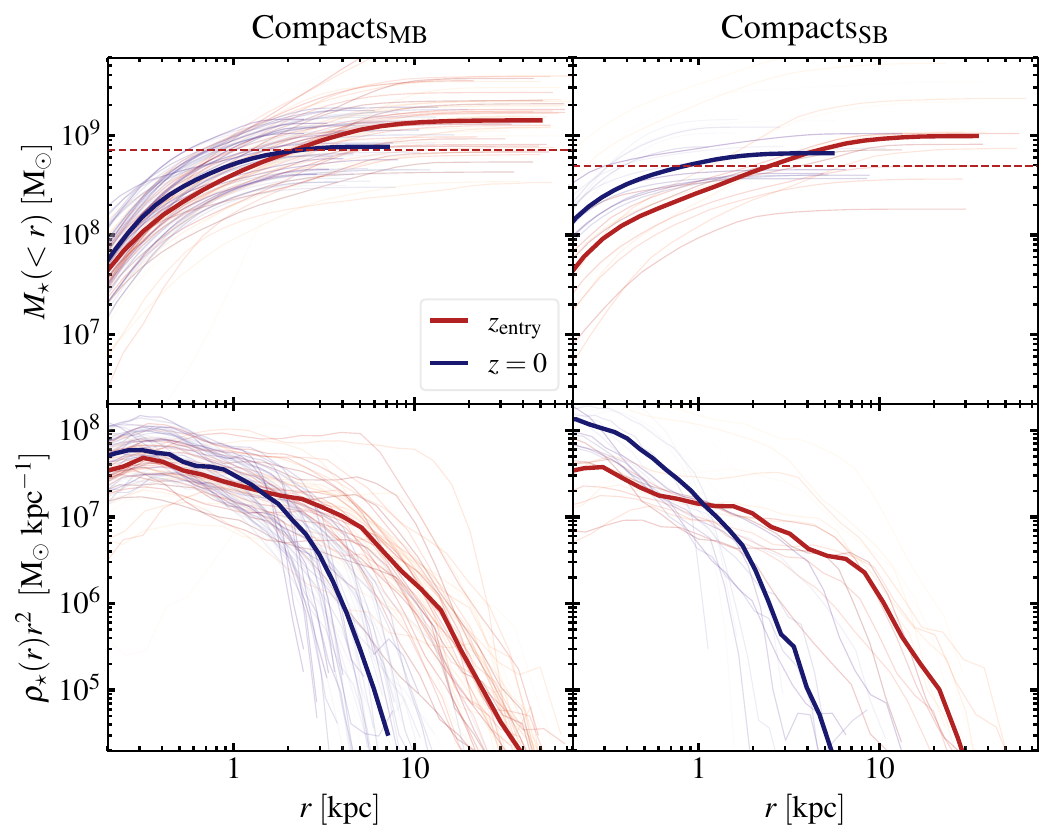}
    \caption{Comparison of radial profiles 
    of the stellar component at entry (dark red) and at $z=0$ (dark blue) for DM-poor \CompactsMB{} (\emph{left}) and \CompactsSB{} (\emph{right}) \Compact{} satellites. \emph{Top}: cumulative stellar mass; \emph{bottom}: stellar density times radius squared. The thick lines show the median radial profiles, while the thin lines show the individual profiles. The dark red dashed lines in the top panel indicate the half stellar mass at entry.
    }
  \label{fig:ProfileAtTime}
\end{figure}

In simplified terms, compaction for satellites can occur either through a greater increase of stellar mass in the inner regions relative to the outer regions or through the removal of stellar material in the outer regions. Figure~\ref{fig:ProfileAtTime} compares the radial profiles for  DM-poor \CompactsMB{} and \CompactsSB{}  of cumulative stellar mass and stellar density (times radius squared) at entry into the first host and at $z = 0$. It shows that DM-poor \Compact{} satellites typically 1), lose total stellar mass; 2) become less extended (bottom panels), and 3) exhibit modest central density enhancement, suggesting a combination of outer stripping and inner growth.

\begin{figure}
    \centering
    \includegraphics[width=\columnwidth]{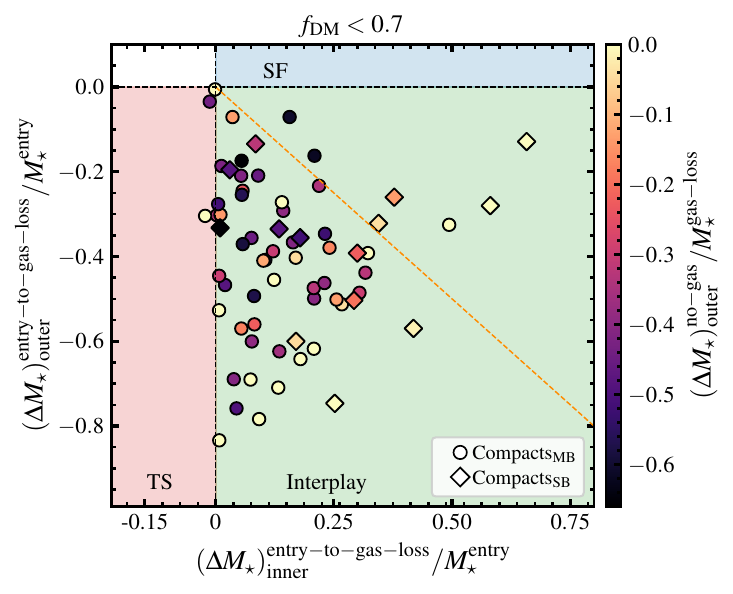}
    \caption{Outer vs. inner relative changes in stellar mass between entry and gas loss. Here, inner and outer regions are relative to the $z$=0 half-stellar-mass radius. The circles and diamonds represent \CompactsMB{} and \CompactsSB{}, respectively. The symbol colours indicate the relative change of outer stellar mass during the subsequent no-gas epoch. Note that `gas-loss' refers to $z=0$ for the quarter of the galaxies whose gas is still present at $z=0$ (black symbols). The pink, light blue, and light green zones delimit our definitions of galaxies that evolve by pure tidal stripping (TS), by star formation (SF), and from an interplay between tidal stripping and gas dynamics, respectively. The dashed black lines represent $y = 0$ and $x = 0$, i.e.  no change of stellar mass in, respectively, outer and inner regions of the galaxy during entry-to-gas-loss epoch; while the dashed orange line represents $y = -x$, i.e.  equal relative change in absolute values between the inner and outer regions of the galaxy.
    }
    \label{fig:MassInAbove}
\end{figure}

To quantify whether compaction is driven primarily by outer stellar mass loss or by inner stellar mass growth, Fig.~\ref{fig:MassInAbove} compares the stellar mass change inside and outside the $z = 0$ half-stellar--mass radius (hereafter, `inner' and `outer' regions), between entry and gas loss (or $z=0$ for the few systems that retain gas). This allows us to disentangle the relative roles of tidal stripping and centrally concentrated star formation in driving size evolution. Positive (negative) values indicate an increase (decrease) in stellar mass.  The colours in Fig.~\ref{fig:MassInAbove} represent the relative change in the outer stellar mass after reaching the minimum gas mass, either at the moment of gas loss or $z = 0$.\footnote{Among DM-poor \CompactMB{} and \CompactSB{} satellites, one quarter (12 among 54 and 4 among 16, respectively), end the simulation with gas (Sec.~\ref{sec:environImpact}).} The inner and outer regions are defined using the $z=0$ half-stellar--mass radius, because our aim is to assess whether the final compact configuration is driven more by stellar mass growth in the region that ends up as the inner component or by stellar mass loss from the region that ends up as the outer component.  Only two DM-poor \CompactsSB{} galaxies retain gas until the end of the simulation, while two others are excluded from this analysis because their histories are not reliable.\footnote{Galaxy \texttt{602133} is excluded due to a glitch in obtaining its particle data related to the subhalo switching problem, while galaxy \texttt{603556} is excluded because it never crosses $R_\mathrm{200}$.} The analysis of Fig.~\ref{fig:MassInAbove} for \CompactsSB{} must thus be treated with caution.

Figure~\ref{fig:MassInAbove} shows three regions: \emph{tidal stripping} (TS) where both inner and outer stellar masses decrease; \emph{star formation} (SF) where both inner and outer masses increase; and \emph{interplay} where the inner stellar mass increases while the outer mass decreases, indicating simultaneous central star formation and tidal removal of the envelope.

More than 90 per cent of DM-poor \Compacts\ are in the interplay region, including those that lose more stellar mass during the no-gas epoch.  All four TS among the DM-poor \Compacts{} are \CompactsMB{}. Additionally, no galaxy is located in the SF region. Most of the interplay galaxies are below the diagonal line of $y = -x$, which shows that the decrease in the outer stellar mass is larger in absolute value than the increase in the inner stellar mass. This suggests that tidal stripping and ram pressure acting in the outer region are more important than star formation in the inner region. 

Quantitatively, the median outer stellar mass loss during the entry-to-gas-loss phase is $40$ per cent for \CompactsMB{} and $34$ per cent for \CompactsSB{} ($P = 0.12$), while the median inner stellar mass increases are only $9$ and $25$ per cent, respectively ($P = 0.0002$). Hence, outer mass loss dominates over inner growth, especially for \CompactsMB{}, whereas the two effects are more comparable for \CompactsSB{}. After gas removal, outer stellar mass loss continues but at a slower rate, with median outer stellar mass loss of $31$ and $14$ per cent (see colour bar of Fig.~\ref{fig:MassInAbove}), respectively for \CompactsMB{} and \CompactsSB{} ($P = 0.07$). This is consistent with the fact that at entry the outer material is only loosely bound to the galaxy, making it easier for tidal forces to strip it away. As this outer material is lost, the galaxy’s remaining stars present a more concentrated, denser distribution, which is more tightly bound, hence more resilient to tidal forces. As a result, the outer stellar mass loss slows down over time.

Figures  \ref{fig:ProfileAtTime} and \ref{fig:MassInAbove} show that DM-poor satellites end up as \Compact{} mainly through tidal stripping of their outer stellar envelopes, with inner star formation providing a secondary, though non-negligible, contribution. 

\subsection{How do DM-poor satellite galaxies form stars?}
\label{sec:SFinSatellite}

The inner star formation observed in DM-poor \Compact{} satellites during the early orbital passages may arise from different gas–dynamical processes. One possibility is that ram pressure first removes or thins the outer gas, causing star formation to persist longer in the central regions. Alternatively, compressive effects, either from the tidal field of the host or from ram pressure itself, may directly enhance the gas density in the inner regions, temporarily boosting central star formation. 

\begin{figure}
    \center
    \includegraphics[width=1\hsize]{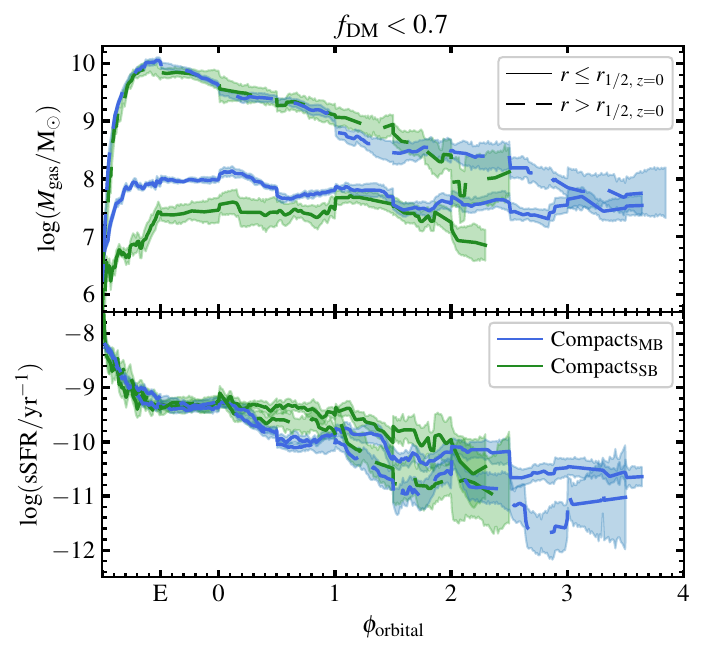}
    \caption{Median evolution, as a function of orbital phase, of gas mass (\emph{top}) and specific star formation rate (\emph{bottom}).  Both parameters are given in terms of inner and outer regions (solid and dashed lines, respectively), split by the $z$=0 half-mass radius. The shaded region around the medians indicates the $1\sigma$  uncertainty on the median, estimated using bootstraps. All DM-poor galaxies are considered here. Phase `E' represents entry time, while integer numbers (starting with zero) correspond to pericentres after entry.}
  \label{fig:sSFRvsPhase}
\end{figure}

To discriminate between these possibilities, Fig.~\ref{fig:sSFRvsPhase} shows the evolution of the gas mass and the sSFR in the inner and outer regions of DM-poor \Compact{} satellites. Median analyses as a function of time obscure the environmental effects since each galaxy experiences different environments at different epochs. We therefore replace time by orbital phase ($\phi_{\mathrm{orbital}}$) since entry to examine the timing of environmental effects along the orbit.

The top panel of Fig.~\ref{fig:sSFRvsPhase} indicates that DM-poor \Compact{} satellites experience a faster decrease in outer gas mass soon after entry, without a corresponding increase in the inner regions. Consequently, these galaxies are preferentially losing outer gas mass rather than accumulating it in the inner region. After the first pericentre following the entry ($\phi_\mathrm{orbital}=0$), DM-poor \CompactMB{} satellites lose their inner gas almost as fast as they lose their outer gas, while \CompactSB{} show a much slower gas loss in the inner regions, which becomes more pronounced after the second pericentre $\phi_\mathrm{orbital}=1$). 

The lack of a significant increase in inner gas mass suggests that the inner star formation does not require more gas material in the inner region (i.e., without a significant role of gas compression by tides). Indeed, the bottom panel of Fig.~\ref{fig:sSFRvsPhase} indicates that the sSFR in DM-poor \Compacts{} remains nearly constant from entry to the first pericentre, and then steadily declines, with quenching occurring faster in the outer regions, as expected from ram pressure stripping. We find no statistically significant enhancement of inner sSFR, which shows that tidal compression or ram pressure compression can only have a marginal impact, except for a $\sim 0.2\,$dex enhancement at entry.

\begin{figure}
    \center
    \includegraphics[width=1\hsize]{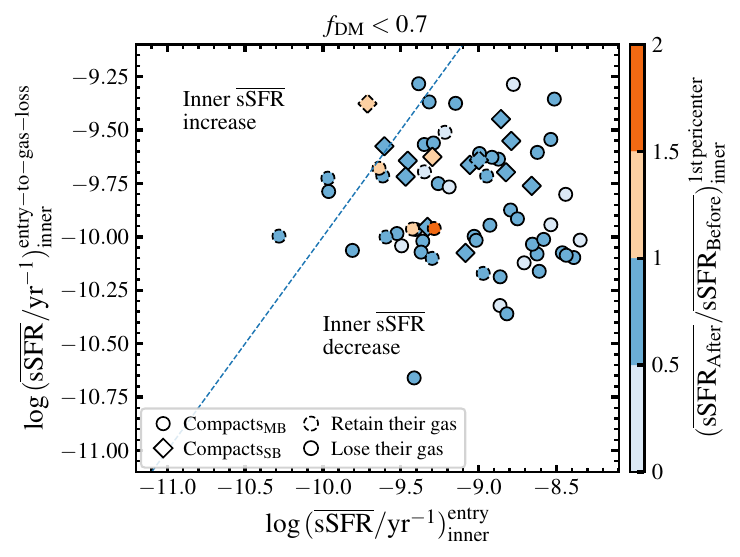}
    \caption{Comparison of inner specific star formation rates at entry and during the subsequent entry-to-gas-loss phase.   The inner sSFR at entry is the average over the three snapshots centred on that of entry, while the inner sSFR in the entry-to-gas-loss-phase is the average over all snapshots  excluding the final two snapshots before full gas loss. Inner is defined by the $z$=0 half-stellar-mass radius. Circles and diamonds represent \CompactsMB{} and \CompactsSB{}, respectively. Solid and dashed symbol edge are respectively for galaxies that lose their gas and those that retain their gas. The colours represent the ratio between the mean inner sSFR two snapshots after and the mean inner sSFR two snapshots before the first pericentre. The dashed blue line marks equality, markers below it represent galaxies whose inner sSFR decreases after entry. 
    }
  \label{fig:sSFRDetail}
\end{figure}

The possible impact of gas compression on inner star formation is assessed  by studying the evolution of sSFR for individual \Compact{} satellites between entry and gas loss. Figure~\ref{fig:sSFRDetail} compares the inner sSFR at entry (averaged over the entry plus and minus one snapshot) and averaged during the entry-to-gas-loss epoch. These choices are adopted to ensure a fair comparison between the different periods. Since the goal is to identify relevant enhancements in sSFR (i.e., longer than one snapshot), we exclude the maximum value to ensure that it does not bias the calculation of the average. The colours in Fig.~\ref{fig:sSFRDetail} indicate the ratio of the mean sSFR after and before the first pericentre, computed using the two snapshots immediately preceding and following it.

Over 90 per cent of  DM-poor \Compact{} satellites see a decline of their inner sSFR between entry and gas loss (most symbols lie below the line of equality). This inner quenching is significant for both \CompactsMB{} and \CompactsSB{}, indicating that inner star formation is generally quenched rather than enhanced during the entry-to-gas-loss phase. The colour coding identifies systems with a short-lived inner sSFR enhancement immediately after first pericentre. Only five galaxies show a temporary increase of sSFR after the first pericentre (three \CompactsMB{} and two \CompactsSB{}),  four of which retain their gas until $z$=0, suggesting that compression–induced star formation is at most a minor, short-lived effect.

During the entry-to-gas-loss phase, \CompactSB{} satellites that lose their gas display slightly higher median inner sSFR than their \CompactMB{} counterparts ($\log{(\rm sSFR/yr^{-1})}\simeq -9.65$ versus $-9.97$ and $P = 0.04$). However, the relative decline of inner sSFR across first pericentre is similar for both populations, with medians of 0.76 and 0.87, respectively ($P = 0.31$).  Systems that retain gas show the same qualitative behaviour, although small-number statistics prevent firm conclusions for \CompactsSB{}. 

Both Figs.~\ref{fig:sSFRvsPhase} and \ref{fig:sSFRDetail} show that the inner sSFR of the great majority of DM-poor \Compacts{} continuously declines after entry, with no significant enhancement due to environmental interaction. The relative persistence of central star formation mainly reflects the faster suppression of star formation in the outer regions, rather than a true enhancement in the inner ones. This behaviour is consistent with outside-in environmental processing, in which the dense inner baryonic component is more resistant to removal than the outskirts due to the balance between external pressure and gravitational restoring pressure (e.g. \citealt{Rhee2026AA}). Therefore, DM-poor Compact satellites lose gas mainly from their outskirts, while their deeper central potential wells, associated with their high inner densities, help to preserve the centre during stripping, with no clear evidence of gas compression.

\subsection{Compaction by tidal stripping or inner star formation?}
\label{subsec:tdorSF}

As shown in Sects.~\ref{sec:DeltaMassInnerOuter} and \ref{sec:SFinSatellite}, DM-poor \Compact{} satellites  simultaneously lose outer stellar mass through tidal stripping while forming inner stellar mass. However, the mass loss outweighs the gain (Fig.~\ref{fig:MassInAbove}), and the inner sSFR declines steadily after entry (Figs.~\ref{fig:sSFRvsPhase} and ~\ref{fig:sSFRDetail}). Rather than being triggered by gas compression, inner sSFR persists mainly because ram pressure affects the outer regions earlier, leaving the inner parts active for a longer time.

To disentangle the roles of tidal stripping and inner star formation,  we use the following two toy models. In the `outer stellar profile evolution (oSPev)' scenario, the inner stellar density profile does not evolve from entry to $z=0$, while the outer component is removed by tidal stripping. In the `inner stellar profile evolution (iSPev)' scenario, the outer stellar density profile does not evolve from entry to $z=0$, while the inner stellar profile is allowed to evolve to its final form, by star formation, perhaps diminished by tidal stripping. Here, we define inner and outer regions by the half-stellar-mass radius at entry. 

\begin{figure}
    \center
    \includegraphics[width=1\hsize]{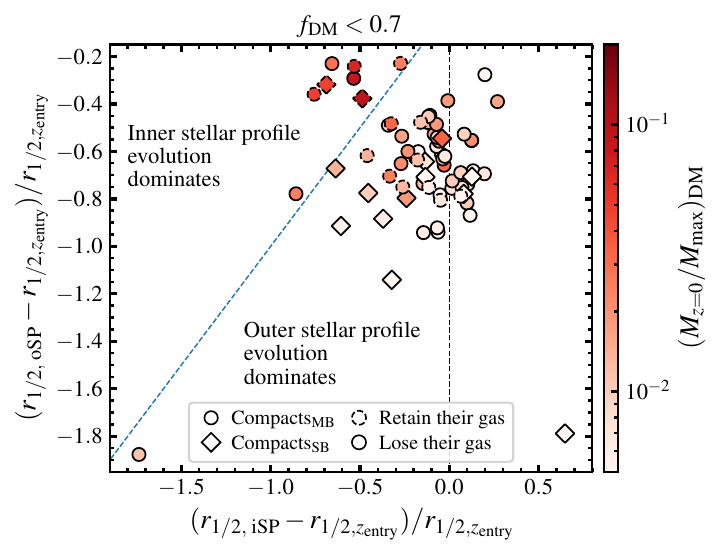}
    \caption{Comparison of relative compaction after entry of DM-poor \Compact{} satellites from two toy models. {Abscissa}: freezing the outer stellar mass density profile to its state at entry (iSPev scenario). Ordinate: freezing the inner stellar mass density profile to its state at entry (oSPev scenario). The markers and their edge lines are the same as in Fig.~\ref{fig:sSFRDetail}, and the  symbol colours represent the fraction of retained DM. The dashed blue line represents equality where, below it, tidal stripping leads to a largest compaction. The vertical dashed black line represents $x = 0$.}
  \label{fig:RhalfDetail}
\end{figure}
Figure~\ref{fig:RhalfDetail} compares the relative compaction from entry to $z=0$ expected from these two toy models. It also shows the fraction of retained DM. A huge majority of DM-poor \Compact{} satellites (62 out of 70) show stronger compaction in the oSPev scenario than in the iSPev one. This holds independently of size class and gas content. The oSPev scenario leads to much greater compactions of DM-poor Compacts that lose their gas with median relative size changes of $-0.62$ and $-0.78$ for \CompactsMB{} and \CompactsSB{}, respectively, versus $-0.06$ and $-0.19$ respectively for the iSPev scenario, which are highly significant ($P = 0.0001$ and $0.003$,  respectively). This is also true for DM-poor \CompactsMB{} that retain their gas, as their median relative size changes are $-0.63$ vs. $-0.27$ ($P=0.009$). Also, the two \CompactsSB{} that retain their gas have a median relative size change of $-0.59$. Therefore, outer tidal stripping is more effective in the compaction compared to inner star formation in nearly 90 per cent of the DM-poor Compacts. In fact, without tidal stripping of the outer stars, one-third of the satellites would grow in size (positive abscissae).

The few  galaxies with larger compaction in the iSPev scenario also retain a larger fraction of their DM (very red star symbols above the equality line of Fig.~\ref{fig:RhalfDetail}), indicating weaker tidal stripping compared to those with larger compaction in the oSPev scenario. This weaker environmental processing naturally explains both their ability to preserve gas until $z=0$ and the relatively larger role of inner stellar-profile evolution in their compaction. 

\begin{figure}
    \centering
    \includegraphics[width=\linewidth]
    {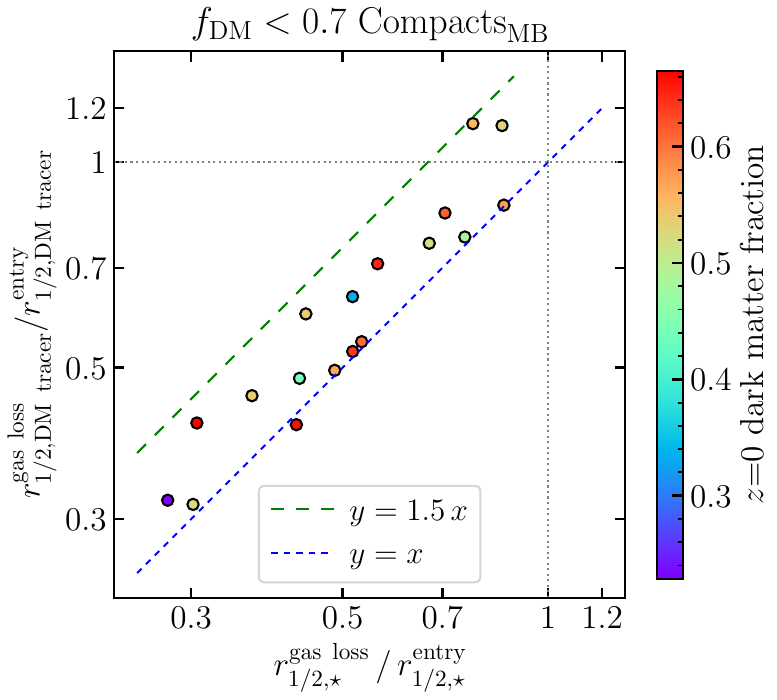}
    \caption{Comparison of entry-to-gas-loss compactions of dark matter tracer particles and stars. The figure is limited to the DM-poor \CompactsMB{} satellites that entered their final host as centrals (no pre-compaction in previous hosts), lost their gas before $z=0$, and for which the DM tracers reproduced well the stellar distribution at entry.}
    \label{fig:DMtracer}
\end{figure}

One may still ask whether the faster size evolution observed while gas is present could partly reflect genuine  gas–dynamical effects rather than a stronger tidal field at early orbital times. To isolate the role of the tidal field alone, we use suitably weighted DM particles as tracers (hereafter 'tracer particles'). Since DM particles respond only to the gravitational field, we assign them weights such that their cumulative mass profile at entry matches that of the stellar component. This allows us to estimate how the half-mass radius of a purely collisionless tracer evolves.

Our assignment of the DM tracer particle weights is described in Appendix~\ref{sec:appDMtracer}. By saving the tracer particle IDs and weights at entry, we estimate the mass profile of the tracer particles that were still present in the subhalo descendant at the epoch of gas loss. This allows us to measure the half-mass radius of the DM tracer component at gas loss. By construction, at entry the half-mass radius of the tracer component matches almost perfectly that of the stellar component.

Figure~\ref{fig:DMtracer} displays the entry-to-gas-loss compaction of the tracer particles compared to that of the stars.  We restrict the figure to \CompactsMB{} to avoid the resolution effects of TNG50, and to those that entered their final host as centrals to avoid those that had pre-compaction in previous hosts. We also discarded those for which we were not able to build DM tracer weights matching the radial distribution of the stars: this represented one-quarter of our sample and corresponded to cases where the stellar component at entry showed a disturbed density profile. 

The compaction of the tracer particles closely follows that of the stars, but is less important, with ratios of DM to stellar compactions of 0.96 to 1.47, and a median ratio of 1.14. This enhancement of the gas dynamics appears to be independent of the amount of compaction and of the final DM fraction. We note, however, that in the TNG simulations, the DM particles are more massive than the stellar particles, so the two components may be affected differently by spurious collisional heating associated with unequal particle masses \citep[e.g.,][]{Ludlow2019MNRAS, Ludlow2023MNRAS, Zeng2024MNRAS}. In particular, dynamical friction should cause the heavier DM particles to undergo faster compaction than they would if they had the same mass as the stellar particles. Our DM-tracer analysis should therefore be interpreted as an approximate test for the role of the tidal field alone, rather than as a perfectly controlled comparison between DM and stars. If anything, this effect would make the median 14 per cent faster compaction of the star particles an underestimation.

Overall, although residual star formation in the inner regions can modestly favour compaction while gas is still present, the toy-model and the dark-matter–tracer analysis jointly demonstrate that tidal stripping of the outer stellar component is the primary physical mechanism driving the compaction of DM-poor \Compact{} satellites. The dark-matter--tracer test applied to the \CompactsMB{} provides the most direct support for this interpretation in \CompactsMB{}, for which it is explicitly performed.

\section{colour and stellar metallicity diagnostics}
\label{sec:color}

\subsection{\textit{\textbf{u}}--\textit{\textbf{r}} colours}

An easily-measured observational diagnostic is the colour of a galaxy.
\begin{figure}
    \centering
\includegraphics[width=\columnwidth]{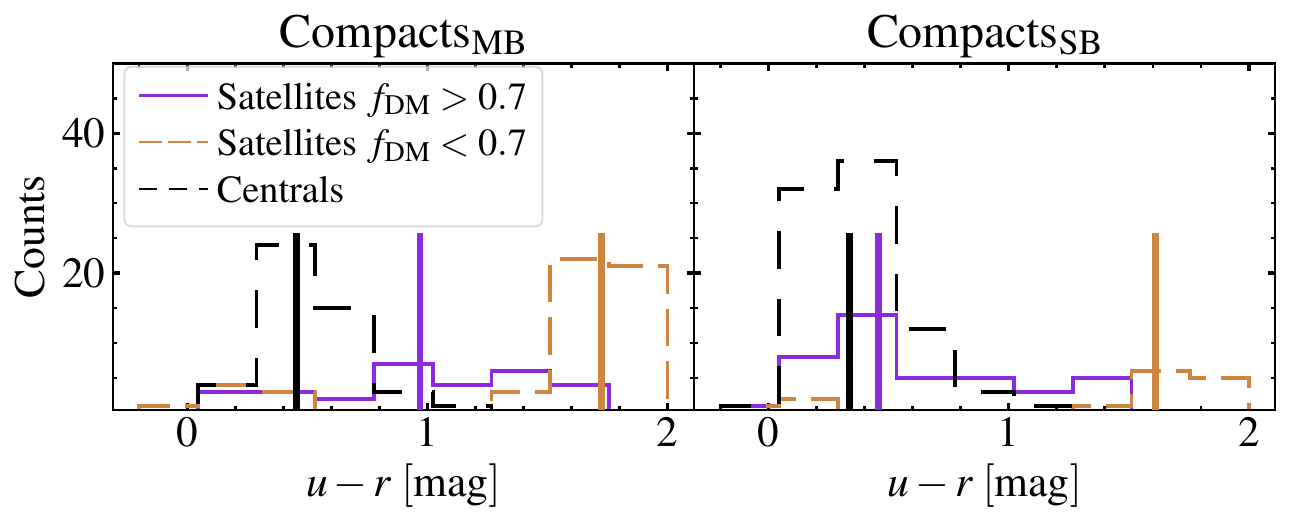}
    \caption{Same as Fig.~\ref{fig:sSFRHist} but for the distributions of the final colour $u-r$.
   }
  \label{fig:HistColor}
\end{figure}
Figure ~\ref{fig:HistColor} shows the distribution of $u$--$r$ colours (measured over the full subhalo), extracted from the TNG database, for \CompactsSB{} and \CompactsMB{}, comparing centrals with DM-rich and DM-poor satellites. Central compact dwarf galaxies are bluer than their satellite counterparts, as expected, since they end up as \Compact{} by continuous concentrated star formation, as discussed in \citetalias{DeAlmeida2024AA}. 
Among \Compact{} satellites, DM-rich \CompactsSB{} and \CompactsMB{}  are bluer than their DM-poor counterparts, while centrals are even bluer. The median colours of DM-rich satellites are $u-r=0.46$ ($0.97$) for \CompactsSB{} (\CompactsMB{}), compared to much redder values for DM-poor satellites, $u-r=1.60$ ($1.72$). Central \Compacts{} are significantly bluer, with median $u-r= 0.34$ ($0.45$) for \CompactsSB{} (\CompactsMB{}). For both size classes, the colour distributions of DM-rich and DM-poor satellites are statistically different from each other and from centrals ($P < 0.01$).

These different distributions of colours agree with our results on star formation in satellite galaxies (Sect.\ref{subsec:dmrichcentral}), highlighting that DM-poor galaxies are more strongly influenced by environmental effects, such as tidal and ram pressure stripping.

\subsection{Stellar mass - metallicity relation}

Previous studies have suggested that compact satellite dwarf galaxies may be the stripped remnants of a more massive progenitor, such as elliptical galaxies (for cEs) and dwarf galaxies (for UCDs) \cite[e.g.,][]{Faber1973ApJ, Drinkwater2000PASA, Misgeld2011MNRAS, Brodie2011AJ,  Kim2020ApJ, Wang2023Nature, Khoperskov+23}. A good indicator for such tidal stripping is a higher stellar metallicity than expected for a given stellar mass, which may suggest that the dwarf galaxy had more stellar mass in the past \cite[e.g.,][]{Chilingarian&Mamon08, Chilingarian+09, Pfeffer2013MNRAS, Janz2016MNRAS}.

\begin{figure}
    \centering
    \includegraphics[width=0.8\columnwidth]{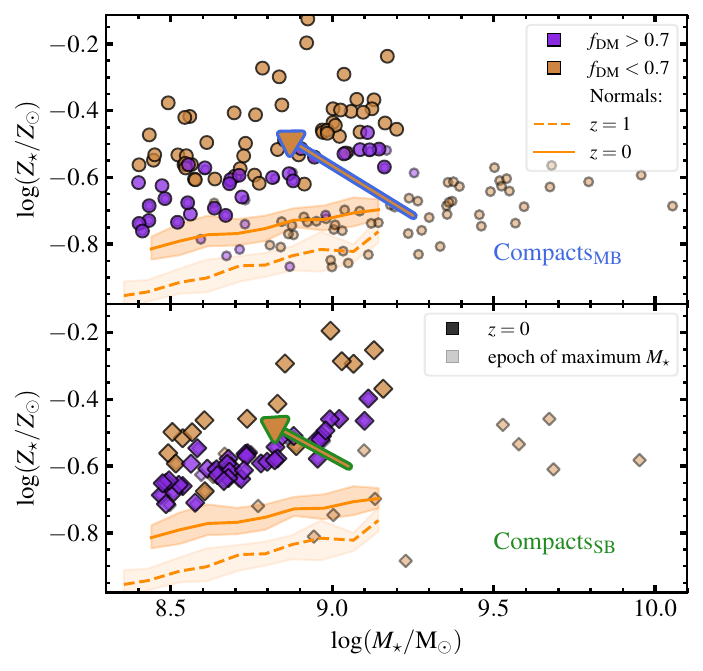}
    \caption{Stellar mass--stellar metallicity relation at $z = 0$ for satellite galaxies:  \CompactsMB{}  (\emph{top}) and \CompactsSB{} (\emph{bottom}) galaxies.  A solar metallicity of $Z_\odot=0.0127$ was assumed \citep{Asplund+09}. Stellar metallicities are shown after applying the uniform $-0.75$ dex shift as described in the text. The small pale symbols represent the maximum stellar mass and corresponding stellar metallicity, while the solid symbols represent the final values. Purple and  light brown markers indicate DM-rich and DM-poor satellites, respectively. The arrows show the median evolutionary shift from the maximum-mass epoch to $z=0$ for the DM-poor population. We do not show equivalent arrows for the DM-rich population because their epoch of maximum stellar mass typically occurs at, or very close to, $z = 0$, so their displacement in the stellar mass--metallicity plane is marginal for most systems. The orange solid (resp. dashed) lines show the median mass-metallicity relation of \Normals{} at $z = 0 \, (\mathrm{resp.}\, 1)$, while the orange regions indicate its $1\,\sigma$ range.
    }
  \label{fig:Zmetallicty}
\end{figure}

Figure~\ref{fig:Zmetallicty} shows the $z=0$ stellar mass--metallicity relation for satellite galaxies. Here, stellar metallicity refers to the stellar metallicity (i.e. the mass fraction in all metals, expressed as  $\log( Z_\star / Z_\odot)$) measured within $2r_{1/2}$, corrected by the uniform $-0.75$ dex shift adopted by \citet{Bian2025ApJ} to better reproduce the observed dwarf-galaxy stellar mass--metallicity relation of \citet{Kirby2013ApJ}.

DM-rich Compact{} satellites (purple symbols) lie on a narrow, more metal-rich sequence compared to Normals. Central \Compacts{} (not shown in the figure) display a similar metallicity excess. The metallicity excess of DM-rich \Compacts{} is consistent with their enhanced late-time central star formation, driven by efficient gas inflow and low-angular-momentum accretion, as discussed in \citetalias{DeAlmeida2024AA}.  This higher star formation in central and DM-rich satellite \Compacts{} leads to more metal enrichment by type~II supernovae of the gas, hence of the new stars formed within.

In contrast, DM-poor \Compact{} satellites (light brown symbols) have even larger metallicity excess at given stellar mass. The arrows in Fig.~\ref{fig:Zmetallicty} indicate the median evolution from the maximal stellar mass,  but are only shown for DM-poor galaxies. DM-rich galaxies typically reach their maximum stellar mass at, or very close to, $z = 0$, and therefore show little or no displacement between these two epochs. DM-poor \Compacts{} evolve by jumping above their DM-rich counterparts (whose lack of arrows indicates little evolution). This behaviour is a signature of tidal stripping. While stellar mass is efficiently removed from the outskirts, the metal-rich inner component is preserved, increasing the mean metallicity of the remnant. 

Fig.~\ref{fig:Zmetallicty} also shows the median stellar mass-metallicity relation of satellite \Normals{} at $z=1$, a redshift close to the median $z_\mathrm{entry}$ of the DM-poor \Compact{} satellites (Table~\ref{tab:SnapEntry}). At early epochs, the DM-poor \Compacts{} are more massive and metal-rich than the \Normals{}, as expected if tidal stripping is the main driver of their subsequent evolution. We also note that there are no significant differences in the $z$=0 metallicities of the pre-processed DM-poor \Compacts{} and those that entered their hosts as centrals, nor for the pre-processed DM-rich ones compared to the other DM-rich \Compacts.

\begin{figure}
    \centering
    \includegraphics[width=0.9\columnwidth]{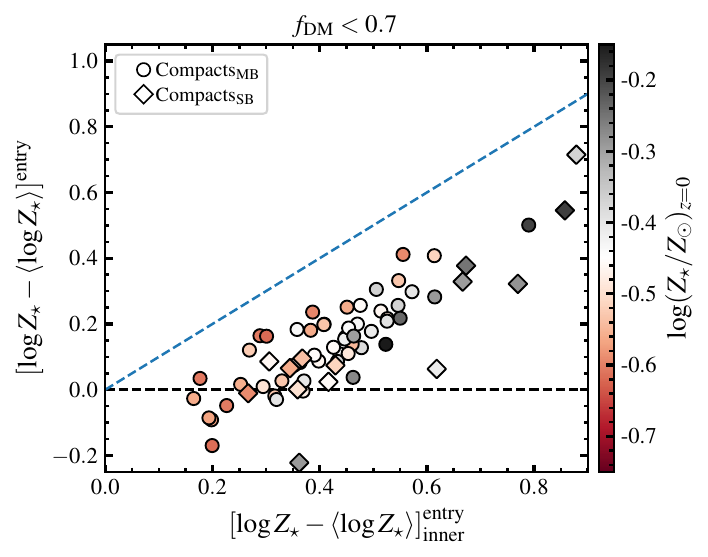}
     \caption{Total vs. inner stellar metallicity offsets relative to the median of the main branch (principally Normals)  of DM-poor satellites at entry. The {diamonds} represent \CompactsSB{} and the {circles} represent \CompactsMB{}. The colours indicate the final global metallicity. The blue and black dashed line indicates equality and $y=0$, respectively.}
  \label{fig:ZEntry}
\end{figure}

Figure~\ref{fig:ZEntry} further clarifies the origin of the metallicity excess by comparing the total and inner metallicity (restricted to stars within $r_{1/2}$)  offsets at entry, relative to the median of the main branch of the size-mass relation at that epoch. Most DM-poor \Compacts{} are already more metal-rich than Normals upon entry, especially in their inner regions (where they are more metal-rich). The subsequent tidal stripping then preferentially removes the metal-poorer outer stars, amplifying the global metallicity offset. Furthermore, we find that galaxies that end up with the highest metallicities (grey to black symbols) almost always enter their group with higher metallicities relative to the Normals (both over the entire galaxy and in the inner region). This means that high-metallicity present-day galaxies almost always experienced an important enrichment prior to entry.

 Figure~\ref{fig:Zmetallicty} thus highlights the importance of the stellar mass-metallicity diagram for identifying dwarf satellite galaxies that are survivors of the tidal field of a massive progenitor. That figure combined with Fig.~\ref{fig:ZEntry} support a scenario where tidal stripping helps the metallicity increase by removing the outer stellar material, thereby increasing the relative contribution from the metal-rich inner region.

\section{Discussion}

\subsection{Limitations of the simulation}
\label{sec:limits}

As with any cosmological simulation, TNG50 has limitations that must be kept in mind: 
\begin{itemize}
\item The Plummer-equivalent softening of 288~pc at $z<1$ limits the interpretation of the half-stellar-mass radii of the most compact systems. This is particularly relevant for the late-time compaction of the smallest \CompactsSB{} (Fig.~\ref{fig:sizeEvol}), but should not affect the earlier, physically driven compaction nor the evolution of \CompactsMB{} and Normals. 
\item Spurious dynamical heating can affect the evolution of the inner regions of galaxies \citep[e.g.,][]{Ludlow2019MNRAS, Ludlow2023MNRAS, Zeng2024MNRAS}.
\item Tidal stripping in state-of-the-art simulations can itself remain significantly resolution dependent \citep[e.g.,][]{, Lovell2025MNRAS,Chiang2026OJAp}. The earlier compaction phases, when galaxies are still larger, are likely less sensitive to these limitations than the late-time evolution of the most compact remnants.
\item Subhalo identification with \subfind is often hampered when satellites probe deep into their hosts. In these regimes, subhalos are more prone to fragmentation in the \sublink merger trees or incomplete tracking, which is reflected in the bad-flag and `young' populations. Excluding these systems therefore helps reduce potential misidentifications. Nevertheless, the classification of the most compact and strongly stripped satellites may still depend in detail on the adopted subhalo finder, and could differ with alternative algorithms such as \rockstar, \citep{Behroozi2013ApJ} or \textsc{HBT-Heron}, \citep{Forouhar2025MNRAS}. 
\item In IllustrisTNG, star-forming gas is treated using the two-phase subgrid ISM model of \cite{Springel2003MNRAS}, with an effective equation of state above the star-formation threshold of $n_{\rm H}\simeq 0.13\,\mathrm{cm}^{-3}$. As a result, star formation may be artificially smoothed or extended into diffuse outer gas. This can reduce the contrast between inner and outer star formation, potentially underestimating the outer quenching by environmental processes. Furthermore, as pointed out by \cite{Hemler2021MNRAS} and \cite{Celiz2025AA}, the stellar winds are decoupled from the interstellar gas until they reach the more diffuse outer regions, and this may affect the gas evolution and the star formation rate. This may bias the inferred evolution of the inner component, making some \CompactsSB{} systems appear more resilient to stripping and quenching than they would otherwise be.
\item The TNG BH model seeds BHs only above a halo-mass threshold and places them at the location of the highest density cell, to mimic unresolved dynamical friction \citep{Vogelsberger2013MNRAS}. In our good-flag sample, the $z$=0 BH occupation fraction decreases from Normals to \Compacts{}, with the strongest decrease among DM-poor satellites, where BHs are rare or absent in the most compact systems (Table~\ref{tab:BHfrac}). This is potentially relevant because BHs are observed in several nearby compact stellar systems around more massive galaxies \citep[e.g.][]{vanderMarel+97,Chilingarian&Mamon08,Afanasiev+18,Tahmasebzadeh+25}. The halo mass threshold is close to the median $z$=0 host mass of centrals and above those of DM-rich satellites. DM-poor Compacts penetrate deep into their host, and their BH experiences dynamical friction before merging with the BH of the central of galaxy of the host halo. If AGN feedback is underestimated in such Compacts (centrals and satellites), TNG may allow excessive central star formation and therefore stronger compaction.
\end{itemize}

It is not at all clear how each of these limitations of the simulation would affect our main conclusions. If anything, the limitations  discussed above  are more likely to blur or weaken some physical trends than to artificially create the main compaction pathways identified here. However, the limited resolution of the simulation should  not qualitatively affect our main conclusions for either DM-rich Compact satellites  or DM-poor \CompactMB{} satellites. On the other hand, our results for \CompactsSB{} are the most sensitive to the simulation's resolution, and should be therefore taken with caution. 

\label{sec:discuss}
\subsection{Comparison to previous work}

Recent studies have explored the origin of compact satellite dwarfs in cosmological and idealized simulations, sometimes reaching different conclusions on the dominant physical channel for compaction.

\cite{Deeley+23} selected galaxies at $z$=0 of TNG50 with stellar masses above $10^9\msun$ and effective radii lower than 1 kpc. Most of their galaxies are part of the main branch of the size-mass relation (right panel of their fig.~1). They split their galaxies into three similar classes as our DM-poor, DM-rich, and Centrals, where their DM-poor called `stripped' were those that lost a fraction greater than $0.23$ of their maximum stellar mass. Our choice of maximum $z$=0 DM fraction of 0.7 for the DM-poor is more restrictive as these all have a final DM mass less than 0.1 times their maximum value (see Table~\ref{tab:SnapEntry}). \citeauthor{Deeley+23} studied the evolutions of these three classes, but discussed little their compaction mechanisms, other than briefly noting that the centrals and DM-rich satellites underwent little merging. This is in agreement with \citetalias{DeAlmeida2024AA} for centrals and with the present work for DM-rich satellites.

\citet{Du+19} ran idealized simulations of a dwarf galaxy falling very deeply into a Milky Way--mass galaxy on an elongated orbit. Their simulation was better resolved than TNG50, with a softening scale of 50 pc for the stars and gas and 100 pc for the DM. They argue that tidal or ram-pressure compression of gas at pericentre can trigger central starbursts that drive compaction. While such bursts are present in their model (their fig.~7), the associated gas-density enhancement is brief and occurs mainly near second pericentre (their fig.~5), and the satellite expands slightly after gas removal. In contrast, our statistical analysis of TNG50 DM-poor satellites shows that, although transient star-formation enhancements can occur, the dominant and sustained size reduction is associated with tidal stripping of the stellar component rather than with gas-driven central contraction. And we see no expansion after the gas loss, aside from size oscillations in sync with the orbit (see Fig.~\ref{fig:EvolwZstarfor7SBs}). 

\cite{Jang+24} analysed the NewHorizon zoomed-in cosmological hydrodynamical simulation \citep{Dubois+21}, which is better resolved than TNG50, but in a volume 33 times smaller. They were able to select compact galaxies down to $M_\star = 10^6 \msun$, with a median stellar mass of $10^7 \msun$ and a median effective radius of 120 pc at $z = 0.17$. Their compacts are much less massive and tinier than those in our study and the two aforementioned ones. They find that roughly one-quarter of their galaxies were born big and underwent compaction by tidal stripping, while the remaining three-quarters were compact since their birth, among which 80 per cent were born within groups of mass above $10^9 \msun$. In contrast, the typical $z=0$ \Compacts{} in our sample present sizes at birth comparable to those of the Normal population and most subsequently shrank by a factor over two (Fig.~\ref{fig:sizeEvol} for DM-rich and poor satellites and fig.~3 of \citetalias{DeAlmeida2024AA} for Centrals). This difference possibly reflects the different ranges of probed stellar masses.

The closest comparison to our work is the study of \citet{Bian2025ApJ}, who identified metal-rich compact satellites in TNG50 ($\log(Z_\star/Z_\odot)>-0.45$ and $r_{1/2}\,$ $<600\,\mathrm{pc}$). They circumvented the issues with the \sublink merger tree (see Sect.~\ref{sec:limits}) using merger trees that they built from stellar particle ID matching. They found that while some of their galaxies became metal-rich for their stellar mass through tidal stripping, most instead gained their metals in starbursts near first pericentric passage, which were caused by ram pressure compression.
In our sample, the most metal-rich DM-poor \Compacts{} are predominantly galaxies that have lost a large fraction of their maximum dark matter and stellar mass (Fig.~\ref{fig:Zmetallicty}), indicating strong tidal stripping. These systems already show elevated metallicities at group entry (Fig.~\ref{fig:ZEntry}), suggesting early star formation followed by substantial mass loss. In comparison, our most metal-rich \CompactsSB{} likewise experienced early starbursts, but many subsequently underwent strong stripping (Appendix~\ref{sec:AppFigs}).

The differences between the two studies mainly reflect the distinct regimes probed by the samples. Our analysis focuses on long-lived, good-flag satellites, whereas \citet{Bian2025ApJ} included bad-flag galaxies, which often have broken or incorrect \sublink histories that are better traced by their own merger tree. As a result, starburst-driven metallicity enhancement is more prominent in their sample, while tidal stripping dominates the long-term evolution of the good-flag satellites considered in our analysis.

However, bad-flag galaxies are ubiquitous among \CompactsSB{}, accounting for two-thirds of the galaxies in the secondary branch (before our random sampling described in Sect.~\ref{sec:sample}). Figure~\ref{fig:histZ} shows that bad-flag and high-metallicity \CompactsSB{} are very strongly linked. Indeed, bad-flag galaxies account for  the great majority (87 per cent) of \CompactsSB{} stellar metallicities above the threshold used by  \citeauthor{Bian2025ApJ} (including all four of their examples of compact dwarfs becoming very metal-rich by starbursts), while only 1 out of 78 galaxies on the secondary branch that are below the metallicity threshold of \citeauthor{Bian2025ApJ} have bad-flags and only 1 out of 230 bad-flag \CompactsSB{} have metallicities below the threshold. The compaction channel by starbursts driven by ram pressure compression, found by \cite{Du+19} and \cite{Bian2025ApJ}, is dominant for \CompactsSB{} if the neglected bad-flag subhalos  turn out to be bona fide galaxies.

Our avoidance of the bad-flag galaxies may thus render our analysis of the \CompactsSB{} incomplete. However, it is not yet clear that most bad-flag galaxies are bona fide systems and not artefacts of the simulation due to the numerical limitations of the TNG50 (see Sects.~\ref{sec:sample} and ~\ref{sec:limits}). On the other hand, the early starburst compaction channel should at best be rare for \CompactsMB{}, where bad-flag galaxies account for only 5 per cent of the subhalos (before our random sampling). Our conclusion that tidal stripping is the major compaction mechanism for DM-poor satellites is therefore robust.

\section{Conclusions}
\label{Conclusion}

After analysing the compaction mechanisms of central compact dwarfs in \citetalias{DeAlmeida2024AA}, we continued this analysis, focusing now on why satellite dwarf galaxies ended up compact by $z = 0$. We selected old satellite dwarf galaxies with $\log (M_\star/{\rm M_\odot})$ between 8.4 and 9.2 from the TNG50 cosmological hydrodynamical simulation, excluding systems flagged as potentially spurious or numerically suspect. We studied their evolution by comparing three different populations based on the size-mass relation: `\CompactsSB{}', which are in the secondary branch of the size-mass scatter plot (Fig.~1 from \citetalias{DeAlmeida2024AA}) with half-stellar-mass radii below 447 pc; `\CompactsMB{}', located on the lower envelope of the main branch of the size-mass scatter plot, and `Normals', which lie on the spine of the main branch of the size-mass relation.

We used the final dark matter fraction as a proxy to distinguish satellites experiencing varying degrees of environmental impact. In addition to being a somewhat observable parameter, the final dark matter fraction correlates well with the final dark matter mass normalised by the maximum dark matter mass of the galaxy during its evolution (Fig.~\ref{fig:DMFracMaxTot}), which is directly related to the decrease in dark matter mass, but is not observable. We classified galaxies with DM fractions above 0.7 at $z = 0$ as DM-rich and those with DM fractions below 0.7 as DM-poor.

The physical mechanism driving the compaction of satellite dwarf galaxies depends on the final DM fraction and stellar metallicity. DM-rich Compact satellites behave very much as centrals \citepalias{DeAlmeida2024AA}. They have a late entry (Table~\ref{tab:SnapEntry}) into less massive hosts, in whose  outer regions they typically reside (Table~\ref{tab:SnapEntry}). Their ex-situ stellar mass stops growing at $z \approx 1$ and ends up less important than for Normals (Fig.~\ref{fig:ExSituMass}). These galaxies exhibit concentrated star formation  (Figs.~\ref{fig:sizeEvol} and \ref{fig:FirstProfileGas})  similar to what is observed for central Compacts in \citetalias{DeAlmeida2024AA}{} (Fig.~\ref{fig:sSFRHist}). This suggests that the primary compaction process for DM-rich Compacts is the same as for central \Compacts{}, occurring before environmental interactions (Fig.~\ref{fig:SizeDelta}): they are the dwarfs ending up as satellites that previously lived in poor environments, thus grew less by mergers, hence not acquiring angular momentum from such events, which enabled them to keep their gas infall to drive inner star formation.

Galaxies that end up as DM-poor Compact satellites enter their hosts earlier than their DM-rich counterparts (Table~\ref{tab:SnapEntry}) and live in the inner regions of more massive hosts  (Table~\ref{tab:SnapEntry}). They typically retain less than half of their maximum stellar mass and lose their gas before $z=0$ (Fig.~\ref{fig:sizeEvolMass}), presenting their main compaction after entering their host group (Fig.~\ref{fig:SizeDelta}). They show a significant cut in their outer stellar density profile (Fig.~\ref{fig:ProfileAtTime}), due to tidal stripping. DM-poor Compacts also exhibit star formation in the inner regions driven by gas dynamics during interactions with the environment, while experiencing stellar mass loss in the outer region (Fig.~\ref{fig:MassInAbove}). Although the compaction rate is usually fastest during the presence of gas, most galaxies lose more stellar mass in the outer region (by tidal stripping) than they gain in the inner region (by inner star formation from gas compression, Figs.~\ref{fig:MassInAbove} and \ref{fig:HistDecreases}).  Quenching occurs more rapidly in the outer regions, where  ram pressure acts first, while the inner sSFR continuously decreases over time until it is eventually suppressed, without showing a significant increase throughout evolution (Figs.~\ref{fig:sSFRvsPhase} and \ref{fig:sSFRDetail}). This indicates that, in most cases, gas dynamics is important to suppress outer star formation without significantly influencing inner star formation. This conclusion is consistent with the absence of a notable increase in the inner gas mass (Fig.~\ref{fig:sSFRvsPhase}). We confirmed this by showing that, for most DM-poor Compacts, the size decrease due to tidal stripping alone  is greater than that caused solely by the evolution of the inner stellar component during their interaction with the environment (Fig.~\ref{fig:RhalfDetail}). Using dark matter tracers, suitably weighted to mimic the stellar mass density profile at entry into the final host, we showed that the presence of gas enhances the compaction, but by only roughly 15 per cent after effectively correcting for the different tidal field when gas is present (Fig.~\ref{fig:DMtracer}).  Finally, we found that DM-poor Compact satellites are  redder (Fig.~\ref{fig:HistColor}), and are high-metallicity outliers in the $z$=0 stellar mass-metallicity relation (Fig.~\ref{fig:Zmetallicty}), as expected from tidal stripping coupled with sustained inner star formation (Fig.~\ref{fig:sSFRvsPhase}). 

We found only minor differences between DM-poor \CompactsMB{} and \CompactsSB{} populations. The most significant one is that the fraction of DM-poor systems among compact satellites is substantially lower for \CompactsSB{} than for \CompactsMB{} (Sect.~\ref{sec:tidalproxies}). In addition, at fixed DM-poor status, \CompactsSB{} exhibit enhanced star formation in their inner regions compared to their \CompactsMB{} counterparts (Figs.~\ref{fig:sSFRvsPhase} and \ref{fig:sSFRDetail}). This indicates that \CompactsSB{} should not be interpreted as the simple extreme continuation of the environmentally stripped \CompactsMB{} population. Rather, the secondary branch appears to define a population whose interpretation requires caution given the known numerical limitations affecting the inner structure of low-mass TNG50 galaxies (Sect.~\ref{sec:limits}).

Overall, our results suggest that, in the TNG50 cosmological hydrodynamical simulation, satellite dwarf galaxies end up as \Compact{} primarily through two robust pathways, with a third more tentative channel. The first is concentrated star formation for centrals and satellites that do not deeply penetrate their hosts. The second is tidal stripping for deeply penetrating satellites, with a minor acceleration from gas dynamical effects: ram pressure stripping and thinning of the outer regions and tidal compression of the inner gas. A third, less secure channel may operate in a small subset of very metal-rich \CompactsSB{}, where starbursts triggered by ram pressure compression appear to contribute to compaction. However, in many of these systems the final compactness also seems to be aided by subsequent tidal stripping (Appendix~\ref{sec:AppFigs}), and this interpretation should therefore be treated with caution. The balance among the three different pathways may nevertheless be affected by the numerical effects discussed in Sect.~\ref{sec:limits}, including the treatment of BH seeding, sinking, and merging, since stronger AGN feedback in compact progenitors could suppress inner star formation and may thus change the relative roles of star formation and tidal stripping.

Future integral field spectroscopy of compact dwarf satellite galaxies can serve as valuable tools to establish their star formation histories and identify the different pathways leading to the formation of compact dwarf satellites. Additionally, tailor-made controlled hydrodynamic simulations could be useful for understanding how various parameters related to galaxy-environment interactions impact the interplay between tidal stripping and gas dynamics in DM-poor galaxies.

\section*{Acknowledgements}

We thank the anonymous referee for the careful reading and constructive comments. We also thank Igor Chilingarian, Avishai Dekel (sadly deceased 17 November 2025), and Jin Koda for useful discussions. APA thanks the São Paulo Research Foundation, FAPESP, for financial support through contracts 2022/05059-2 and  2020/16152-8. GBLN acknowledges partial financial support from CNPq grant 314528/2023-7 and FAPESP  grant 2024/06400-5. We thank the builders and maintainers of the IllustrisTNG collaboration. We also thank the developers of Python and its scientific computing ecosystem, particularly NumPy \citep{Harris2020}, SciPy \citep{Virtanen2020}, Pandas \citep{McKinney2010} and Matplotlib \citep{Hunter2007}. 

\section*{Data Availability}

All analyses were performed using publicly available data from the IllustrisTNG collaboration, accessible at \href{www.tng-project.org}{www.tng-project.org}. The analysis scripts used to produce the figures and tables will be shared upon reasonable request to the corresponding author.



\bibliographystyle{mnras}
\bibliography{main} 

@ARTICLE{DeAlmeida2024AA,
       author = {{De Almeida}, Abhner Pinto and {Mamon}, Gary A. and {Dekel}, Avishai and {Lima Neto}, Gast{\~a}o B.},
        title = "{What drives the corpulence of galaxies?. I. The formation of central compact dwarf galaxies in TNG50}",
      journal = {\aap},
         year = 2024,
        month = jul,
       volume = {687},
          eid = {A131},
        pages = {A131},
         note = {(Paper I)},
          doi = {10.1051/0004-6361/202449939},
archivePrefix = {arXiv},
       eprint = {2404.15482},
 primaryClass = {astro-ph.GA},
       adsurl = {https://ui.adsabs.harvard.edu/abs/2024A&A...687A.131D}
}

@ARTICLE{Afanasiev+18,
   author = {{Afanasiev}, Anton V. and {Chilingarian}, Igor V. and {Mieske}, Steffen and {Voggel}, Karina T. and {Picotti}, Arianna and {Hilker}, Michael and {Seth}, Anil and {Neumayer}, Nadine and {Frank}, Matthias and {Romanowsky}, Aaron J. and {Hau}, George and {Baumgardt}, Holger and {Ahn}, Christopher and {Strader}, Jay and {den Brok}, Mark and {McDermid}, Richard and {Spitler}, Lee and {Brodie}, Jean and {Walsh}, Jonelle L.},
        title = {A 3.5 million Solar masses black hole in the centre of the ultracompact dwarf galaxy fornax UCD3},
      journal = {\mnras},
         year = 2018,
        month = jul,
       volume = {477},
       number = {4},
        pages = {4856-4865},
          doi = {10.1093/mnras/sty913},
archivePrefix = {arXiv},
       eprint = {1804.02938},
 primaryClass = {astro-ph.GA},
       adsurl = {https://ui.adsabs.harvard.edu/abs/2018MNRAS.477.4856A}
}

@ARTICLE{Asplund+09,
   author = {{Asplund}, Martin and {Grevesse}, Nicolas and {Sauval}, A. Jacques and {Scott}, Pat},
        title = {The Chemical Composition of the Sun},
      journal = {\araa},
         year = 2009,
        month = sep,
       volume = {47},
       number = {1},
        pages = {481-522},
          doi = {10.1146/annurev.astro.46.060407.145222},
archivePrefix = {arXiv},
       eprint = {0909.0948},
 primaryClass = {astro-ph.SR},
       adsurl = {https://ui.adsabs.harvard.edu/abs/2009ARA&A..47..481A}
}

@ARTICLE{Weinberger2017MNRAS,
       author = {{Weinberger}, Rainer and {Springel}, Volker and {Hernquist}, Lars and {Pillepich}, Annalisa and {Marinacci}, Federico and {Pakmor}, R{\"u}diger and {Nelson}, Dylan and {Genel}, Shy and {Vogelsberger}, Mark and {Naiman}, Jill and {Torrey}, Paul},
        title = "{Simulating galaxy formation with black hole driven thermal and kinetic feedback}",
      journal = {\mnras},
         year = 2017,
        month = mar,
       volume = {465},
       number = {3},
        pages = {3291-3308},
          doi = {10.1093/mnras/stw2944},
archivePrefix = {arXiv},
       eprint = {1607.03486},
 primaryClass = {astro-ph.GA},
       adsurl = {https://ui.adsabs.harvard.edu/abs/2017MNRAS.465.3291W}
}

@ARTICLE{Vogelsberger2013MNRAS,
       author = {{Vogelsberger}, Mark and {Genel}, Shy and {Sijacki}, Debora and {Torrey}, Paul and {Springel}, Volker and {Hernquist}, Lars},
        title = "{A model for cosmological simulations of galaxy formation physics}",
      journal = {\mnras},
         year = 2013,
        month = dec,
       volume = {436},
       number = {4},
        pages = {3031-3067},
          doi = {10.1093/mnras/stt1789},
archivePrefix = {arXiv},
       eprint = {1305.2913},
 primaryClass = {astro-ph.CO},
       adsurl = {https://ui.adsabs.harvard.edu/abs/2013MNRAS.436.3031V}
}

@ARTICLE{Chilingarian+09,
   author = {{Chilingarian}, Igor and {Cayatte}, V{\'e}ronique and {Revaz}, Yves and {Dodonov}, Serguei and {Durand }, Daniel and {Durret}, Florence and {Micol}, Alberto and {Slezak}, Eric},
        title = {A Population of Compact Elliptical Galaxies Detected with the Virtual Observatory},
      journal = {Science},
         year = 2009,
        month = dec,
       volume = {326},
       number = {5958},
        pages = {1379},
          doi = {10.1126/science.1175930},
archivePrefix = {arXiv},
       eprint = {0910.0293},
 primaryClass = {astro-ph.CO},
       adsurl = {https://ui.adsabs.harvard.edu/abs/2009Sci...326.1379C}
}

@ARTICLE{Balogh+97,
   author = {{Balogh}, Mike L. and {Morris}, Simon L. and {Yee}, H.~K.~C. and {Carlberg}, R.~G. and {Ellingson}, Erica},
        title = {Star Formation in Cluster Galaxies at 0.2 < Z < 0.55},
      journal = {\apjl},
         year = 1997,
        month = oct,
       volume = {488},
       number = {2},
        pages = {L75-L78},
          doi = {10.1086/310927},
archivePrefix = {arXiv},
       eprint = {astro-ph/9707339},
 primaryClass = {astro-ph},
       adsurl = {https://ui.adsabs.harvard.edu/abs/1997ApJ...488L..75B}
}

@ARTICLE{Steyrleithner2020MNRAS,
       author = {{Steyrleithner}, P. and {Hensler}, G. and {Boselli}, A.},
        title = "{The effect of ram-pressure stripping on dwarf galaxies}",
      journal = {\mnras},
         year = 2020,
        month = may,
       volume = {494},
       number = {1},
        pages = {1114-1127},
          doi = {10.1093/mnras/staa775},
archivePrefix = {arXiv},
       eprint = {2003.09591},
 primaryClass = {astro-ph.GA},
       adsurl = {https://ui.adsabs.harvard.edu/abs/2020MNRAS.494.1114S}
}

@ARTICLE{Harris2020,
       author = {{Harris}, Charles R. and {Millman}, K. Jarrod and {van der Walt}, St{\'e}fan J. and {Gommers}, Ralf and {Virtanen}, Pauli and {Cournapeau}, David and {Wieser}, Eric and {Taylor}, Julian and {Berg}, Sebastian and {Smith}, Nathaniel J. and {Kern}, Robert and {Picus}, Matti and {Hoyer}, Stephan and {van Kerkwijk}, Marten H. and {Brett}, Matthew and {Haldane}, Allan and {del R{\'\i}o}, Jaime Fern{\'a}ndez and {Wiebe}, Mark and {Peterson}, Pearu and {G{\'e}rard-Marchant}, Pierre and {Sheppard}, Kevin and {Reddy}, Tyler and {Weckesser}, Warren and {Abbasi}, Hameer and {Gohlke}, Christoph and {Oliphant}, Travis E.},
        title = "{Array programming with NumPy}",
      journal = {\nat},
         year = 2020,
        month = sep,
       volume = {585},
       number = {7825},
        pages = {357-362},
          doi = {10.1038/s41586-020-2649-2},
archivePrefix = {arXiv},
       eprint = {2006.10256},
 primaryClass = {cs.MS},
       adsurl = {https://ui.adsabs.harvard.edu/abs/2020Natur.585..357H}
}

@ARTICLE{Virtanen2020,
       author = {{Virtanen}, Pauli and {Gommers}, Ralf and {Oliphant}, Travis E. and {Haberland}, Matt and {Reddy}, Tyler and {Cournapeau}, David and {Burovski}, Evgeni and {Peterson}, Pearu and {Weckesser}, Warren and {Bright}, Jonathan and {van der Walt}, St{\'e}fan J. and {Brett}, Matthew and {Wilson}, Joshua and {Millman}, K. Jarrod and {Mayorov}, Nikolay and {Nelson}, Andrew R.~J. and {Jones}, Eric and {Kern}, Robert and {Larson}, Eric and {Carey}, C.~J. and {Polat}, {\.I}lhan and {Feng}, Yu and {Moore}, Eric W. and {VanderPlas}, Jake and {Laxalde}, Denis and {Perktold}, Josef and {Cimrman}, Robert and {Henriksen}, Ian and {Quintero}, E.~A. and {Harris}, Charles R. and {Archibald}, Anne M. and {Ribeiro}, Ant{\^o}nio H. and {Pedregosa}, Fabian and {van Mulbregt}, Paul and {SciPy 1. 0 Contributors}},
        title = "{SciPy 1.0: fundamental algorithms for scientific computing in Python}",
      journal = {Nature Methods},
         year = 2020,
        month = feb,
       volume = {17},
        pages = {261-272},
          doi = {10.1038/s41592-019-0686-2},
archivePrefix = {arXiv},
       eprint = {1907.10121},
 primaryClass = {cs.MS},
       adsurl = {https://ui.adsabs.harvard.edu/abs/2020NatMe..17..261V}
}

@InProceedings{McKinney2010,
  author    = { {W}es {M}c{K}inney },
  title     = { {D}ata {S}tructures for {S}tatistical {C}omputing in {P}ython },
  booktitle = { {P}roceedings of the 9th {P}ython in {S}cience {C}onference },
  pages     = { 56 - 61 },
  year      = { 2010 },
  editor    = { {S}t\'efan van der {W}alt and {J}arrod {M}illman },
  doi       = { 10.25080/Majora-92bf1922-00a }
}

@ARTICLE{Nelson2019ComAC,
       author = {{Nelson}, Dylan and {Springel}, Volker and {Pillepich}, Annalisa and {Rodriguez-Gomez}, Vicente and {Torrey}, Paul and {Genel}, Shy and {Vogelsberger}, Mark and {Pakmor}, Ruediger and {Marinacci}, Federico and {Weinberger}, Rainer and {Kelley}, Luke and {Lovell}, Mark and {Diemer}, Benedikt and {Hernquist}, Lars},
        title = "{The IllustrisTNG simulations: public data release}",
      journal = {Computational Astrophysics and Cosmology},
         year = 2019,
        month = may,
       volume = {6},
       number = {1},
          eid = {2},
        pages = {2},
          doi = {10.1186/s40668-019-0028-x},
archivePrefix = {arXiv},
       eprint = {1812.05609},
 primaryClass = {astro-ph.GA},
       adsurl = {https://ui.adsabs.harvard.edu/abs/2019ComAC...6....2N}
}

@ARTICLE{Hunter2007,
       author = {{Hunter}, John D.},
        title = "{Matplotlib: A 2D Graphics Environment}",
      journal = {Computing in Science and Engineering},
         year = 2007,
        month = may,
       volume = {9},
       number = {3},
        pages = {90-95},
          doi = {10.1109/MCSE.2007.55},
       adsurl = {https://ui.adsabs.harvard.edu/abs/2007CSE.....9...90H}
}

@ARTICLE{Whitmore&Gilmore91,
   author = {{Whitmore}, Bradley C. and {Gilmore}, Diane M.},
        title = {On the Interpretation of the Morphology-Density Relation for Galaxies in Clusters},
      journal = {\apj},
         year = 1991,
        month = jan,
       volume = {367},
        pages = {64},
          doi = {10.1086/169602},
       adsurl = {https://ui.adsabs.harvard.edu/abs/1991ApJ...367...64W}
}

@ARTICLE{Jang+24,
   author = {{Jang}, J.~K. and {Yi}, Sukyoung K. and {Rey}, Soo-Chang and {Rhee}, Jinsu and {Dubois}, Yohan and {Kimm}, Taysun and {Pichon}, Christophe and {Kraljic}, Katarina and {Kim}, Suk},
        title = {Formation Pathways of the Compact Stellar Systems},
      journal = {\apj},
         year = 2024,
        month = jul,
       volume = {969},
       number = {1},
          eid = {59},
        pages = {59},
          doi = {10.3847/1538-4357/ad4d8a},
archivePrefix = {arXiv},
       eprint = {2405.10195},
 primaryClass = {astro-ph.GA},
       adsurl = {https://ui.adsabs.harvard.edu/abs/2024ApJ...969...59J}
}

@ARTICLE{Khoperskov+23,
       author = {{Khoperskov}, Alexander V. and {Khrapov}, Sergey S. and {Sirotin}, Danila S.},
        title = "{Formation of Transitional cE/UCD Galaxies through Massive/Dwarf Disc Galaxy Mergers}",
      journal = {Galaxies},
         year = 2023,
        month = dec,
       volume = {12},
       number = {1},
          eid = {1},
        pages = {1},
          doi = {10.3390/galaxies12010001},
archivePrefix = {arXiv},
       eprint = {2412.03100},
 primaryClass = {astro-ph.GA},
       adsurl = {https://ui.adsabs.harvard.edu/abs/2023Galax..12....1K}
}

@ARTICLE{Dubois+21,
   author = {{Dubois}, Yohan and {Beckmann}, Ricarda and {Bournaud}, Fr{\'e}d{\'e}ric and {Choi}, Hoseung and {Devriendt}, Julien and {Jackson}, Ryan and {Kaviraj}, Sugata and {Kimm}, Taysun and {Kraljic}, Katarina and {Laigle}, Clotilde and {Martin}, Garreth and {Park}, Min-Jung and {Peirani}, S{\'e}bastien and {Pichon}, Christophe and {Volonteri}, Marta and {Yi}, Sukyoung K.},
        title = {Introducing the NEWHORIZON simulation: Galaxy properties with resolved internal dynamics across cosmic time},
      journal = {\aap},
         year = 2021,
        month = jul,
       volume = {651},
          eid = {A109},
        pages = {A109},
          doi = {10.1051/0004-6361/202039429},
archivePrefix = {arXiv},
       eprint = {2009.10578},
 primaryClass = {astro-ph.GA},
       adsurl = {https://ui.adsabs.harvard.edu/abs/2021A&A...651A.109D}
}

@ARTICLE{Dressler+80,
       author = {{Dressler}, A.},
        title = "{Galaxy morphology in rich clusters: implications for the formation and evolution of galaxies.}",
      journal = {\apj},
         year = 1980,
        month = mar,
       volume = {236},
        pages = {351-365},
          doi = {10.1086/157753},
       adsurl = {https://ui.adsabs.harvard.edu/abs/1980ApJ...236..351D}
}

@ARTICLE{Melnick+77,
       author = {{Melnick}, J. and {Sargent}, W.~L.~W.},
        title = "{The radial distribution of morphological types of galaxies in X-ray clusters.}",
      journal = {\apj},
         year = 1977,
        month = jul,
       volume = {215},
        pages = {401-407},
          doi = {10.1086/155369},
       adsurl = {https://ui.adsabs.harvard.edu/abs/1977ApJ...215..401M}
}

@ARTICLE{Dressler+99,
       author = {{Dressler}, Alan and {Smail}, Ian and {Poggianti}, Bianca M. and {Butcher}, Harvey and {Couch}, Warrick J. and {Ellis}, Richard S. and {Oemler}, Augustus, Jr.},
        title = "{A Spectroscopic Catalog of 10 Distant Rich Clusters of Galaxies}",
      journal = {\apjs},
         year = 1999,
        month = may,
       volume = {122},
       number = {1},
        pages = {51-80},
          doi = {10.1086/313213},
archivePrefix = {arXiv},
       eprint = {astro-ph/9901263},
 primaryClass = {astro-ph},
       adsurl = {https://ui.adsabs.harvard.edu/abs/1999ApJS..122...51D}
}

@ARTICLE{DeLucia+04,
       author = {{De Lucia}, Gabriella and {Kauffmann}, Guinevere and {White}, Simon D.~M.},
        title = "{Chemical enrichment of the intracluster and intergalactic medium in a hierarchical galaxy formation model}",
      journal = {\mnras},
         year = 2004,
        month = apr,
       volume = {349},
       number = {3},
        pages = {1101-1116},
          doi = {10.1111/j.1365-2966.2004.07584.x},
archivePrefix = {arXiv},
       eprint = {astro-ph/0310268},
 primaryClass = {astro-ph},
       adsurl = {https://ui.adsabs.harvard.edu/abs/2004MNRAS.349.1101D}
}

@ARTICLE{Balogh1999ApJ,
       author = {{Balogh}, Michael L. and {Morris}, Simon L. and {Yee}, H.~K.~C. and {Carlberg}, R.~G. and {Ellingson}, Erica},
        title = "{Differential Galaxy Evolution in Cluster and Field Galaxies at z\raisebox{-0.5ex}\textasciitilde0.3}",
      journal = {\apj},
         year = 1999,
        month = dec,
       volume = {527},
       number = {1},
        pages = {54-79},
          doi = {10.1086/308056},
archivePrefix = {arXiv},
       eprint = {astro-ph/9906470},
 primaryClass = {astro-ph},
       adsurl = {https://ui.adsabs.harvard.edu/abs/1999ApJ...527...54B}
}

@ARTICLE{Balogh2000,
       author = {{Balogh}, Michael L. and {Navarro}, Julio F. and {Morris}, Simon L.},
        title = "{The Origin of Star Formation Gradients in Rich Galaxy Clusters}",
      journal = {\apj},
         year = 2000,
        month = sep,
       volume = {540},
       number = {1},
        pages = {113-121},
          doi = {10.1086/309323},
archivePrefix = {arXiv},
       eprint = {astro-ph/0004078},
 primaryClass = {astro-ph},
       adsurl = {https://ui.adsabs.harvard.edu/abs/2000ApJ...540..113B}
}

@ARTICLE{Larson+80,
       author = {{Larson}, R.~B. and {Tinsley}, B.~M. and {Caldwell}, C.~N.},
        title = "{The evolution of disk galaxies and the origin of S0 galaxies}",
      journal = {\apj},
         year = 1980,
        month = may,
       volume = {237},
        pages = {692-707},
          doi = {10.1086/157917},
       adsurl = {https://ui.adsabs.harvard.edu/abs/1980ApJ...237..692L}
}

@ARTICLE{Pfeffer2014,
       author = {{Pfeffer}, J. and {Griffen}, B.~F. and {Baumgardt}, H. and {Hilker}, M.},
       adstitle = "{Contribution of stripped nuclear clusters to globular cluster and ultracompact dwarf galaxy populations}",
       journal = {MNRAS},
       year = 2014,
       volume = {444},
       pages = {3670},
       adsurl = {https://ui.adsabs.harvard.edu/abs/2014MNRAS.444.3670P}
}

@ARTICLE{Contini+14,
       author = {{Contini}, E. and {De Lucia}, G. and {Villalobos}, {\'A}. and {Borgani}, S.},
        title = "{On the formation and physical properties of the intracluster light in hierarchical galaxy formation models}",
      journal = {\mnras},
         year = 2014,
        month = feb,
       volume = {437},
       number = {4},
        pages = {3787-3802},
          doi = {10.1093/mnras/stt2174},
archivePrefix = {arXiv},
       eprint = {1311.2076},
 primaryClass = {astro-ph.CO},
       adsurl = {https://ui.adsabs.harvard.edu/abs/2014MNRAS.437.3787C}
}

@ARTICLE{Fattahi2018MNRAS,
       author = {{Fattahi}, Azadeh and {Navarro}, Julio F. and {Frenk}, Carlos S. and {Oman}, Kyle A. and {Sawala}, Till and {Schaller}, Matthieu},
        title = "{Tidal stripping and the structure of dwarf galaxies in the Local Group}",
      journal = {\mnras},
         year = 2018,
        month = may,
       volume = {476},
       number = {3},
        pages = {3816-3836},
          doi = {10.1093/mnras/sty408},
archivePrefix = {arXiv},
       eprint = {1707.03898},
 primaryClass = {astro-ph.GA},
       adsurl = {https://ui.adsabs.harvard.edu/abs/2018MNRAS.476.3816F}
}

@proceedings{Zinnecker1988,
        author = {{Zinnecker}, H. and {Keable}, C.~J. and {Dunlop}, J.~S. and {Cannon}, R.~D. and {Griffiths}, W.~K.},
        title = "{The Nuclei of Nucleated Dwarf Elliptical Galaxies - are they Globular Clusters?}",
    booktitle = {The Harlow-Shapley Symposium on Globular Cluster Systems in Galaxies},
         year = 1988,
       editor = {{Grindlay}, Jonathan E. and {Philip}, A.~G. Davis},
       series = {IAU Symposium},
       volume = {126},
        month = jan,
        pages = {603},
       adsurl = {https://ui.adsabs.harvard.edu/abs/1988IAUS..126..603Z}
}

@ARTICLE{DOnghia2016,
       author = {{D'Onghia}, Elena and {Fox}, Andrew J.},
       adstitle = "{The Magellanic Stream: Circumnavigating the Galaxy}",
       journal = {ARA\&A},
       year = 2016,
       volume = {54},
       pages = {363},
       adsurl = {https://ui.adsabs.harvard.edu/abs/2016ARA&A..54..363D}
}

@ARTICLE{Ibata2019,
       author = {{Ibata}, Rodrigo A. and {Bellazzini}, Michele and {Malhan}, Khyati and {Martin}, Nicolas and {Bianchini}, Paolo},
        title = "{Identification of the long stellar stream of the prototypical massive globular cluster {\ensuremath{\omega}} Centauri}",
      journal = {Nature Astronomy},
         year = 2019,
        month = apr,
       volume = {3},
        pages = {667-672},
          doi = {10.1038/s41550-019-0751-x},
archivePrefix = {arXiv},
       eprint = {1902.09544},
 primaryClass = {astro-ph.GA},
       adsurl = {https://ui.adsabs.harvard.edu/abs/2019NatAs...3..667I}
}

@ARTICLE{Bekki2001,
       author = {{Bekki}, Kenji and {Couch}, Warrick J. and {Drinkwater}, Michael J.},
       adstitle = "{Galaxy Threshing and the Formation of Ultracompact Dwarf Galaxies}",
       journal = {ApJL},
       year = 2001,
       volume = {552},
       pages = {L105},
       adsurl = {https://ui.adsabs.harvard.edu/abs/2001ApJ...552L.105B}
}

@ARTICLE{Bassino1994ApJ,
       author = {{Bassino}, Lilia P. and {Muzzio}, Juan C. and {Rabolli}, Monica},
        title = "{Are Globular Clusters the Nuclei of Cannibalized Dwarf Galaxies?}",
      journal = {\apj},
         year = 1994,
        month = aug,
       volume = {431},
        pages = {634},

          doi = {10.1086/174514},
       adsurl = {https://ui.adsabs.harvard.edu/abs/1994ApJ...431..634B}
}

@ARTICLE{Klimentowski2009,
   author = {{Klimentowski}, Jaros{\l}aw and {{\L}okas}, Ewa L. and {Kazantzidis}, Stelios and {Mayer}, Lucio and {Mamon}, Gary A.},
        title = {Tidal evolution of discy dwarf galaxies in the Milky Way potential: the formation of dwarf spheroidals},
      journal = {\mnras},
         year = 2009,
        month = aug,
       volume = {397},
       number = {4},
        pages = {2015-2029},
          doi = {10.1111/j.1365-2966.2009.15046.x},
archivePrefix = {arXiv},
       eprint = {0803.2464},
 primaryClass = {astro-ph},
       adsurl = {https://ui.adsabs.harvard.edu/abs/2009MNRAS.397.2015K}
}

@ARTICLE{Brodie2011AJ,
       author = {{Brodie}, Jean P. and {Romanowsky}, Aaron J. and {Strader}, Jay and {Forbes}, Duncan A.},
        title = "{The Relationships among Compact Stellar Systems: A Fresh View of Ultracompact Dwarfs}",
      journal = {\aj},
         year = 2011,
        month = dec,
       volume = {142},
       number = {6},
          eid = {199},
        pages = {199},
          doi = {10.1088/0004-6256/142/6/199},
archivePrefix = {arXiv},
       eprint = {1109.5696},
 primaryClass = {astro-ph.CO},
       adsurl = {https://ui.adsabs.harvard.edu/abs/2011AJ....142..199B}
}

@ARTICLE{Du+19,
   author = {{Du}, Min and {Debattista}, Victor P. and {Ho}, Luis C. and {C{\^o}t{\'e}}, Patrick and {Spengler}, Chelsea and {Erwin}, Peter and {Wadsley}, James W. and {Norris}, Mark A. and {Earp}, Samuel W.~F. and {Quinn}, Thomas R. and {Fiteni}, Karl and {Caruana}, Joseph},
        title = {The Formation of Compact Elliptical Galaxies in the Vicinity of a Massive Galaxy: The Role of Ram-pressure Confinement},
      journal = {\apj},
         year = 2019,
        month = apr,
       volume = {875},
       number = {1},
          eid = {58},
        pages = {58},
          doi = {10.3847/1538-4357/ab0e0c},
archivePrefix = {arXiv},
       eprint = {1811.06778},
 primaryClass = {astro-ph.GA},
       adsurl = {https://ui.adsabs.harvard.edu/abs/2019ApJ...875...58D}
}

@ARTICLE{Bian2025ApJ,
    author = {{Bian}, Yuan and {Du}, Min and {Debattista}, Victor P. and {Nelson}, Dylan and {Norris}, Mark A. and {Ho}, Luis C. and {Lu}, Shuai and {Cen}, Renyue and {Ma}, Shuo and {Ge}, Chong and {Fang}, Taotao and {Li}, Hui},
    title = "{Two Channels of Metal-rich Compact Stellar System Formation: Starbursts under High Ram Pressure versus Tidal Stripping}",
      journal = {\apjl},
         year = 2025,
        month = feb,
       volume = {979},
       number = {2},
          eid = {L33},
        pages = {L33},
          doi = {10.3847/2041-8213/ada912},
archivePrefix = {arXiv},
       eprint = {2409.05229},
 primaryClass = {astro-ph.GA},
       adsurl = {https://ui.adsabs.harvard.edu/abs/2025ApJ...979L..33B}
}

@ARTICLE{Chilingarian&Mamon08,
   author = {{Chilingarian}, Igor V. and {Mamon}, Gary A.},
        title = {SDSSJ124155.33+114003.7 - a missing link between compact elliptical and ultracompact dwarf galaxies},
      journal = {\mnras},
         year = 2008,
        month = mar,
       volume = {385},
       number = {1},
        pages = {L83-L87},
          doi = {10.1111/j.1745-3933.2008.00438.x},
archivePrefix = {arXiv},
       eprint = {0712.2724},
 primaryClass = {astro-ph},
       adsurl = {https://ui.adsabs.harvard.edu/abs/2008MNRAS.385L..83C}
}

@ARTICLE{Dashyan+19,
       author = {{Dashyan}, Gohar and {Choi}, Ena and {Somerville}, Rachel S. and {Naab}, Thorsten and {Quirk}, Amanda C.~N. and {Hirschmann}, Michaela and {Ostriker}, Jeremiah P.},
        title = "{AGN-driven quenching of satellite galaxies}",
      journal = {\mnras},
         year = 2019,
        month = aug,
       volume = {487},
       number = {4},
        pages = {5889-5901},
          doi = {10.1093/mnras/stz1697},
archivePrefix = {arXiv},
       eprint = {1906.07431},
 primaryClass = {astro-ph.GA},
       adsurl = {https://ui.adsabs.harvard.edu/abs/2019MNRAS.487.5889D}
}

@ARTICLE{Davis1985ApJ,
       author = {{Davis}, M. and {Efstathiou}, G. and {Frenk}, C.~S. and {White}, S.~D.~M.},
        title = "{The evolution of large-scale structure in a universe dominated by cold dark matter}",
      journal = {\apj},
         year = 1985,
        month = may,
       volume = {292},
        pages = {371-394},
          doi = {10.1086/163168},
       adsurl = {https://ui.adsabs.harvard.edu/abs/1985ApJ...292..371D}
}

@ARTICLE{Deeley+23,
       author = {{Deeley}, Simon and {Drinkwater}, Michael J. and {Sweet}, Sarah M. and {Bekki}, Kenji and {Couch}, Warrick J. and {Forbes}, Duncan A.},
        title = "{The formation pathways of compact elliptical galaxies}",
      journal = {\mnras},
         year = 2023,
        month = oct,
       volume = {525},
       number = {1},
        pages = {1192-1209},
          doi = {10.1093/mnras/stad2313},
archivePrefix = {arXiv},
       eprint = {2308.00305},
 primaryClass = {astro-ph.GA},
       adsurl = {https://ui.adsabs.harvard.edu/abs/2023MNRAS.525.1192D}
}

@ARTICLE{Dekel+03,
   author = {{Dekel}, Avishai and {Devor}, Jonathan and {Hetzroni}, Guy},
        title = {Galactic halo cusp-core: tidal compression in mergers},
      journal = {\mnras},
         year = 2003,
        month = may,
       volume = {341},
       number = {1},
        pages = {326-342},
          doi = {10.1046/j.1365-8711.2003.06432.x},
archivePrefix = {arXiv},
       eprint = {astro-ph/0204452},
 primaryClass = {astro-ph},
       adsurl = {https://ui.adsabs.harvard.edu/abs/2003MNRAS.341..326D}
}

@ARTICLE{Drinkwater2000PASA,
       author = {{Drinkwater}, M.~J. and {Jones}, J.~B. and {Gregg}, M.~D. and {Phillipps}, S.},
        title = "{Compact Stellar Systems in the Fornax Cluster: Super-massive Star Clusters or Extremely Compact Dwarf Galaxies?}",
      journal = {\pasa},
         year = 2000,
        month = dec,
       volume = {17},
       number = {3},
        pages = {227-233},
          doi = {10.1071/AS00034},
archivePrefix = {arXiv},
       eprint = {astro-ph/0002003},
 primaryClass = {astro-ph},
       adsurl = {https://ui.adsabs.harvard.edu/abs/2000PASA...17..227D}
}

@ARTICLE{Faber1973ApJ,
       author = {{Faber}, S.~M.},
        title = "{Tidal Origin of Elliptical Galaxies of High Surface Brightness}",
      journal = {\apj},
         year = 1973,
        month = jan,
       volume = {179},
        pages = {423-426},
          doi = {10.1086/151881},
       adsurl = {https://ui.adsabs.harvard.edu/abs/1973ApJ...179..423F}
}

@ARTICLE{Gunn1972ApJ,
       author = {{Gunn}, James E. and {Gott}, J. R. {\rm III}},
        title = "{On the Infall of Matter Into Clusters of Galaxies and Some Effects on Their Evolution}",
      journal = {\apj},
         year = 1972,
        month = aug,
       volume = {176},
        pages = {1},
          doi = {10.1086/151605},
       adsurl = {https://ui.adsabs.harvard.edu/abs/1972ApJ...176....1G}
}

@ARTICLE{Hayashi+03,
   author = {{Hayashi}, Eric and {Navarro}, Julio F. and {Taylor}, James E. and {Stadel}, Joachim and {Quinn}, Thomas},
        title = {The Structural Evolution of Substructure},
      journal = {\apj},
         year = 2003,
        month = feb,
       volume = {584},
       number = {2},
        pages = {541-558},
          doi = {10.1086/345788},
archivePrefix = {arXiv},
       eprint = {astro-ph/0203004},
 primaryClass = {astro-ph},
       adsurl = {https://ui.adsabs.harvard.edu/abs/2003ApJ...584..541H}
}

@ARTICLE{Janz2016MNRAS,
       author = {{Janz}, J. and {Laurikainen}, E. and {Laine}, J. and {Salo}, H. and {Lisker}, T.},
        title = "{How similar is the stellar structure of low-mass late-type galaxies to that of early-type dwarfs?}",
      journal = {\mnras},
         year = 2016,
        month = sep,
       volume = {461},
       number = {1},
        pages = {L82-L86},
          doi = {10.1093/mnrasl/slw104},
archivePrefix = {arXiv},
       eprint = {1605.06189},
 primaryClass = {astro-ph.GA},
       adsurl = {https://ui.adsabs.harvard.edu/abs/2016MNRAS.461L..82J}
}

@ARTICLE{Kim2020ApJ,
       author = {{Kim}, Suk and {Jeong}, Hyunjin and {Rey}, Soo-Chang and {Lee}, Youngdae and {Lee}, Jaehyun and {Joo}, Seok-Joo and {Kim}, Hak-Sub},
        title = "{Compact Elliptical Galaxies in Different Local Environments: A Mixture of Galaxies with Different Origins?}",
      journal = {\apj},
         year = 2020,
        month = nov,
       volume = {903},
       number = {1},
          eid = {65},
        pages = {65},
          doi = {10.3847/1538-4357/abaef5},
archivePrefix = {arXiv},
       eprint = {2008.10686},
 primaryClass = {astro-ph.GA},
       adsurl = {https://ui.adsabs.harvard.edu/abs/2020ApJ...903...65K}
}

@ARTICLE{Kormendy2012ApJS,
       author = {{Kormendy}, John and {Bender}, Ralf},
        title = "{A Revised Parallel-sequence Morphological Classification of Galaxies: Structure and Formation of S0 and Spheroidal Galaxies}",
      journal = {\apjs},
         year = 2012,
        month = jan,
       volume = {198},
       number = {1},
          eid = {2},
        pages = {2},
          doi = {10.1088/0067-0049/198/1/2},
archivePrefix = {arXiv},
       eprint = {1110.4384},
 primaryClass = {astro-ph.CO},
       adsurl = {https://ui.adsabs.harvard.edu/abs/2012ApJS..198....2K}
}

@article{GonzalezJara2025,
	author = {{Gonzalez-Jara}, Jenny and {Tissera}, Patricia B. and {Monachesi}, Antonela and {Sillero}, Emanuel and {Pallero}, Diego and {Pedrosa}, Susana and {Tau}, Elisa A. and {Tapia-Contreras}, Brian and {Bignone}, Lucas},
        title = "{Unveiling the formation channels of stellar halos through their chemical fingerprints}",
      journal = {\aap},
         year = 2025,
        month = jan,
       volume = {693},
          eid = {A282},
        pages = {A282},
          doi = {10.1051/0004-6361/202452639},
archivePrefix = {arXiv},
       eprint = {2412.13483},
 primaryClass = {astro-ph.GA},
       adsurl = {https://ui.adsabs.harvard.edu/abs/2025A&A...693A.282G}
}

@article{Celiz2025cAA,
	author = {{Celiz}, Bruno M. and {Navarro}, Julio F. and {Abadi}, Mario G.},
        title = "{Accreted stars and stellar haloes of simulated galaxies in TNG50}",
      journal = {\aap},
         year = 2025,
        month = nov,
       volume = {704},
          eid = {A57},
        pages = {A57},
          doi = {10.1051/0004-6361/202556633},
archivePrefix = {arXiv},
       eprint = {2510.18971},
 primaryClass = {astro-ph.GA},
       adsurl = {https://ui.adsabs.harvard.edu/abs/2025A&A...704A..57C}
}

@article{Tau2025a,
	author = {{Tau}, Elisa A. and {Monachesi}, Antonela and {Gomez}, Facundo A. and {Grand}, Robert J.~J. and {Pakmor}, R{\"u}diger and {van de Voort}, Freeke and {Gonzalez-Jara}, Jenny and {Tissera}, Patricia B. and {Marinacci}, Federico and {Bieri}, Rebekka},
        title = "{The role of accreted and in situ populations in shaping the stellar halos of low-mass galaxies}",
      journal = {\aap},
         year = 2025,
        month = jul,
       volume = {699},
          eid = {A93},
        pages = {A93},
          doi = {10.1051/0004-6361/202453488},
archivePrefix = {arXiv},
       eprint = {2412.13807},
 primaryClass = {astro-ph.GA},
       adsurl = {https://ui.adsabs.harvard.edu/abs/2025A&A...699A..93T}
}

@ARTICLE{Ludlow2019MNRAS,
       author = {{Ludlow}, Aaron D. and {Schaye}, Joop and {Bower}, Richard},
        title = "{Numerical convergence of simulations of galaxy formation: the abundance and internal structure of cold dark matter haloes}",
      journal = {\mnras},
         year = 2019,
        month = sep,
       volume = {488},
       number = {3},
        pages = {3663-3684},
          doi = {10.1093/mnras/stz1821},
archivePrefix = {arXiv},
       eprint = {1812.05777},
 primaryClass = {astro-ph.CO},
       adsurl = {https://ui.adsabs.harvard.edu/abs/2019MNRAS.488.3663L}
}

@ARTICLE{Ludlow2023MNRAS,
       author = {{Ludlow}, Aaron D. and {Fall}, S. Michael and {Wilkinson}, Matthew J. and {Schaye}, Joop and {Obreschkow}, Danail},
        title = "{Spurious heating of stellar motions by dark matter particles in cosmological simulations of galaxy formation}",
      journal = {\mnras},
         year = 2023,
        month = nov,
       volume = {525},
       number = {4},
        pages = {5614-5630},
          doi = {10.1093/mnras/stad2615},
archivePrefix = {arXiv},
       eprint = {2306.05753},
 primaryClass = {astro-ph.GA},
       adsurl = {https://ui.adsabs.harvard.edu/abs/2023MNRAS.525.5614L}
}

@ARTICLE{Kirby2013ApJ,
       author = {{Kirby}, Evan N. and {Cohen}, Judith G. and {Guhathakurta}, Puragra and {Cheng}, Lucy and {Bullock}, James S. and {Gallazzi}, Anna},
        title = "{The Universal Stellar Mass-Stellar Metallicity Relation for Dwarf Galaxies}",
      journal = {\apj},
         year = 2013,
        month = dec,
       volume = {779},
       number = {2},
          eid = {102},
        pages = {102},
          doi = {10.1088/0004-637X/779/2/102},
archivePrefix = {arXiv},
       eprint = {1310.0814},
 primaryClass = {astro-ph.GA},
       adsurl = {https://ui.adsabs.harvard.edu/abs/2013ApJ...779..102K}
}

@ARTICLE{Zeng2024MNRAS,
       author = {{Zeng}, Guangquan and {Wang}, Lan and {Gao}, Liang and {Yang}, Hang},
        title = "{Kinematic morphology of low-mass galaxies in IllustrisTNG}",
      journal = {\mnras},
         year = 2024,
        month = aug,
       volume = {532},
       number = {2},
        pages = {2558-2569},
          doi = {10.1093/mnras/stae1651},
archivePrefix = {arXiv},
       eprint = {2404.14184},
 primaryClass = {astro-ph.GA},
       adsurl = {https://ui.adsabs.harvard.edu/abs/2024MNRAS.532.2558Z}
}

@ARTICLE{Chiang2026OJAp,
       author = {{Chiang}, Barry T. and {van den Bosch}, Frank C. and {Schive}, Hsi-Yu},
        title = "{Universal numerical convergence criteria for subhalo tidal evolution}",
      journal = {The Open Journal of Astrophysics},
         year = 2026,
        month = jan,
       volume = {9},
        pages = {55367},
          doi = {10.33232/001c.155367},
archivePrefix = {arXiv},
       eprint = {2510.26901},
 primaryClass = {astro-ph.CO},
       adsurl = {https://ui.adsabs.harvard.edu/abs/2026OJAp....955367C}
}

@ARTICLE{Hemler2021MNRAS,
       author = {{Hemler}, Z.~S. and {Torrey}, Paul and {Qi}, Jia and {Hernquist}, Lars and {Vogelsberger}, Mark and {Ma}, Xiangcheng and {Kewley}, Lisa J. and {Nelson}, Dylan and {Pillepich}, Annalisa and {Pakmor}, R{\"u}diger and {Marinacci}, Federico},
        title = "{Gas-phase metallicity gradients of TNG50 star-forming galaxies}",
      journal = {\mnras},
         year = 2021,
        month = sep,
       volume = {506},
       number = {2},
        pages = {3024-3048},
          doi = {10.1093/mnras/stab1803},
archivePrefix = {arXiv},
       eprint = {2007.10993},
 primaryClass = {astro-ph.GA},
       adsurl = {https://ui.adsabs.harvard.edu/abs/2021MNRAS.506.3024H}
}

@ARTICLE{Sharonova2025ApJ,
       author = {{Sharonova}, Aleksandra V. and {Grishin}, Kirill A. and {Chilingarian}, Igor V. and {Mamon}, Gary A. and {Caldwell}, Nelson and {Fabricant}, Daniel},
        title = "{A Census of Compact Elliptical Galaxies in the Coma Cluster}",
      journal = {\apj},
         year = 2025,
        month = nov,
       volume = {993},
       number = {2},
          eid = {229},
        pages = {229},
          doi = {10.3847/1538-4357/ae070d},
archivePrefix = {arXiv},
       eprint = {2505.14772},
 primaryClass = {astro-ph.GA},
       adsurl = {https://ui.adsabs.harvard.edu/abs/2025ApJ...993..229S}
}

@ARTICLE{Rhee2026AA,
       author = {{Rhee}, Jinsu and {Pichon}, Christophe and {Dubois}, Yohan and {Yi}, Sukyoung K. and {Ko}, Jongwan and {Sheen}, Yun-Kyeong and {Han}, San and {Jeon}, Seyoung and {Jang}, J.~K. and {Lee}, Wonki and {Contini}, Emanuele and {Lee}, Bumhyun and {Lee}, Jaehyun and {Kraljic}, Katarina and {Peirani}, S{\'e}bastien},
        title = "{The origin of gas stripping of galaxies in group environments}",
      journal = {\aap},
         year = 2026,
        month = jan,
       volume = {705},
          eid = {A106},
        pages = {A106},
          doi = {10.1051/0004-6361/202556771},
archivePrefix = {arXiv},
       eprint = {2512.03145},
 primaryClass = {astro-ph.GA},
       adsurl = {https://ui.adsabs.harvard.edu/abs/2026A&A...705A.106R}
}

@ARTICLE{Forouhar2025MNRAS,
       author = {{Forouhar Moreno}, Victor J. and {Helly}, John and {McGibbon}, Robert and {Schaye}, Joop and {Schaller}, Matthieu and {Han}, Jiaxin and {Kugel}, Roi and {Bah{\'e}}, Yannick M.},
        title = "{Assessing subhalo finders in cosmological hydrodynamical simulations}",
      journal = {\mnras},
         year = 2025,
        month = oct,
       volume = {543},
       number = {2},
        pages = {1339-1372},
          doi = {10.1093/mnras/staf1478},
archivePrefix = {arXiv},
       eprint = {2502.06932},
 primaryClass = {astro-ph.CO},
       adsurl = {https://ui.adsabs.harvard.edu/abs/2025MNRAS.543.1339F}
}

@ARTICLE{Behroozi2013ApJ,
       author = {{Behroozi}, Peter S. and {Wechsler}, Risa H. and {Wu}, Hao-Yi and {Busha}, Michael T. and {Klypin}, Anatoly A. and {Primack}, Joel R.},
        title = "{Gravitationally Consistent Halo Catalogs and Merger Trees for Precision Cosmology}",
      journal = {\apj},
         year = 2013,
        month = jan,
       volume = {763},
       number = {1},
          eid = {18},
        pages = {18},
          doi = {10.1088/0004-637X/763/1/18},
archivePrefix = {arXiv},
       eprint = {1110.4370},
 primaryClass = {astro-ph.CO},
       adsurl = {https://ui.adsabs.harvard.edu/abs/2013ApJ...763...18B}
}

@ARTICLE{Springel2003MNRAS,
       author = {{Springel}, Volker and {Hernquist}, Lars},
        title = "{Cosmological smoothed particle hydrodynamics simulations: a hybrid multiphase model for star formation}",
      journal = {\mnras},
         year = 2003,
        month = feb,
       volume = {339},
       number = {2},
        pages = {289-311},
          doi = {10.1046/j.1365-8711.2003.06206.x},
archivePrefix = {arXiv},
       eprint = {astro-ph/0206393},
 primaryClass = {astro-ph},
       adsurl = {https://ui.adsabs.harvard.edu/abs/2003MNRAS.339..289S}
}

@ARTICLE{Lovell2025MNRAS,
       author = {{Lovell}, Mark R. and {Pillepich}, Annalisa and {Engler}, Christoph and {Nelson}, Dylan and {Ramesh}, Rahul and {Springel}, Volker and {Hernquist}, Lars},
        title = "{Numerical effects on the stripping of dark matter and stars in IllustrisTNG galaxy groups and clusters}",
      journal = {\mnras},
         year = 2025,
        month = dec,
       volume = {544},
       number = {4},
        pages = {4367-4389},
          doi = {10.1093/mnras/staf2012},
archivePrefix = {arXiv},
       eprint = {2509.07078},
 primaryClass = {astro-ph.GA},
       adsurl = {https://ui.adsabs.harvard.edu/abs/2025MNRAS.544.4367L}
}

@ARTICLE{Mamon87,
   author = {{Mamon}, Gary A.},
        title = {The Dynamics of Small Groups of Galaxies. I. Virialized Groups},
      journal = {\apj},
         year = 1987,
        month = oct,
       volume = {321},
        pages = {622},
          doi = {10.1086/165658},
       adsurl = {https://ui.adsabs.harvard.edu/abs/1987ApJ...321..622M}
}

@ARTICLE{Marinacci2018MNRAS,
       author = {{Marinacci}, Federico and {Vogelsberger}, Mark and {Pakmor}, R{\"u}diger and {Torrey}, Paul and {SpringMarinacciel}, Volker and {Hernquist}, Lars and {Nelson}, Dylan and {Weinberger}, Rainer and {Pillepich}, Annalisa and {Naiman}, Jill and {Genel}, Shy},
        title = "{First results from the IllustrisTNG simulations: radio haloes and magnetic fields}",
      journal = {\mnras},
         year = 2018,
        month = nov,
       volume = {480},
       number = {4},
        pages = {5113-5139},
          doi = {10.1093/mnras/sty2206},
archivePrefix = {arXiv},
       eprint = {1707.03396},
 primaryClass = {astro-ph.CO},
       adsurl = {https://ui.adsabs.harvard.edu/abs/2018MNRAS.480.5113M}
}

@ARTICLE{Merritt1983ApJ,
       author = {{Merritt}, D.},
        title = "{Relaxation and tidal stripping in rich clusters of galaxies. I. Evolution of the mass distribution.}",
      journal = {\apj},
         year = 1983,
        month = jan,
       volume = {264},
        pages = {24-48},
          doi = {10.1086/160571},
       adsurl = {https://ui.adsabs.harvard.edu/abs/1983ApJ...264...24M}
}

@ARTICLE{Misgeld2011MNRAS,
       author = {{Misgeld}, I. and {Hilker}, M.},
        title = "{Families of dynamically hot stellar systems over 10 orders of magnitude in mass}",
      journal = {\mnras},
         year = 2011,
        month = jul,
       volume = {414},
       number = {4},
        pages = {3699-3710},
          doi = {10.1111/j.1365-2966.2011.18669.x},
archivePrefix = {arXiv},
       eprint = {1103.1628},
 primaryClass = {astro-ph.CO},
       adsurl = {https://ui.adsabs.harvard.edu/abs/2011MNRAS.414.3699M}
}

@ARTICLE{Moore1996Natur,
       author = {{Moore}, Ben and {Katz}, Neal and {Lake}, George and {Dressler}, Alan and {Oemler}, Augustus},
        title = "{Galaxy harassment and the evolution of clusters of galaxies}",
      journal = {\nat},
         year = 1996,
        month = feb,
       volume = {379},
       number = {6566},
        pages = {613-616},
          doi = {10.1038/379613a0},
archivePrefix = {arXiv},
       eprint = {astro-ph/9510034},
 primaryClass = {astro-ph},
       adsurl = {https://ui.adsabs.harvard.edu/abs/1996Natur.379..613M}
}

@ARTICLE{Dolag2009MNRAS,
       author = {{Dolag}, K. and {Borgani}, S. and {Murante}, G. and {Springel}, V.},
        title = "{Substructures in hydrodynamical cluster simulations}",
      journal = {\mnras},
         year = 2009,
        month = oct,
       volume = {399},
       number = {2},
        pages = {497-514},
          doi = {10.1111/j.1365-2966.2009.15034.x},
archivePrefix = {arXiv},
       eprint = {0808.3401},
 primaryClass = {astro-ph},
       adsurl = {https://ui.adsabs.harvard.edu/abs/2009MNRAS.399..497D}
}

@ARTICLE{Naiman2018MNRAS,
       author = {{Naiman}, Jill P. and {Pillepich}, Annalisa and {Springel}, Volker and {Ramirez-Ruiz}, Enrico and {Torrey}, Paul and {Vogelsberger}, Mark and {Pakmor}, R{\"u}diger and {Nelson}, Dylan and {Marinacci}, Federico and {Hernquist}, Lars and {Weinberger}, Rainer and {Genel}, Shy},
        title = "{First results from the IllustrisTNG simulations: a tale of two elements - chemical evolution of magnesium and europium}",
      journal = {\mnras},
         year = 2018,
        month = jun,
       volume = {477},
       number = {1},
        pages = {1206-1224},
          doi = {10.1093/mnras/sty618},
archivePrefix = {arXiv},
       eprint = {1707.03401},
 primaryClass = {astro-ph.GA},
       adsurl = {https://ui.adsabs.harvard.edu/abs/2018MNRAS.477.1206N}
}

@ARTICLE{Nelson2018MNRAS,
       author = {{Nelson}, Dylan and {Pillepich}, Annalisa and {Springel}, Volker and {Weinberger}, Rainer and {Hernquist}, Lars and {Pakmor}, R{\"u}diger and {Genel}, Shy and {Torrey}, Paul and {Vogelsberger}, Mark and {Kauffmann}, Guinevere and {Marinacci}, Federico and {Naiman}, Jill},
        title = "{First results from the IllustrisTNG simulations: the galaxy colour bimodality}",
      journal = {\mnras},
         year = 2018,
        month = mar,
       volume = {475},
       number = {1},
        pages = {624-647},
          doi = {10.1093/mnras/stx3040},
archivePrefix = {arXiv},
       eprint = {1707.03395},
 primaryClass = {astro-ph.GA},
       adsurl = {https://ui.adsabs.harvard.edu/abs/2018MNRAS.475..624N}
}

@ARTICLE{Nelson2019MNRAS,
       author = {{Nelson}, Dylan and {Pillepich}, Annalisa and {Springel}, Volker and {Pakmor}, R{\"u}diger and {Weinberger}, Rainer and {Genel}, Shy and {Torrey}, Paul and {Vogelsberger}, Mark and {Marinacci}, Federico and {Hernquist}, Lars},
        title = "{First results from the TNG50 simulation: galactic outflows driven by supernovae and black hole feedback}",
      journal = {\mnras},
         year = 2019,
        month = dec,
       volume = {490},
       number = {3},
        pages = {3234-3261},
          doi = {10.1093/mnras/stz2306},
archivePrefix = {arXiv},
       eprint = {1902.05554},
 primaryClass = {astro-ph.GA},
       adsurl = {https://ui.adsabs.harvard.edu/abs/2019MNRAS.490.3234N}
}

@ARTICLE{Pfeffer2013MNRAS,
       author = {{Pfeffer}, J. and {Baumgardt}, H.},
        title = "{Ultra-compact dwarf galaxy formation by tidal stripping of nucleated dwarf galaxies}",
      journal = {\mnras},
         year = 2013,
        month = aug,
       volume = {433},
       number = {3},
        pages = {1997-2005},
          doi = {10.1093/mnras/stt867},
archivePrefix = {arXiv},
       eprint = {1305.3656},
 primaryClass = {astro-ph.GA},
       adsurl = {https://ui.adsabs.harvard.edu/abs/2013MNRAS.433.1997P}
}

@ARTICLE{Pillepich2018MNRAS,
       author = {{Pillepich}, Annalisa and {Springel}, Volker and {Nelson}, Dylan and {Genel}, Shy and {Naiman}, Jill and {Pakmor}, R{\"u}diger and {Hernquist}, Lars and {Torrey}, Paul and {Vogelsberger}, Mark and {Weinberger}, Rainer and {Marinacci}, Federico},
        title = "{Simulating galaxy formation with the IllustrisTNG model}",
      journal = {\mnras},
         year = 2018,
        month = jan,
       volume = {473},
       number = {3},
        pages = {4077-4106},
          doi = {10.1093/mnras/stx2656},
archivePrefix = {arXiv},
       eprint = {1703.02970},
 primaryClass = {astro-ph.GA},
       adsurl = {https://ui.adsabs.harvard.edu/abs/2018MNRAS.473.4077P}
}

@ARTICLE{Pillepich2018bMNRAS,
       author = {{Pillepich}, Annalisa and {Nelson}, Dylan and {Hernquist}, Lars and {Springel}, Volker and {Pakmor}, R{\"u}diger and {Torrey}, Paul and {Weinberger}, Rainer and {Genel}, Shy and {Naiman}, Jill P. and {Marinacci}, Federico and {Vogelsberger}, Mark},
        title = "{First results from the IllustrisTNG simulations: the stellar mass content of groups and clusters of galaxies}",
      journal = {\mnras},
         year = 2018,
        month = mar,
       volume = {475},
       number = {1},
        pages = {648-675},
          doi = {10.1093/mnras/stx3112},
archivePrefix = {arXiv},
       eprint = {1707.03406},
 primaryClass = {astro-ph.GA},
       adsurl = {https://ui.adsabs.harvard.edu/abs/2018MNRAS.475..648P}
}

@ARTICLE{Benavides2020MNRAS,
       author = {{Benavides}, Jos{\'e} A. and {Sales}, Laura V. and {Abadi}, Mario G.},
        title = "{Accretion of galaxy groups into galaxy clusters}",
      journal = {\mnras},
         year = 2020,
        month = nov,
       volume = {498},
       number = {3},
        pages = {3852-3862},
          doi = {10.1093/mnras/staa2636},
archivePrefix = {arXiv},
       eprint = {2005.05344},
 primaryClass = {astro-ph.GA},
       adsurl = {https://ui.adsabs.harvard.edu/abs/2020MNRAS.498.3852B}
}

@ARTICLE{Celiz2025AA,
       author = {{Celiz}, Bruno M. and {Navarro}, Julio F. and {Abadi}, Mario G. and {Springel}, Volker},
        title = "{Mass-morphology relation of TNG50 galaxies}",
      journal = {\aap},
         year = 2025,
        month = jul,
       volume = {699},
          eid = {A12},
        pages = {A12},
          doi = {10.1051/0004-6361/202554847},
archivePrefix = {arXiv},
       eprint = {2505.01620},
 primaryClass = {astro-ph.GA},
       adsurl = {https://ui.adsabs.harvard.edu/abs/2025A&A...699A..12C}
}

@ARTICLE{Donnari2021MNRAS,
       author = {{Donnari}, Martina and {Pillepich}, Annalisa and {Joshi}, Gandhali D. and {Nelson}, Dylan and {Genel}, Shy and {Marinacci}, Federico and {Rodriguez-Gomez}, Vicente and {Pakmor}, R{\"u}diger and {Torrey}, Paul and {Vogelsberger}, Mark and {Hernquist}, Lars},
        title = "{Quenched fractions in the IllustrisTNG simulations: the roles of AGN feedback, environment, and pre-processing}",
      journal = {\mnras},
         year = 2021,
        month = jan,
       volume = {500},
       number = {3},
        pages = {4004-4024},
          doi = {10.1093/mnras/staa3006},
archivePrefix = {arXiv},
       eprint = {2008.00005},
 primaryClass = {astro-ph.GA},
       adsurl = {https://ui.adsabs.harvard.edu/abs/2021MNRAS.500.4004D}
}

@ARTICLE{Pillepich2019MNRAS,
       author = {{Pillepich}, Annalisa and {Nelson}, Dylan and {Springel}, Volker and {Pakmor}, R{\"u}diger and {Torrey}, Paul and {Weinberger}, Rainer and {Vogelsberger}, Mark and {Marinacci}, Federico and {Genel}, Shy and {van der Wel}, Arjen and {Hernquist}, Lars},
        title = "{First results from the TNG50 simulation: the evolution of stellar and gaseous discs across cosmic time}",
      journal = {\mnras},
         year = 2019,
        month = dec,
       volume = {490},
       number = {3},
        pages = {3196-3233},
          doi = {10.1093/mnras/stz2338},
archivePrefix = {arXiv},
       eprint = {1902.05553},
 primaryClass = {astro-ph.GA},
       adsurl = {https://ui.adsabs.harvard.edu/abs/2019MNRAS.490.3196P}
}

@ARTICLE{Planck2016AA,
       author = {{Planck Collaboration} and {Ade}, P.~A.~R. and {Aghanim}, N. and {Arnaud}, M. and {Ashdown}, M. and {Aumont}, J. and {Baccigalupi}, C. and {Banday}, A.~J. and {Barreiro}, R.~B. and {Bartlett}, J.~G. and {Bartolo}, N. and {Battaner}, E. and {Battye}, R. and {Benabed}, K. and {Beno{\^\i}t}, A. and {Benoit-L{\'e}vy}, A. and {Bernard}, J. -P. and {Bersanelli}, M. and {Bielewicz}, P. and {Bock}, J.~J. and {Bonaldi}, A. and {Bonavera}, L. and {Bond}, J.~R. and {Borrill}, J. and {Bouchet}, F.~R. and {Boulanger}, F. and {Bucher}, M. and {Burigana}, C. and {Butler}, R.~C. and {Calabrese}, E. and {Cardoso}, J. -F. and {Catalano}, A. and {Challinor}, A. and {Chamballu}, A. and {Chary}, R. -R. and {Chiang}, H.~C. and {Chluba}, J. and {Christensen}, P.~R. and {Church}, S. and {Clements}, D.~L. and {Colombi}, S. and {Colombo}, L.~P.~L. and {Combet}, C. and {Coulais}, A. and {Crill}, B.~P. and {Curto}, A. and {Cuttaia}, F. and {Danese}, L. and {Davies}, R.~D. and {Davis}, R.~J. and {de Bernardis}, P. and {de Rosa}, A. and {de Zotti}, G. and {Delabrouille}, J. and {D{\'e}sert}, F. -X. and {Di Valentino}, E. and {Dickinson}, C. and {Diego}, J.~M. and {Dolag}, K. and {Dole}, H. and {Donzelli}, S. and {Dor{\'e}}, O. and {Douspis}, M. and {Ducout}, A. and {Dunkley}, J. and {Dupac}, X. and {Efstathiou}, G. and {Elsner}, F. and {En{\ss}lin}, T.~A. and {Eriksen}, H.~K. and {Farhang}, M. and {Fergusson}, J. and {Finelli}, F. and {Forni}, O. and {Frailis}, M. and {Fraisse}, A.~A. and {Franceschi}, E. and {Frejsel}, A. and {Galeotta}, S. and {Galli}, S. and {Ganga}, K. and {Gauthier}, C. and {Gerbino}, M. and {Ghosh}, T. and {Giard}, M. and {Giraud-H{\'e}raud}, Y. and {Giusarma}, E. and {Gjerl{\o}w}, E. and {Gonz{\'a}lez-Nuevo}, J. and {G{\'o}rski}, K.~M. and {Gratton}, S. and {Gregorio}, A. and {Gruppuso}, A. and {Gudmundsson}, J.~E. and {Hamann}, J. and {Hansen}, F.~K. and {Hanson}, D. and {Harrison}, D.~L. and {Helou}, G. and {Henrot-Versill{\'e}}, S. and {Hern{\'a}ndez-Monteagudo}, C. and {Herranz}, D. and {Hildebrandt}, S.~R. and {Hivon}, E. and {Hobson}, M. and {Holmes}, W.~A. and {Hornstrup}, A. and {Hovest}, W. and {Huang}, Z. and {Huffenberger}, K.~M. and {Hurier}, G. and {Jaffe}, A.~H. and {Jaffe}, T.~R. and {Jones}, W.~C. and {Juvela}, M. and {Keih{\"a}nen}, E. and {Keskitalo}, R. and {Kisner}, T.~S. and {Kneissl}, R. and {Knoche}, J. and {Knox}, L. and {Kunz}, M. and {Kurki-Suonio}, H. and {Lagache}, G. and {L{\"a}hteenm{\"a}ki}, A. and {Lamarre}, J. -M. and {Lasenby}, A. and {Lattanzi}, M. and {Lawrence}, C.~R. and {Leahy}, J.~P. and {Leonardi}, R. and {Lesgourgues}, J. and {Levrier}, F. and {Lewis}, A. and {Liguori}, M. and {Lilje}, P.~B. and {Linden-V{\o}rnle}, M. and {L{\'o}pez-Caniego}, M. and {Lubin}, P.~M. and {Mac{\'\i}as-P{\'e}rez}, J.~F. and {Maggio}, G. and {Maino}, D. and {Mandolesi}, N. and {Mangilli}, A. and {Marchini}, A. and {Maris}, M. and {Martin}, P.~G. and {Martinelli}, M. and {Mart{\'\i}nez-Gonz{\'a}lez}, E. and {Masi}, S. and {Matarrese}, S. and {McGehee}, P. and {Meinhold}, P.~R. and {Melchiorri}, A. and {Melin}, J. -B. and {Mendes}, L. and {Mennella}, A. and {Migliaccio}, M. and {Millea}, M. and {Mitra}, S. and {Miville-Desch{\^e}nes}, M. -A. and {Moneti}, A. and {Montier}, L. and {Morgante}, G. and {Mortlock}, D. and {Moss}, A. and {Munshi}, D. and {Murphy}, J.~A. and {Naselsky}, P. and {Nati}, F. and {Natoli}, P. and {Netterfield}, C.~B. and {N{\o}rgaard-Nielsen}, H.~U. and {Noviello}, F. and {Novikov}, D. and {Novikov}, I. and {Oxborrow}, C.~A. and {Paci}, F. and {Pagano}, L. and {Pajot}, F. and {Paladini}, R. and {Paoletti}, D. and {Partridge}, B. and {Pasian}, F. and {Patanchon}, G. and {Pearson}, T.~J. and {Perdereau}, O. and {Perotto}, L. and {Perrotta}, F. and {Pettorino}, V. and {Piacentini}, F. and {Piat}, M. and {Pierpaoli}, E. and {Pietrobon}, D. and {Plaszczynski}, S. and {Pointecouteau}, E. and {Polenta}, G. and {Popa}, L. and {Pratt}, G.~W. and {Pr{\'e}zeau}, G. and {Prunet}, S. and {Puget}, J. -L. and {Rachen}, J.~P. and {Reach}, W.~T. and {Rebolo}, R. and {Reinecke}, M. and {Remazeilles}, M. and {Renault}, C. and {Renzi}, A. and {Ristorcelli}, I. and {Rocha}, G. and {Rosset}, C. and {Rossetti}, M. and {Roudier}, G. and {Rouill{\'e} d'Orfeuil}, B. and {Rowan-Robinson}, M. and {Rubi{\~n}o-Mart{\'\i}n}, J.~A. and {Rusholme}, B. and {Said}, N. and {Salvatelli}, V. and {Salvati}, L. and {Sandri}, M. and {Santos}, D. and {Savelainen}, M. and {Savini}, G. and {Scott}, D. and {Seiffert}, M.~D. and {Serra}, P. and {Shellard}, E.~P.~S. and {Spencer}, L.~D. and {Spinelli}, M. and {Stolyarov}, V. and {Stompor}, R. and {Sudiwala}, R. and {Sunyaev}, R. and {Sutton}, D. and {Suur-Uski}, A. -S. and {Sygnet}, J. -F. and {Tauber}, J.~A. and {Terenzi}, L. and {Toffolatti}, L. and {Tomasi}, M. and {Tristram}, M. and {Trombetti}, T. and {Tucci}, M. and {Tuovinen}, J. and {T{\"u}rler}, M. and {Umana}, G. and {Valenziano}, L. and {Valiviita}, J. and {Van Tent}, F. and {Vielva}, P. and {Villa}, F. and {Wade}, L.~A. and {Wandelt}, B.~D. and {Wehus}, I.~K. and {White}, M. and {White}, S.~D.~M. and {Wilkinson}, A. and {Yvon}, D. and {Zacchei}, A. and {Zonca}, A.},
        title = "{Planck 2015 results. XIII. Cosmological parameters}",
      journal = {\aap},
         year = 2016,
        month = sep,
       volume = {594},
          eid = {A13},
        pages = {A13},
          doi = {10.1051/0004-6361/201525830},
archivePrefix = {arXiv},
       eprint = {1502.01589},
 primaryClass = {astro-ph.CO},
       adsurl = {https://ui.adsabs.harvard.edu/abs/2016A&A...594A..13P}
}

@ARTICLE{Richstone1976ApJ,
       author = {{Richstone}, D.~O.},
        title = "{Collisions of galaxies in dense clusters. II. Dynamical evolution of cluster galaxies.}",
      journal = {\apj},
         year = 1976,
        month = mar,
       volume = {204},
        pages = {642-648},
          doi = {10.1086/154213},
       adsurl = {https://ui.adsabs.harvard.edu/abs/1976ApJ...204..642R}
}

@ARTICLE{Rodriguez2017MNRAS,
       author = {{Rodriguez-Gomez}, Vicente and {Sales}, Laura V. and {Genel}, Shy and {Pillepich}, Annalisa and {Zjupa}, Jolanta and {Nelson}, Dylan and {Griffen}, Brendan and {Torrey}, Paul and {Snyder}, Gregory F. and {Vogelsberger}, Mark and {Springel}, Volker and {Ma}, Chung-Pei and {Hernquist}, Lars},
        title = "{The role of mergers and halo spin in shaping galaxy morphology}",
      journal = {\mnras},
         year = 2017,
        month = may,
       volume = {467},
       number = {3},
        pages = {3083-3098},
          doi = {10.1093/mnras/stx305},
archivePrefix = {arXiv},
       eprint = {1609.09498},
 primaryClass = {astro-ph.GA},
       adsurl = {https://ui.adsabs.harvard.edu/abs/2017MNRAS.467.3083R}
}

@ARTICLE{RodriguezGomez2016MNRAS,
       author = {{Rodriguez-Gomez}, Vicente and {Pillepich}, Annalisa and {Sales}, Laura V. and {Genel}, Shy and {Vogelsberger}, Mark and {Zhu}, Qirong and {Wellons}, Sarah and {Nelson}, Dylan and {Torrey}, Paul and {Springel}, Volker and {Ma}, Chung-Pei and {Hernquist}, Lars},
        title = "{The stellar mass assembly of galaxies in the Illustris simulation: growth by mergers and the spatial distribution of accreted stars}",
      journal = {\mnras},
         year = 2016,
        month = may,
       volume = {458},
       number = {3},
        pages = {2371-2390},
          doi = {10.1093/mnras/stw456},
archivePrefix = {arXiv},
       eprint = {1511.08804},
 primaryClass = {astro-ph.GA},
       adsurl = {https://ui.adsabs.harvard.edu/abs/2016MNRAS.458.2371R}
}

@ARTICLE{Sales2007,
       author = {{Sales}, Laura V. and {Navarro}, Julio F. and {Abadi}, Mario G. and {Steinmetz}, Matthias},
        title = "{Cosmic m{\'e}nage {\`a} trois: the origin of satellite galaxies on extreme orbits}",
      journal = {\mnras},
         year = 2007,
        month = aug,
       volume = {379},
       number = {4},
        pages = {1475-1483},
          doi = {10.1111/j.1365-2966.2007.12026.x},
archivePrefix = {arXiv},
       eprint = {0704.1773},
 primaryClass = {astro-ph},
       adsurl = {https://ui.adsabs.harvard.edu/abs/2007MNRAS.379.1475S}
}

@ARTICLE{Springel10,
   author = {{Springel}, Volker},
        title = {E pur si muove: Galilean-invariant cosmological hydrodynamical simulations on a moving mesh},
      journal = {\mnras},
         year = 2010,
        month = jan,
       volume = {401},
       number = {2},
        pages = {791-851},
          doi = {10.1111/j.1365-2966.2009.15715.x},
archivePrefix = {arXiv},
       eprint = {0901.4107},
 primaryClass = {astro-ph.CO},
       adsurl = {https://ui.adsabs.harvard.edu/abs/2010MNRAS.401..791S}
}

@ARTICLE{Springel2001MNRAS,
       author = {{Springel}, Volker and {White}, Simon D.~M. and {Tormen}, Giuseppe and {Kauffmann}, Guinevere},
        title = "{Populating a cluster of galaxies - I. Results at [formmu2]z=0}",
      journal = {\mnras},
         year = 2001,
        month = dec,
       volume = {328},
       number = {3},
        pages = {726-750},
          doi = {10.1046/j.1365-8711.2001.04912.x},
archivePrefix = {arXiv},
       eprint = {astro-ph/0012055},
 primaryClass = {astro-ph},
       adsurl = {https://ui.adsabs.harvard.edu/abs/2001MNRAS.328..726S}
}

@ARTICLE{Springel2018MNRAS,
       author = {{Springel}, Volker and {Pakmor}, R{\"u}diger and {Pillepich}, Annalisa and {Weinberger}, Rainer and {Nelson}, Dylan and {Hernquist}, Lars and {Vogelsberger}, Mark and {Genel}, Shy and {Torrey}, Paul and {Marinacci}, Federico and {Naiman}, Jill},
        title = "{First results from the IllustrisTNG simulations: matter and galaxy clustering}",
      journal = {\mnras},
         year = 2018,
        month = mar,
       volume = {475},
       number = {1},
        pages = {676-698},
          doi = {10.1093/mnras/stx3304},
archivePrefix = {arXiv},
       eprint = {1707.03397},
 primaryClass = {astro-ph.GA},
       adsurl = {https://ui.adsabs.harvard.edu/abs/2018MNRAS.475..676S}
}

@ARTICLE{Tahmasebzadeh+25,
   author = {{Tahmasebzadeh}, Behzad and {Taylor}, Matthew A. and {Valluri}, Monica and {Yoshino}, Haruka and {Vasiliev}, Eugene and {Drinkwater}, Michael J. and {Thompson}, Solveig and {Dage}, Kristen and {C{\^o}t{\'e}}, Patrick and {Ferrarese}, Laura and {Akiba}, Tatsuya and {Baldassare}, Vivienne and {Bentz}, Misty C. and {Blakeslee}, John P. and {Baumgardt}, Holger and {Ko}, Youkyung and {Liu}, Chengze and {Madigan}, Ann-Marie and {Peng}, Eric W. and {Roediger}, Joel and {Wang}, Kaixiang and {Woods}, Tyrone E.},
        title = {A JWST View of the Overmassive Black Hole in NGC 4486B},
      journal = {\apjl},
         year = 2025,
        month = aug,
       volume = {989},
       number = {2},
          eid = {L42},
        pages = {L42},
          doi = {10.3847/2041-8213/adf728},
archivePrefix = {arXiv},
       eprint = {2505.14676},
 primaryClass = {astro-ph.GA},
       adsurl = {https://ui.adsabs.harvard.edu/abs/2025ApJ...989L..42T}
}

@ARTICLE{vanderMarel+97,
   author = {{van der Marel}, Roeland P. and {de Zeeuw}, P. Tim and {Rix}, Hans-Walter and {Quinlan}, Gerald D.},
        title = {A massive black hole at the centre of the quiescent galaxy M32},
      journal = {\nat},
         year = 1997,
        month = feb,
       volume = {385},
       number = {6617},
        pages = {610-612},
          doi = {10.1038/385610a0},
archivePrefix = {arXiv},
       eprint = {astro-ph/9702106},
 primaryClass = {astro-ph},
       adsurl = {https://ui.adsabs.harvard.edu/abs/1997Natur.385..610V}
}

@ARTICLE{Wang2023Nature,
       author = {{Wang}, Kaixiang and {Peng}, Eric W. and {Liu}, Chengze and {Mihos}, J. Christopher and {C{\^o}t{\'e}}, Patrick and {Ferrarese}, Laura and {Taylor}, Matthew A. and {Blakeslee}, John P. and {Cuillandre}, Jean-Charles and {Duc}, Pierre-Alain and {Guhathakurta}, Puragra and {Gwyn}, Stephen and {Ko}, Youkyung and {Lan{\c{c}}on}, Ariane and {Lim}, Sungsoon and {MacArthur}, Lauren A. and {Puzia}, Thomas and {Roediger}, Joel and {Sales}, Laura V. and {S{\'a}nchez-Janssen}, Rub{\'e}n and {Spengler}, Chelsea and {Toloba}, Elisa and {Zhang}, Hongxin and {Zhu}, Mingcheng},
        title = "{An evolutionary continuum from nucleated dwarf galaxies to star clusters}",
      journal = {\nat},
         year = 2023,
        month = nov,
       volume = {623},
       number = {7986},
        pages = {296-300},
          doi = {10.1038/s41586-023-06650-z},
archivePrefix = {arXiv},
       eprint = {2311.05448},
 primaryClass = {astro-ph.GA},
       adsurl = {https://ui.adsabs.harvard.edu/abs/2023Natur.623..296W}
}

@ARTICLE{deLucia2007MNRAS,
       author = {{De Lucia}, Gabriella and {Blaizot}, J{\'e}r{\'e}my},
        title = "{The hierarchical formation of the brightest cluster galaxies}",
      journal = {\mnras},
         year = 2007,
        month = feb,
       volume = {375},
       number = {1},
        pages = {2-14},
          doi = {10.1111/j.1365-2966.2006.11287.x},
archivePrefix = {arXiv},
       eprint = {astro-ph/0606519},
 primaryClass = {astro-ph},
       adsurl = {https://ui.adsabs.harvard.edu/abs/2007MNRAS.375....2D}
}



\appendix

\section{Bad-flag and young subhalos}
\label{sec:samplebad}

\begin{figure}
    \centering
    \includegraphics[width=\columnwidth]{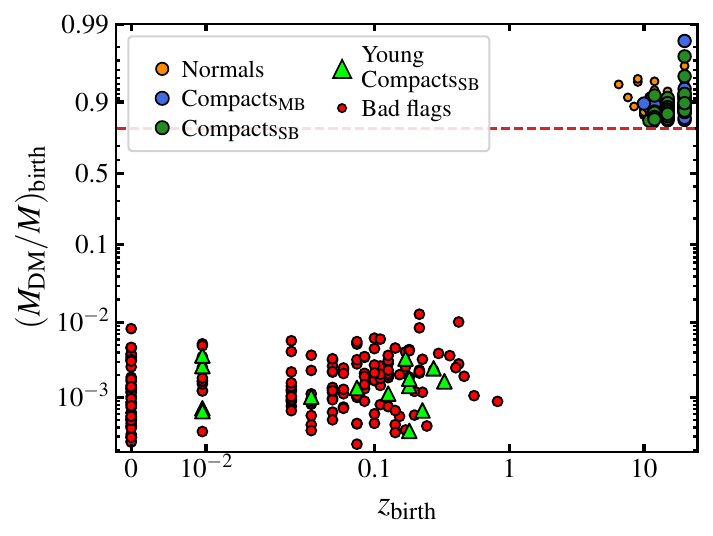}
    \caption{Dark matter fraction at birth versus birth redshift, for satellite dwarfs of different classes. The birth epochs are those obtained with the \sublink main progenitor merger tree. The dashed red line indicates the dark matter fraction at birth threshold (0.8) used to define bad-flag galaxies.
   }
  \label{fig:histBorn}
\end{figure}

In this appendix, we analyse the bad-flag and young galaxies. We restrict ourselves to the global subhalo properties and merger histories provided by the main progenitor branch of the \sublink merger tree of the TNG database. In TNG, a subhalo is classified as non-cosmological if it is a satellite at its time of formation, it forms within the virial radius of its parent halo, and its dark matter fraction at birth is below 0.8. Young galaxies represent one quarter of \CompactsSB{} satellites, but are negligible for all other dwarf populations. Figure~\ref{fig:histBorn} shows a striking bimodality between bad-flag and good-flag young galaxies on one hand and the good-flag old galaxies on the other. First, the bad-flag and young \CompactSB{} galaxies were recently formed, all at $z<1$, with median birth epochs of $z=0$ and 0.1 for the bad-flag and young galaxies, respectively. Second, both populations were formed with very low DM fractions: all less than 2 per cent, which is much lower than the 0.80 maximum DM fraction required to be classified as bad-flag galaxies. In contrast, all the old good-flag galaxies had much higher DM fractions at birth, with all having DM fractions at birth above 0.80, which was not obvious from the bad-flag selection criteria.  This suggests that the nature of the bad-flag and young galaxy populations differs from that of the old good-flag one, unless there are issues with the \sublink algorithm. Interestingly, the young and bad-flag galaxies are similar: both classes follow a similar distribution in (DM fraction, birth epoch) space. Additionally, they both have higher gas fractions at birth, 0.21 and 0.36 for young and bad-flag galaxies, respectively, compared to 0.13 for all three size classes of old galaxies.

\begin{figure}
\includegraphics[width=\columnwidth]{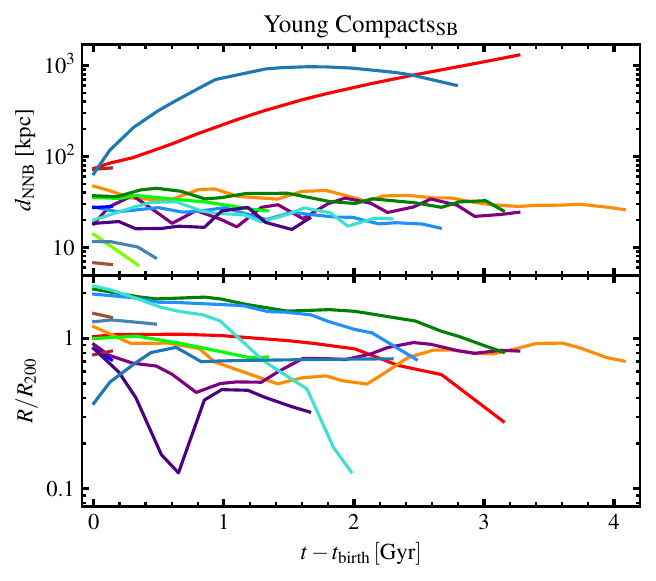}
\caption{Evolution of the distance to the host galaxy at birth (\emph{top}) and to the host group (\emph{bottom}) in virial units for individual galaxies. The origin of the time axis is the epoch of birth.}
 \label{fig:BornEnvironment}
\end{figure}

Finally, in terms of local environment at birth, by construction, bad-flag galaxies are formed as satellites, while  98 per cent of the good-flag satellite \CompactsSB{} are  centrals at birth. In contrast, all young \CompactsSB{} are satellites at birth. Figure \ref{fig:BornEnvironment}  shows the evolution of the distances to the nearest more massive neighbour at birth ($d_{\rm NNB}$) and to the group centre for young \CompactsSB{}. Most of these galaxies are located near their nearest neighbour at birth, which  suggests that these galaxies are H\,{\sc ii} regions or star clusters incorrectly extracted by \subfind. However, two young \CompactsSB{} ($z$=0 IDs {\tt 327} and {\tt 923816}) are moving far away from the nearest neighbour at birth, indicating they may be galaxies born through fragmentation, similar to tidal dwarf galaxies (TDGs). However, while at birth, the mass of galaxy {\tt 327} is  $20$ per cent gas, $80$ per cent  stars and $0.1$ per cent  DM, at $z=0$ its composition changed to  $99.6$ per cent stars and $0.4$ per cent DM. Galaxy {\tt 923816} was born with $100$ per cent of stellar mass and maintained this composition at $z=0$, resembling more a star cluster than a TDG. 

\section{Robustness to the adopted threshold for the dark matter fraction}
\label{sec:fdmthreshold}

In this Appendix, we test how sensitive our conclusions are to the present-day DM fraction threshold of 0.7 used to separate DM-rich from DM-poor satellites. The value of this threshold was not set by physical considerations; it is merely a  practical working definition designed to isolate satellites that experienced substantial present-day DM depletion while still preserving sufficient sample sizes for a meaningful population-level comparison.

\begin{figure}
    \centering
    \includegraphics[width=\linewidth]{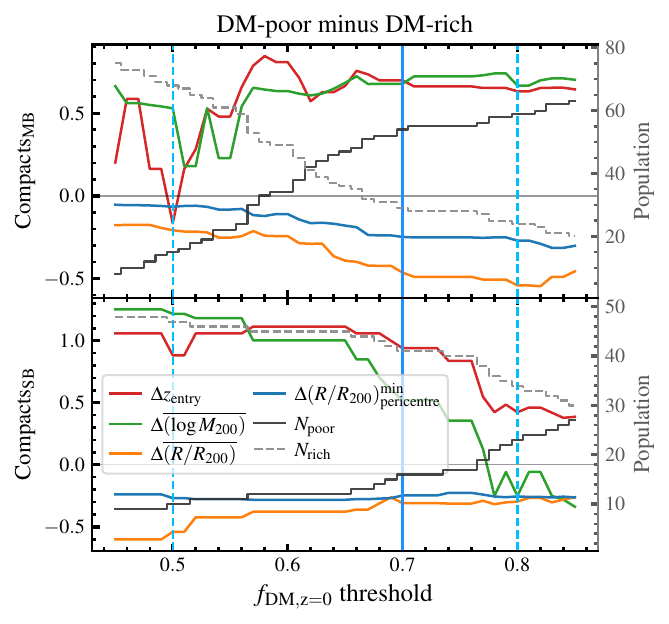}
    \caption{Robustness of the environmental signal to the adopted present-day DM fraction threshold for \CompactsMB{} (\emph{top}) and \CompactsSB{} (\emph{bottom}) satellites. In each panel, coloured curves show the difference between the medians of the DM-poor and DM-rich populations for the main environmental diagnostics discussed in the paper, while the grey curves show the corresponding numbers of DM-poor and DM-rich objects. The solid vertical line shows our fiducial threshold, $f_{\rm DM,z=0}=0.7$, and the dashed vertical lines indicate the alternative reference values $0.5$ and $0.8$.}
    \label{fig:fdm_threshold_scan}
\end{figure}

To address this explicitly, Figure~\ref{fig:fdm_threshold_scan} shows how the main environmental contrasts between DM-poor and DM-rich satellites vary when the threshold is changed continuously over the range $0.45 \leq f_{\rm DM,z=0} \leq 0.85$. For each galaxy class, we plot the difference between the DM-poor and DM-rich populations for the median entry redshift into the first host, the median time-averaged host mass, the median time-averaged orbital radius in units of the host virial radius, and the minimum orbital radius. The same figure also shows the corresponding evolution of $N_{\rm DM-poor}$ and $N_{\rm DM-rich}$, which makes it possible to assess whether apparent changes in the signal are driven by genuinely different population splits or simply by the rapid collapse of the sample size.

\CompactsMB{} present environmental parameters remarkably stable across a broad range of values, with DM-poor systems consistently entering earlier in more massive hosts, and orbiting at smaller radii than DM-rich systems. In other words, the qualitative interpretation is preserved whether the threshold is moved below or above the fiducial choice. What changes is mostly the balance between sufficient sample size and purity. At low thresholds, the DM-poor subset becomes too restrictive, while at higher thresholds the split becomes increasingly permissive.

\CompactsSB{} in general show the same qualitative picture. However, the environmental contrast is strongest at intermediate thresholds and becomes progressively weaker as the cut approaches $f_{\rm DM,z=0}\sim0.8$, especially for the host-mass difference. This behaviour is important since it indicates that although a threshold near $0.8$ still identifies a population that is, on average, more environmentally processed, it also begins to include systems with milder present-day DM depletion, thereby diluting the physical contrast that is more evident at lower values.

Taken together, the two panels of Figure~\ref{fig:fdm_threshold_scan} shows that $f_{\rm DM,z=0}=0.7$ is not a unique or fundamental value, but it is a sensible fiducial choice. A lower value such as $0.5$ is too restrictive and quickly drives some DM-poor subsets to very small numbers, whereas a higher value such as $0.8$ is more inclusive but begins to wash out part of the environmental contrast. We therefore retain $f_{\rm DM,z=0}=0.7$ because it provides a conservative compromise between physical selectiveness and continuity of the inferred trends.

\section{Robustness of the inner sSFR trends to the aperture definition}
\label{sec:sizeTest}

In Sect.~\ref{sec:MassSize}, we showed that DM-rich Compact satellites exhibit enhanced inner sSFR relative to DM-rich Normals when the inner region is defined through the running half-stellar-mass radius. Because this aperture evolves with time, one may worry that part of the trend could be driven by the shrinking measurement radius rather than by a genuine concentration of star formation. To test this explicitly, we repeated the analysis using fixed physical apertures.

\begin{figure}
    \centering
    \includegraphics[width=\columnwidth]{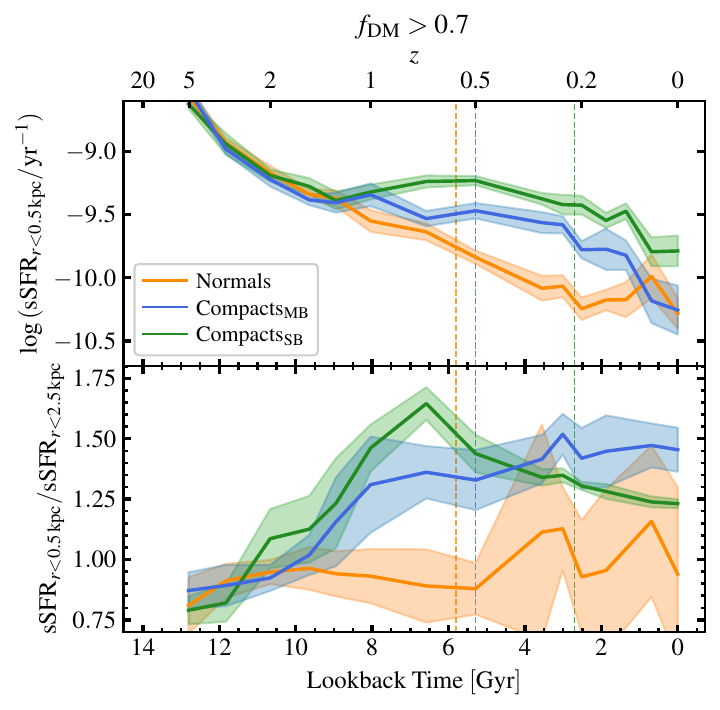}
    \caption{Same as Fig.~\ref{fig:sizeEvol}, but for the sSFR measured within $0.5$ kpc (\emph{top}) and the ratio between the sSFR measured within different fixed physical apertures (\emph{bottom}) for DM-rich satellites. A higher ratio indicates that the ongoing star formation is more centrally concentrated.}
    \label{fig:fixed_aperture_ssfr_ratio}
\end{figure}

Figure~\ref{fig:fixed_aperture_ssfr_ratio} shows the evolution of sSFR measured within $r<0.5\,\mathrm{kpc}$ and the ratio between this and the sSFR measured within $r<2.5\,\mathrm{kpc}$ for DM-rich satellites. We  find that both \Compact{} populations maintain systematically higher inner sSFR than the Normals since $z \sim 1$, which is similar trend as observed in Fig.~\ref{fig:sizeEvol}. Fig.~\ref{fig:fixed_aperture_ssfr_ratio} also shows that \Compacts{} have nearly $50$ per cent higher sSFR within $r<0.5\,\mathrm{kpc}$ than that measured within $r<2.5\,\mathrm{kpc}$. In contrast, for Normals the ratio between the different sSFRs remain closer to unity. Therefore, the main trend discussed in Sect.~\ref{sec:MassSize} is not primarily caused by the use of a time-dependent aperture.

\section{Estimating the weights of the dark matter tracer particles}
\label{sec:appDMtracer}

We discuss here how to build the DM tracer particle weights (or masses). The simple way to assign weights is to correct the DM normalised mass profile, i.e. its cumulative distribution function (CDF), by the ratio of the CDF of the stars (weighted by the stellar mass particles, which are not equal) to that of the DM (where no weighting is required since DM particles all have the same mass). 
This can be written 
\begin{flalign}
    m_\mathrm{DM-tracer}(r_\mathrm{DM}) & \propto 
     \frac{\widehat{\mathrm{CDF}_\star}(r_\mathrm{DM})} {\mathrm{CDF}_\mathrm{DM}(r_\mathrm{DM})}\,\widehat{\mathrm{CDF}_\star}(r_\mathrm{DM})  =
    \frac{\widehat{\mathrm{CDF}_\star}^2(r_\mathrm{DM})} {\mathrm{CDF}_\mathrm{DM}(r_\mathrm{DM})} \ ,
    \label{mDMtr_simple}
\end{flalign}
where $r_\mathrm{DM}$ is the DM radial distance, while $\widehat{\mathrm{CDF}_\star}$ is a numerical interpolation of the CDF of stellar radial distances.
The method of Eq.~(\ref{mDMtr_simple}) works well when the two CDFs are similar. However DM profiles are more extended, leading to half-mass radii of the DM tracers that are typically three times greater than that of the stars.

\begin{figure}
    \centering
    \includegraphics[width=\linewidth]{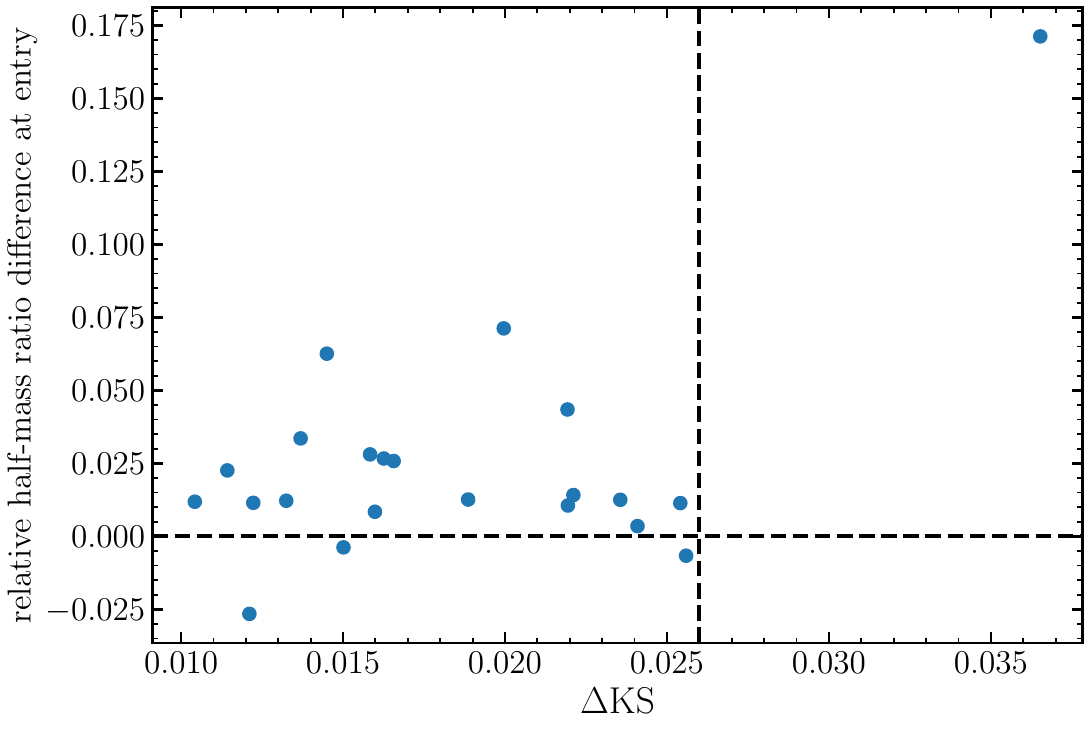}
    \caption{Ability to build the dark matter tracer particles at the epoch of entry.  The relative bias on the half-mass radius is plotted against the absolute value of the maximum difference in the cumulative distribution functions.}
    \label{fig:rhalfratiovsDeltaCDF}
\end{figure}
Our solution is to start with Eq.~(\ref{mDMtr_simple}) and iterate, using again Eq.~(\ref{mDMtr_simple}), but where $\mathrm{CDF}_\star$ is estimated by replacing $r_\star$ by $(r_{1/2,\star}/r_\mathrm{1/2,DM-tracer})\,r_\star/f$, where $f$ is slightly greater than unity to allow faster convergence. In practice, we find that 20 iterations with $f=1.1$ allow us to build the tracer mass profile with a CDF that matches the CDF of the stellar component to better than 0.02 at all radii, for about 80 per cent of the DM-poor \CompactsMB{} that enter their final host as centrals and eventually lose their gas.

The precision is shown in Fig.~\ref{fig:rhalfratiovsDeltaCDF}. We chose to cut our sample for DM tracers for $\mathrm{max}(|\Delta \mathrm{CDF}|) < 0.026$ (this removed only one of the DM-poor \CompactsMB{} that entered their final hosts as centrals and later lost their gas). This subsample has a median $\mathrm{max}(|\Delta \mathrm{CDF}|)$ of 0.016, and a relative bias on the half-mass radius of 1.8$\pm$2.2 per cent. 

\section{Black hole occupation fractions at redshift zero}
\label{sec:BH}

Table~\ref{tab:BHfrac} displays the black hole occupation fractions for the different size classes and environmental classes. One notices that the BH occupation fraction decreases with size for given environment, and decreases from centrals to DM-rich satellites to DM-poor satellites for Compacts.
\begin{table}
\caption{Black hole occupation fractions at $z=0$}
    \centering
\begin{tabular}{lccc}
\hline
\hline
Size class & Centrals & \multicolumn{2}{c}{Satellites} \\
\cline{3-4}
 & & DM-rich & DM-poor \\
 \hline
 Normals & 0.76$\pm$0.02 & 0.51$\pm$0.03 & 0.50$\pm$0.35 \\
 \CompactsMB{} & 0.59$\pm$0.06 & 0.31$\pm$0.08 & 0.16$\pm$0.05 \\
 \CompactsSB{} & 0.22$\pm$0.03 & 0.11$\pm$0.04 & $<0.03$ \\
 \hline
 \end{tabular}
\begin{minipage}{\linewidth}
\footnotesize
\raggedright
    Notes: Only good-flag galaxies are considered. The error bars are binomial and the upper limit is at Wilson 95 per cent confidence.
    \end{minipage}
 \label{tab:BHfrac}
\end{table}

\section{Metal-rich DM-poor Compacts SB}
\label{sec:AppFigs}

\begin{figure}
    \centering
    \includegraphics[width=1.0\linewidth]{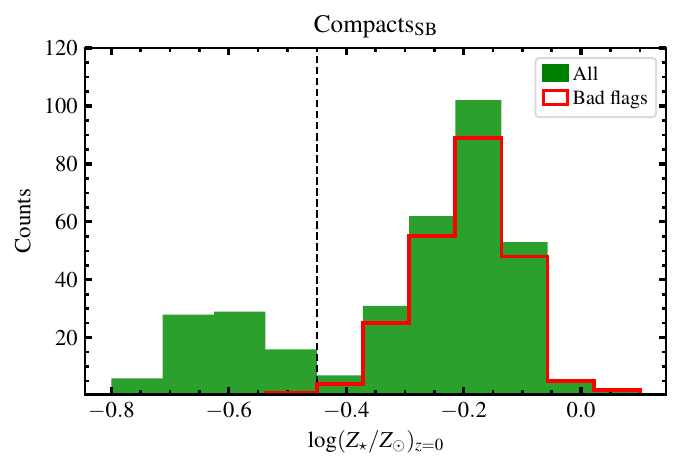}
    \caption{Distribution of $z$=0 stellar metallicities for all (shaded green) and bad-flag (red) satellite \CompactsSB. The vertical line indicates the metallicity threshold $\log(Z_\star/Z_\odot) > -0.45$.
    }
    \label{fig:histZ}
\end{figure}

In this appendix, we analyse the most metal-rich \CompactsSB{} to compare with the results of \cite{Bian2025ApJ}.  We start by studying the link between stellar metallicity and the {\tt SubhaloFlag}. Figure~\ref{fig:histZ} shows the distribution of stellar metallicities for all and bad-flag satellite \CompactsSB. One first notices that the distribution of metallicities is strongly bimodal. Furthermore, there is a striking, nearly one-to-one, match between bad-flag galaxies and the high metallicity mode. Only 1 of the 230 bad-flag galaxies is not in the high-metallicity regime ($\log(Z_\star/Z_\odot)>-0.45$).

\begin{figure*}
    \centering
\includegraphics[width=\hsize]{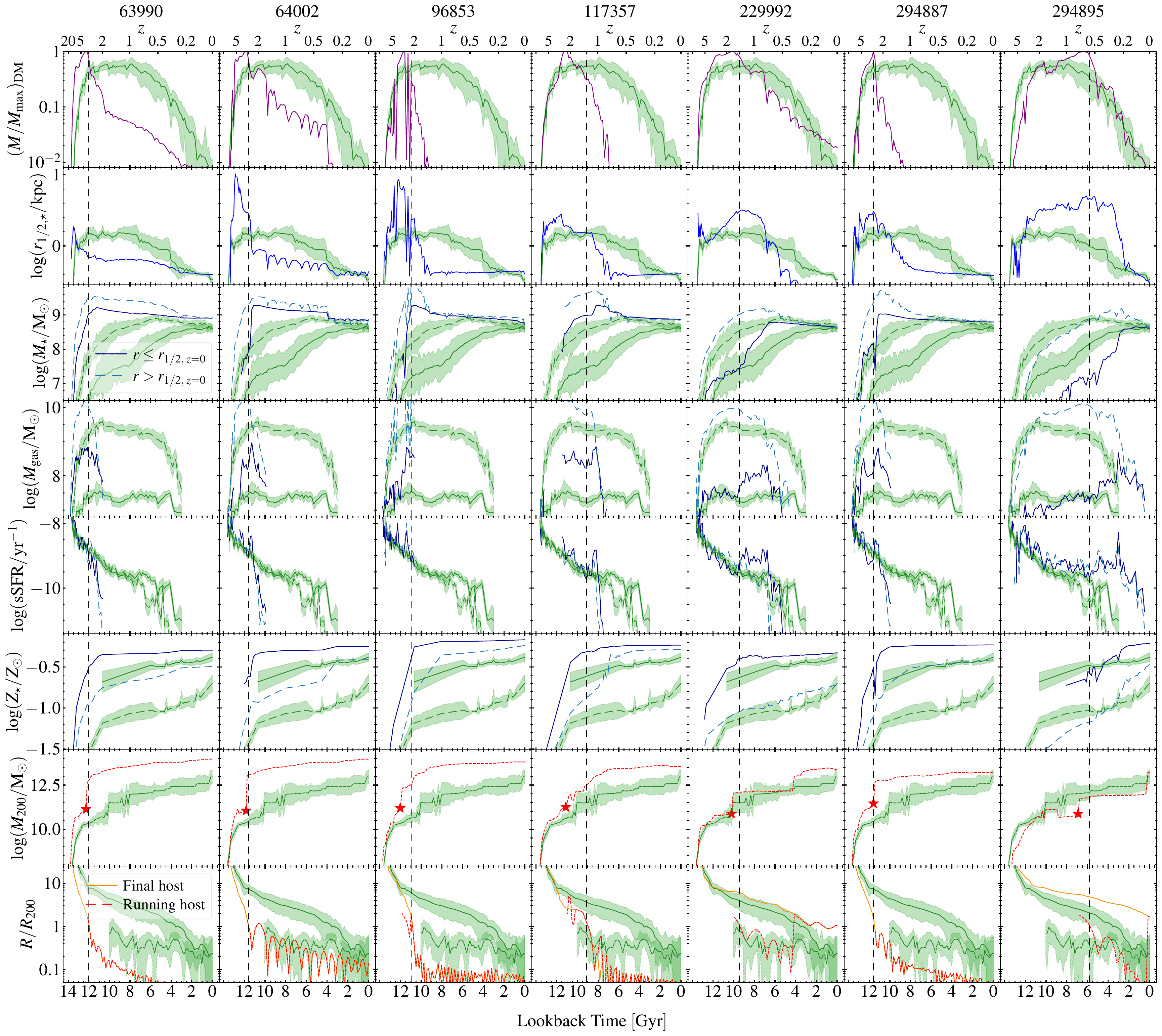}
\caption{Evolution of the seven galaxies that end up as metal-rich \CompactsSB{} satellites compared with the median evolutionary histories of their nine less-metal-rich DM-poor counterparts (green). The dashed vertical line shows the epoch of entry into the running host. The star marker indicates when the galaxy become a satellite. The solid and dashed lines in rows 3 to 6 respectively display the evolution for the inner and outer regions (delimited by their $z$=0 half-stellar--mass radius), while those in the last row respectively display the orbit relative to the final and first host. The galaxies are ordered in decreasing final host group mass.
    }
    \label{fig:EvolwZstarfor7SBs}
\end{figure*}

We next study the evolutionary history of individual \CompactsSB{} that end up very metal poor. Figure~\ref{fig:EvolwZstarfor7SBs} shows the evolution of the seven most metal-rich, long-lived, good-flag, non-young  DM-poor \CompactsSB{} as a reference. The figure also displays the median evolution of the lower metallicity DM-poor \CompactsSB{}.  The two metallicity classes display markedly different histories. The orbital histories (bottom row) show that two of the seven were pre-processed in previous hosts, while the others had early entries and a higher number of orbits in their host. More importantly, the metal-rich DM-poor \CompactsSB{} penetrate much deeper into more massive hosts and typically enter earlier than their metal-poor counterparts. Furthermore, the metal-rich ones suffer earlier compaction (second row) and gas loss (fourth row). Their sSFRs are similar to their metal-poor counterparts, except for occasional, usually short-lived bursts.

In more detail,  galaxy \texttt{63990}\footnote{The galaxy IDs refer to the $z$=0 snapshot.} suffers its strongest and rapid compaction episode well before entry, at $z\approx5$ , synchronous with a short starburst. This may well be a case of ram pressure compaction. Galaxy \texttt{64002} (in the same group)  has rapid compaction between entry and first pericentre, at which point the sSFR is maximum. Since its outer stellar mass is not decreasing, this may also be a case of compaction by ram pressure, here at first pericentre. Galaxy \texttt{96853} falls deeply into its host, causing full gas loss by ram pressure stripping right after first pericentre. Its half-stellar--mass radius fluctuates wildly during a few snapshots before entry (subhalo switching). Its compaction occurs over several short orbits after entry into its final host, without slowing down after gas loss. It had a short burst of star formation a few snapshots before entry into its running host. This does not appear to be a case of ram pressure compaction. The decrease of its outer stellar mass from the onset of compaction suggests that tidal stripping is the main driver of its compaction. Galaxy \texttt{117357} also falls deeply into its host and its gas is lost by 2 Gyr after entry. The phase of strongest compaction is gradual between first pericentre and gas loss. There is a starburst halfway through the compaction. The outer stellar mass is slowly increasing in the first phase and decreasing in the second. This appears to be a mixed case of ram pressure compaction that continues as tidal stripping. Galaxy \texttt{229992} enters its host on a shallow orbit (that deepens later). Its compaction is strongest after second pericentre, connected to an unusually long starburst between entry and soon after the second pericentre, combined with roughly constant outer stellar mass until second pericentre and decreasing thereafter. This compaction also appears to be first driven by ram pressure and then by tidal stripping. Galaxy \texttt{294887} has gradual compaction from entry to after gas loss. It goes through a moderately long starburst at first pericentre. This compaction may be first driven by ram pressure and later by tidal stripping. The outer stellar mass grows slowly at first, then substantially decreases. This compaction again appears to be first driven by ram pressure and then by tidal stripping. Galaxy \texttt{294895} also has gradual compaction between first and third pericentre. It has enhanced star formation from first to second pericentre, followed by a strong short starburst at second pericentre. Since its outer stellar mass is not decreasing during the fastest compaction, the compaction is probably driven by ram pressure.

This indicates that, among the seven metal-rich DM-poor \CompactsSB{}, three appear to be driven by ram pressure compression, one by tidal stripping, and the remaining three by ram pressure followed by tidal stripping. Nevertheless, the tidal field is present during the fastest compaction episodes of all six of the seven metal-rich galaxies (not the first one), as seen in the rapid decrease of the DM mass in lockstep with the half-stellar--mass radius. However the tidal field is usually too weak at the start of compaction to affect the outer stars, hence the half-stellar--mass radius.

\bsp	
\label{lastpage}
\end{document}